\documentclass[aps,pra,reprint,superscriptaddress,floatfix]{revtex4-2}

\usepackage{graphicx}
\usepackage{amsmath}
\usepackage{braket}
\usepackage{xcolor}
\usepackage[nameinlink,capitalise]{cleveref}
\usepackage{notes2bib}
\bibnotesetup{
note-name = ,
use-sort-key = false
}

\begin{document}

\title{Universal thermodynamics of an SU($N$) Fermi-Hubbard Model}

% List of authors 
\author{Eduardo Ibarra-Garc\'ia-Padilla}
\email[]{eibarragp@rice.edu}
\author{Sohail Dasgupta}
\author{Hao-Tian Wei}
\affiliation{Rice Center for Quantum Materials, Department of Physics, Rice University, Houston, TX 77005, USA.}
\author{Shintaro Taie}
\author{Yoshiro Takahashi}
\affiliation{Department of Physics, Graduate School of Science, Kyoto University, Japan 606-8502}
\author{Richard T. Scalettar}
\affiliation{Department of Physics, University of California, Davis, CA 95616, USA.}
\author{Kaden R. A. Hazzard}
\affiliation{Rice Center for Quantum Materials, Department of Physics, Rice University, Houston, TX 77005, USA.}

\date{\today}

\begin{abstract}
The SU(2) symmetric Fermi-Hubbard model (FHM) plays an essential role in strongly correlated fermionic many-body systems. In the one particle per site and strongly interacting limit ${U/t \gg 1}$, it is effectively described by the Heisenberg Hamiltonian. In this limit, enlarging the spin and extending the typical SU(2) symmetry to  SU($N$) has been predicted to give exotic phases of matter in the ground state, with a complicated dependence on $N$. This raises the question of what --- if any --- are the finite-temperature signatures of these phases, especially in the currently experimentally relevant regime near or above the superexchange energy. We explore this question for thermodynamic observables by numerically calculating the thermodynamics of the SU($N$) FHM in the two-dimensional square lattice near densities of one particle per site, using determinant Quantum Monte Carlo and Numerical Linked Cluster Expansion. Interestingly, we find that for temperatures above the superexchange energy, where the correlation length is short, the energy, number of on-site pairs, and kinetic energy are universal functions of $N$. Although the physics in the regime studied is well beyond what can be captured by low-order high-temperature series, we show that an analytic description of the scaling is possible in terms of only one- and two-site calculations. 
\end{abstract}

\maketitle

\section{Introduction}

The Fermi-Hubbard model (FHM), in its original spin-$1/2$, SU(2) symmetric form \cite{Montorsi1992,Tasaki1998,Arovas2021,Bohrdt2021}, plays a central role in the understanding of strongly correlated fermionic many-body systems. This is in part because it is one of the simplest models that captures essential features of real materials, and in part because it exhibits a variety of canonical correlated phases of matter. In the two-dimensional (2D) square lattice, it displays a metal-to-insulator crossover as well as magnetic order, and it is widely studied in the context of $d$-wave superconductivity~\cite{Imada1998,White1989,Schafer2021,Qin2020,Qin2021}.

Its generalization, the SU($N$) FHM, features larger spins and enhanced symmetry, and it provides insight into important strongly correlated systems. First, it is a simple limit of multi-orbital models such as those used to describe transition metal oxides~\cite{Li1998,Tokura2000,Dagotto2001}, graphene's SU(4) spin-valley symmetry~\cite{Goerbig2011}, and twisted-bilayer graphene~\cite{Xu2018_Kekule,You2019,Natori2019,DaLiao2021,Liao2021,Chichinadze2021}. Second, the SU($N$) FHM is predicted to display a variety of interesting and exotic phases even in very special limits, such as: the conventional $N=2$ FHM, the $N=3$ FHM ~\cite{Yamashita2013,Titvinidze2011,Sotnikov2014,Sotnikov2015,Hafez2018,Hafez2019,Hafez2020,Nie2017,Honerkamp2004,Hofstetter_chapter}, the $N=4$ FHM at quarter filling \cite{Unukovych2021,Chen2016}, even values of $N$ at half-filling \cite{Wang2014,Zhou2014,Wang2019,Goubeva2017,Xu2019,Zhou2018,Zhou2017,Zhou2016,Ouyang2021} special $N \to \infty$ limits \cite{Read1983,Affleck1985,Bickers1987,Affleck1988,Auerbach2012}, 1D chains \cite{Yamashita1998,Assaraf1999,Buchta2007,Manmana2011,Bonnes2012,Messio2012,Xu2018}, and the Heisenberg limit for $N=3,4,5$~\cite{Tokura2000,Honerkamp2004,Hermele2009,Toth2010,Hermele2011,Nataf2014,Corboz2011,Bauer2012,Yamamoto2020,Keselman2020,Yao2021}. This richness is well illustrated by numerical studies of the Heisenberg limit, which describes the situation where the average number of particles per site is $\langle n \rangle=1$ and the interactions dominate the kinetic energy ($U\gg t$, with notation discussed below). Already in this simple limit and additionally in the simple 2D square lattice, the model is predicted to exhibit several phases of matter with novel and difficult-to-explain properties depending on the value of $N$. The dependence of the ground state order with $N$ does not follow a simple pattern. This raises the question of whether and how this complicated $N$-dependence manifests in the finite-temperature properties.

Although the SU($N$) FHM is a crude approximation to real materials, it has been realized to high precision by loading alkaline earth-like atoms (AEAs) into an optical lattice (OL). Fermionic AEAs (such as $^{173}$Yb and $^{87}$Sr) feature an almost perfect decoupling of the nuclear spin $I$ from the electronic structure in the ground state, which gives rise to SU(${N = 2I +1}$) symmetric interactions with deviations predicted to be of order $\mathcal{O}(10^{-9})$ \cite{Wu2006,Cazalilla2009,Gorshkov2010,Cazalilla2014,Stellmer2014}. For that reason, by selectively populating nuclear spin projection states $m_I$ of AEAs and loading them into an OL, experiments can engineer the SU($N$) FHM with $N$ tunable, from $2,3,\ldots,10$.

In recent years, experiments with $^{173}$Yb in OLs have probed the SU($N$) FHM's interesting physics: The Mott insulator state for SU(6) in three dimensions~\cite{Taie2012}, the equation of state for SU(3) and SU(6) in three dimensions~\cite{Hofrichter2016}, nearest-neighbor antiferromagnetic (AFM) correlations in an SU(4) system with a dimerized OL \cite{Ozawa2018}, nearest-neighbor SU(6) AFM correlations in OLs with uniform tunneling matrix elements in one, two, and three dimensions~\cite{Taie2020}, and recently a flavor-selective Mott insulator for SU(3)~\cite{Tusi2021}. Furthermore, employing quantum gas microscopy~\cite{Altman2021,Gross2017,Bloch2012,Parsons2016,Mazurenko2017,Koepsell2020,Ji2021} to discriminate finite temperature analogs of the variety of proposed ground states \cite{Hermele2009,Toth2010,Hermele2011,Wang2014,Zhou2014,Nataf2014,Corboz2011,Bauer2012} via direct observation of long-ranged correlations~\cite{Yamamoto2016,Taie2020,Schafer2020,Okuno2020} is expected to reveal a wealth of physics. All of these experimental efforts make an understanding of the 2D square lattice thermodynamics urgent.

In contrast with most previous work that focused on the Heisenberg limit, in this work we study the SU($N$) FHM at finite temperature and for a range of interaction parameters, including far from the Heisenberg limit, a regime that is both interesting and experimentally important. We calculate and analyze thermodynamic properties of the model as a function of $N$, the interaction strength $U$, and the temperature $T$. We numerically explore the evolution of the energy, number of on-site pairs%~\bibnote{The number of on-site pairs and the number of doublons are equivalent for $N=2$, but are not for $N>2$. While the number of on-site pairs is a measure of the interaction energy, with proportionality factor $U$, the number of double occupancies includes configurations where only two particles on a given site are allowed, but higher occupancies are not. Computing the number of double occupancies thus requires the calculation of density fluctuations, i.e. terms of the order $\langle n^x\rangle$ with $x \in [2,N]$, which are computationally more expensive and experimentally harder to access.}
~\bibnote{The number of on-site pairs and the double occupancy are equivalent in the SU(2) Fermi-Hubbard model due to Pauli exclusion principle. However, for $N>2$ this is not the case. The number of on-site pairs $\mathcal{D}$ counts the pairs of particles per site, and is what controls the interaction energy $U \mathcal{D}$, while the double occupancy most naturally refers to the probability of configurations with exactly two particles per site, or to summing probabilities of all configurations with two or more particles per site. Either one of these is distinct from the number of on-site pairs $\mathcal{D}$. Computing the number of double occupancies thus requires the calculation of density fluctuations, i.e. terms of the order $\langle n^x\rangle$ with $x \in [2,N]$, which are computationally more expensive and experimentally harder to access.}, and kinetic energy, as well as their derivatives in the 2D square lattice SU($N$) FHM at $1/N$ filling, i.e., one particle per site on average. 

Some of the quantities we compute, such as the number of on-site pairs are immediately measurable in experiment, while others such as the kinetic energy and total energy are of fundamental importance and may also become accessible. For example, Ref.~\cite{Nakamura2019} experimentally determined the energies of a Bose-Hubbard model. In that work the kinetic energy was measured by analyzing time-of-flight images and the interaction energy was measured by site-resolved high-resolution spectroscopy. These techniques can be also used for the FHM. Additionally, in a quantum gas microscope the number of on-site pairs can be spatially resolved by generalizing the technique used in Ref.~\cite{Koepsell2020PRL} to AEAs. This would require employing an optical (rather than magnetic) Stern-Gerlach technique to split the different spin flavors into different layers, followed by detection by single-site fluorescence. Additionally, access to total density fluctuations in a bilayer quantum gas microscope, as done in Refs.~\cite{Zhou2011,Hartke2020}, provides a route to realize thermometry without the need to comparison with numerical simulations.

We also present some selected results as a function of chemical potential $\mu$. Results are obtained using the determinant Quantum Monte Carlo (DQMC) and Numerical Linked Cluster Expansion (NLCE) methods. Here and throughout we set Boltzmann's constant to $k_B=1$.

Although the ground state has a complicated $N$-dependence, we find that for temperatures above the superexchange energy $T \gtrsim J=4t^2/U$, the energy, the number of on-site pairs, and the kinetic energy depend on $N$ in a particularly simple way, obeying a simple, analytic dependence on $N$.

Even though a simple scaling at very high temperatures would be unsurprising --- since a low-order high-temperature series expansion (HTSE) would be expected to be accurate and to produce analytic expressions that plausibly would show simple $N$-dependence --- such expansions are insufficient to explain our findings. The HTSE is accurate only for $T\gtrsim 4t$, while the universal scaling persists to temperatures $T\gtrsim 4t^2/U$ that are much lower when $U\gg t$. At such temperatures the HTSE is not only inaccurate, but diverges. 

Despite the failure of the HTSE to fully explain the observations, a simple explanation is possible by recognizing that correlations are short-ranged in this temperature regime. We show that in this limit, the second order NLCE accurately reproduces the results and the $N$ scaling relation. Furthermore, under controlled approximations in the $J \ll T \ll U$ regime one can analytically evaluate the pertinent contributions based on the NLCE, and with this explain the observed universal scaling with $N$ to zeroth order in $\beta J$. This demonstrates the utility of the NLCE framework for analytic calculations, beyond its typical application in numerical calculations. These observations show that the one- and two-site correlations control the physics deep in this regime.

The remainder of this paper is organized as follows: Section \ref{sec::Model_Methods} presents the SU($N$) Hubbard Hamiltonian, defines the observables we consider, and presents details of the numerical and analytical methods used. Section \ref{sec::Results} presents the main results, and Section \ref{sec::Conclusion} concludes.

\section{Model and Methods}\label{sec::Model_Methods} 

\subsection{The SU($N$) Hubbard Hamiltonian and observables}\label{sec::Hamiltonian}

The SU($N$) FHM is defined by the grand canonical Hamiltonian
\begin{equation}\label{eq:Hubbard_N1}
H = -t \sum_{\langle i,j \rangle, \sigma} \left( c_{i \sigma}^\dagger c_{j \sigma}^{\phantom{\dagger}} + \mathrm{h.c.} \right) + \frac{U}{2} \sum_{i,\sigma \neq \tau} n_{i \sigma} n_{i \tau} - \mu \sum_{i,\sigma} n_{i \sigma},
\end{equation} 
where $c_{i \sigma}^\dagger$ ($c_{i \sigma}^{\phantom{\dagger}} $) is the creation (annihilation) operator for a fermion with spin flavor $\sigma = 1,2,...,N$ on site $i = 1,2,...,N_s$ in a 2D square lattice, $N_s$ denotes the number of lattice sites, $n_{i \sigma} = c_{i \sigma}^\dagger c_{i \sigma}^{\phantom{\dagger}}$ is the number operator for flavor $\sigma$, $t$ is the nearest-neighbor hopping amplitude, $U$ is the interaction strength, and $\mu$ is the chemical potential that controls the fermion density.

We are interested in thermodynamic quantities such as the number of on-site pairs per site
\begin{equation}
    \mathcal{D} = \frac{1}{N_s} \sum_i \left[\frac{1}{2}\sum_{\sigma \neq \tau} \langle n_{i \sigma} n_{i \tau} \rangle\right],
\end{equation}
the kinetic energy per site
\begin{equation}
    K = \frac{1}{N_s} \langle -t \sum_{\langle i,j \rangle, \sigma} \left( c_{i \sigma}^\dagger c_{j \sigma}^{\phantom{\dagger}} + c_{j \sigma}^\dagger c_{i \sigma}^{\phantom{\dagger}} \right)\rangle,
\end{equation}
the energy per site $E= \langle H/N_s + \mu n \rangle$ (where $n=(1/N_s)\sum_{i,\sigma} n_{i \sigma} $), and the entropy $S$. We present these observables and the derivatives $dE/dT$, $dK/dT$, and $U d\mathcal{D}/dT$ as functions of $T/t$ for different values of the interaction strength $U/t$ either as a function of chemical potential $\mu/t$ or at fixed density $\langle n \rangle =(1/N_s) \sum_{i,\sigma} \langle n_{i \sigma} \rangle =1$. We also show the compressibility $\kappa = d\langle n \rangle/d\mu$ as a function of $\mu$ for various $T/t$, $U/t$, and $N$. These observables provide valuable knowledge about the physics: the number of on-site pairs is a useful measure of the Mott insulating nature of the system, the kinetic energy of its spatial coherence, and the entropy and specific heat provide information about the temperature scales at which various degrees of freedom cease to fluctuate.

\subsection{Numerical methods}\label{sec::num_methods}

To calculate the thermodynamic observables, we employ two numerical techniques, DQMC~\cite{Blankenbecler1981,Sorella1989} and NLCE~\cite{rigol2006numerical,tang2013short}, which have complementary strengths, and compare in some cases with low-order analytic HTSE and the non-interacting limit. The DQMC and NLCE are often the numerical methods of choice for the SU(2) FHM in the finite-temperature regime studied in ultracold matter~\cite{Hart2015,Greif2013,Cheuk2016,Brown2017,Brown2018}, and we use our extensions of these methods to SU($N$) systems~\cite{Taie2020}. Generally speaking, the DQMC will perform best at weak to intermediate interactions, while the NLCE performs best at strong interactions; we present both methods where both are viable.

\subsubsection{Determinant Quantum Monte Carlo (DQMC)}

Averages of the thermal equilibrium observables are evaluated with DQMC on $6 \times 6$ lattices by introducing $N(N-1)/2$ auxiliary Hubbard-Stratonovich fields, one for each interaction term \bibnote{Previous work applied Determinant Quantum Monte Carlo for the SU($2N$) Fermi-Hubbard model at half-filling, i.e. $\langle n \rangle=N/2$, using an alternative Hubbard-Stratonovitch decomposition. This alternative decomposition is free of the sign problem at half-fillling for SU($2N$)~\cite{Assaad1998,Wang2014,Zhou2014,Assaad2005}}. In this method, the inverse temperature $\beta$ is discretized in steps of $\Delta \tau$ with a Trotter step $\Delta \tau = 0.05/t$ for $U/t=4,8$ and $\Delta \tau = 0.04/t$ for $U/t=12$. In order obtain accurate results, we obtain DQMC data for $40-60$ different random seeds for $T/t \leq 4$ and for $2-10$ different random seeds for $T/t>4$. For each Monte Carlo trajectory we perform $2000$ warm up sweeps and $8000$ sweeps for measurements~\bibnote{A sweep updates all the auxiliary fields at every lattice site and imaginary time slice.}. In addition, the number of global moves per sweep to mitigate possible ergodicity issues~\cite{Scalettar1991} is set to 2 for $U/t=4,8$ and to 4 for $U/t=12$. These global moves update, at a given lattice site, all the imaginary time slices that couple two spin flavors. DQMC results presented in the paper are obtained by computing the weighted average and weighted standard error of the mean of the results obtained by using different random seeds. We use the inverse squared error of each measurement as their weight. Results obtained using a uniform weight for all measurements yield consistent results but with larger error bars ($\sim$2-4 times larger). Estimates of systematic errors are obtained for $N=6$ at $U/t=12$ (Trotter) and $N=6$ at $U/t=4$ (finite-size), where they are expected to be worst. We estimate the Trotter error by comparing the results obtained with $\Delta \tau = 0.04/t$ and $\Delta \tau = 0.05/t$. Their difference is below $4\%$ for all observables of interest at $T/t=0.5$. This discretization error is even smaller at higher temperatures and for the other two values of $U/t$ considered. Finite-size errors are estimated by comparing results for different thermodynamic quantities in $4 \times 4$ and $6 \times 6$ lattices. At $T/t=0.5$ their differences are $\lesssim 6.5\%$ for $U/t=4$ and $\lesssim 5.7\%$ for $U/t=12$.

\subsubsection{Calculation of specific heat and entropy in DQMC}
For DQMC data we evaluate the specific heat and entropy in two ways. In the first approach, we numerically differentiate the energy to obtain the specific heat (see footnote~\bibnote{We used the three-point differentiation rule
\begin{align*}\label{eq:Deriv}
f'(x) =& \left[\frac{x_i - x_{i+1}}{(x_{i-1} - x_i)(x_{i-1} - x_{i+1})} \right] f(x_{i-1}) \\ 
 + &\left[\frac{2x_i - x_{i-1} - x_{i+1}}{(x_i - x_{i-1})(x_i - x_{i+1})} \right] f(x_{i}) \\
 + &\left[\frac{ x_i - x_{i-1} }{(x_{i+1} - x_{i-1})(x_{i+1} - x_i)} \right] f(x_{i+1}),
\end{align*}
with error $\mathcal{O}(h^2)$ where $h$ is the maximum spacing of adjacent $x_i$. Statistical error bars are obtained by error propagation.} for details on the differentiation procedure), and we compute the entropy by integrating ${dS = dQ/T = C/T \, dT}$, with ${C = dE /dT}$ the specific heat. Integrating by parts, $S$ can be rewritten in terms of the energy $E$,
\begin{equation}\label{eq:S_DQMC}
S(T) = S_\infty + \frac{E(T)}{T} - \int_T^{\infty} \frac{E(T')}{{T'}^2} dT',
\end{equation}
where $S_\infty$ is the entropy at fixed density in the limit when $T \to \infty$ (see Appendix~\ref{App::S_inf} for more details).

%The DQMC becomes unreliable at $T$ much below the superexchange scale $J$. 
The DQMC starts becoming unreliable at $T$ below the superexchange scale $J$. In this regime the statistical noise increases due to the sign problem, severely limiting calculations. In addition to presenting the DQMC calculations directly, we also show results obtained from fitting and from differentiating this smooth fit function, which can reduce the noise at the cost of potentially biasing the data. For the energy, we fit to the simple functional form \cite{Paiva2001,McMahan1998},
\begin{equation}\label{eq::E_fit}
    E(T) = E(0) + \sum_{k=1}^M c_k e^{- \beta k \Delta},
\end{equation}
with fitting parameters $c_k$, $\Delta$, and $E(0)$. The number of parameters $c_k$, $M$, is chosen to be around 6-12 (slightly less than one-third of the data points to be fit), which is similar to Refs.~\cite{Paiva2001,McMahan1998}. We smooth the 10 lowest temperature data points using a moving average with a 3-point window fitted with a local first order polynomial (Savitzky–Golay filter). Then the data is fit with Eq.~(\ref{eq::E_fit}), by choosing the fitting parameters that minimize
\begin{align}\label{eq::chi}
   \Xi^2 = \frac{1}{N_p +1} \left( \sum_{n=1}^{N_p} \bigg[E(T_n) - E_n\bigg]^2  + \bigg[S_\infty - \sum_{k=1}^M \frac{c_k}{k \Delta} \bigg]^2 \right),
\end{align}
where $N_p$ is the number of data points, and $E_n$ is the DQMC energy at $T_n$. The first term ensures a good fit of the data, while the second term regularizes the fit and ensures that $S\to 0$ as $T\to 0$ by enforcing the constraint $S_\infty = \int_{0}^{\infty} \frac{C(T')}{T'}dT' = \sum_{k=1}^M \frac{c_k}{k \Delta}$~\bibnote{An equal weight on regularization and fitting terms is enough to ensure that $S\to 0$ as $T\to 0$ with an error $\lesssim 10^{-2}$ for all $N$ and $U/t$.}. A similar procedure is used to obtain fits for the number of on-site pairs and the kinetic energy: Each dataset is fit using the same form as Eq.~(\ref{eq::E_fit}), subject to the constraint that the derivative of their sum obeys the specific heat sum rule.

Results obtained from fitting remove the noise providing smooth guides to the eye. By construction they also satisfy important physical features such as sum rules. However, fitting necessarily biases the results, and should be interpreted with caution. Care is especially warranted in the high-noise regimes (mainly occurring in the derivative data at the lowest temperatures presented) where the fits are used to extrapolate the data. Nevertheless, the fits suggest interesting features and trends that may help guide future low-temperature calculations and experiments.

\subsubsection{Numerical Linked Cluster Expansion (NLCE)}
Thermodynamic observables are computed using a fifth-order site expansion NLCE. We briefly derive and present this algorithm, which is reviewed in Ref.~\cite{tang2013short}. Extensive properties in a lattice are evaluated by performing a weighted sum of their value in all possible clusters $c$ embeddable in the lattice; specifically,
\begin{equation}\label{eq:nlce_sum}
P(\mathcal{L})/N_s = \sum_{c \in \mathcal{L}} L(c) W_P(c)
\end{equation}
where $P(\mathcal{L})$ is the property evaluated on the entire lattice $\mathcal{L}$, $N_s$ is the number of lattice sites, $L(c)$ is the number of ways that the cluster $c$ can be embedded in the lattice (up to translation invariance), and $W_P(c)$ is defined as
\begin{equation}
W_P(c) = P(c) - \sum_{s\subset c}W_P(s).
\end{equation}
Eq.~\eqref{eq:nlce_sum} follows directly from the definition of the $W_P(c)$. Eq.~\eqref{eq:nlce_sum} is an infinite sum over all clusters, and the key idea of the NLCE is to truncate this sum to clusters of small size (different variants use different measures of size) and evaluate properties on each cluster using exact diagonalization (ED). Here we truncate the sum over clusters based on the number of sites, performing calculations up to five site clusters, which shows good convergence (see Appendix~\ref{App::ed_compare}).

The Hilbert space dimension increases rapidly with $N$, limiting the size of clusters that can be included in the expansion, and we use multiple methods to reduce the computational cost in order to reach 5-site clusters for SU(6). The most straightforward is to account for the SU($N$) symmetry, in particular its abelian symmetries (the $N$ conserved flavor numbers) and the flavor permutation symmetry. Additionally, for $N=6$, we truncate the Hilbert space in the Fock basis using two criteria: (1) We include only basis states with a number of particles below a cutoff value (chosen to be 6, which is one larger than the number of sites in the largest cluster), and (2) We include only basis states whose interactions energy is less than a cutoff value (chosen to be $3U$). These choices provide highly accurate (several decimal places) results over the temperature and density ranges of interest in this paper, though at high temperatures or densities they can break down. Appendix~\ref{App::nlce_trunc} provides details of these  truncations and the calculations' convergence~\bibnote{The idea of truncating the Hilbert space is not new, see for example, Ref.~\cite{Bhattaram2019}, where authors limit the thermal averages to a fraction of low-lying states and use Lanczos algorithm to reduce the computational cost of full diagonalization.}.

The NLCE is much more accurate than an exact diagonalization (ED) that uses the same number of (or even more) sites. At all temperatures considered, the 5-site NLCE calculations are dramatically more accurate than $3\times 2$ ED calculations in either periodic or open boundary conditions to quite low temperatures. In fact, at least for temperatures where the NLCE is convergent and the density $\langle n \rangle=1$ case that is our main focus, even a 2-site NLCE calculation outperforms the $3 \times 2$ ED, despite requiring enormously less computational resources. We note that this is, to our knowledge, the first application of NLCE to the SU($N$) FHM. The convergence with expansion order and comparisons with ED are discussed in Appendix~\ref{App::ed_compare}. 

The NLCE self-diagnoses its accuracy, with converged results expected when adjacent orders give nearly the same answer. Results in the main text are presented for the highest order computed and the NLCE data is cutoff at temperatures where the three highest consecutive orders deviate more than 2\%. 

\subsubsection{Low-order high-temperature series expansions (HTSE)}

It is useful to compare computed observables against simple analytic zeroth and second order high temperature series in $t/T$~\cite{Hazzard2012}. The region of validity of the HTSE to any order is $T \gtrsim t$, yielding unphysical results for $T \lesssim t$. 

\section{Results}\label{sec::Results}

This section presents our main results, the calculation of several thermodynamic observables and analysis of features observed in them, especially their striking universal $N$-dependence. Specifically, we calculate the number of on-site pairs $\mathcal{D}$, the kinetic energy $K$, the energy $E$, the entropy $S$, the specific heat $C$ and the contributions to it from the interaction and kinetic energies $Ud \mathcal{D}/dT$, and $dK/dT$, respectively, all defined previously. Mostly we focus results at a density $\langle n \rangle =1$, but some results are also presented as a function of chemical potential $\mu/t$ which causes the density to vary.

This section is organized as follows: Section \ref{sub::rhovsmu} presents the $\mu/t$-dependence of $\langle n \rangle$, $\mathcal{D}$, the compressibility $\kappa = \partial \langle n \rangle/ \partial \mu$, and the determinantal sign. The following subsections present the $U/t$, $T/t$, and $N$ dependence of $\mathcal{D}$ (Section~\ref{sub::DvsT}), $K$ (Section~\ref{sub::KvsT}), and $E$ (Section~\ref{sub::EvsT}). Section \ref{sub::universalN} presents the scaling collapse demonstrating the universal $N$-dependence of $E$, $\mathcal{D}$, and $K$. Section \ref{sub::DerivativevsT} presents the temperature derivatives. Finally Section \ref{sub::SvsT} presents the $U/t$, $T/t$, and $N$ dependence of $S$. Results in Sections \ref{sub::DvsT} to \ref{sub::SvsT} are all at unit density.

\subsection{Density, number of on-site pairs, compressibility, and determinantal sign dependence on chemical potential $\mu/t$}\label{sub::rhovsmu}

\begin{figure*}[tbp!]
\includegraphics[width=\linewidth]{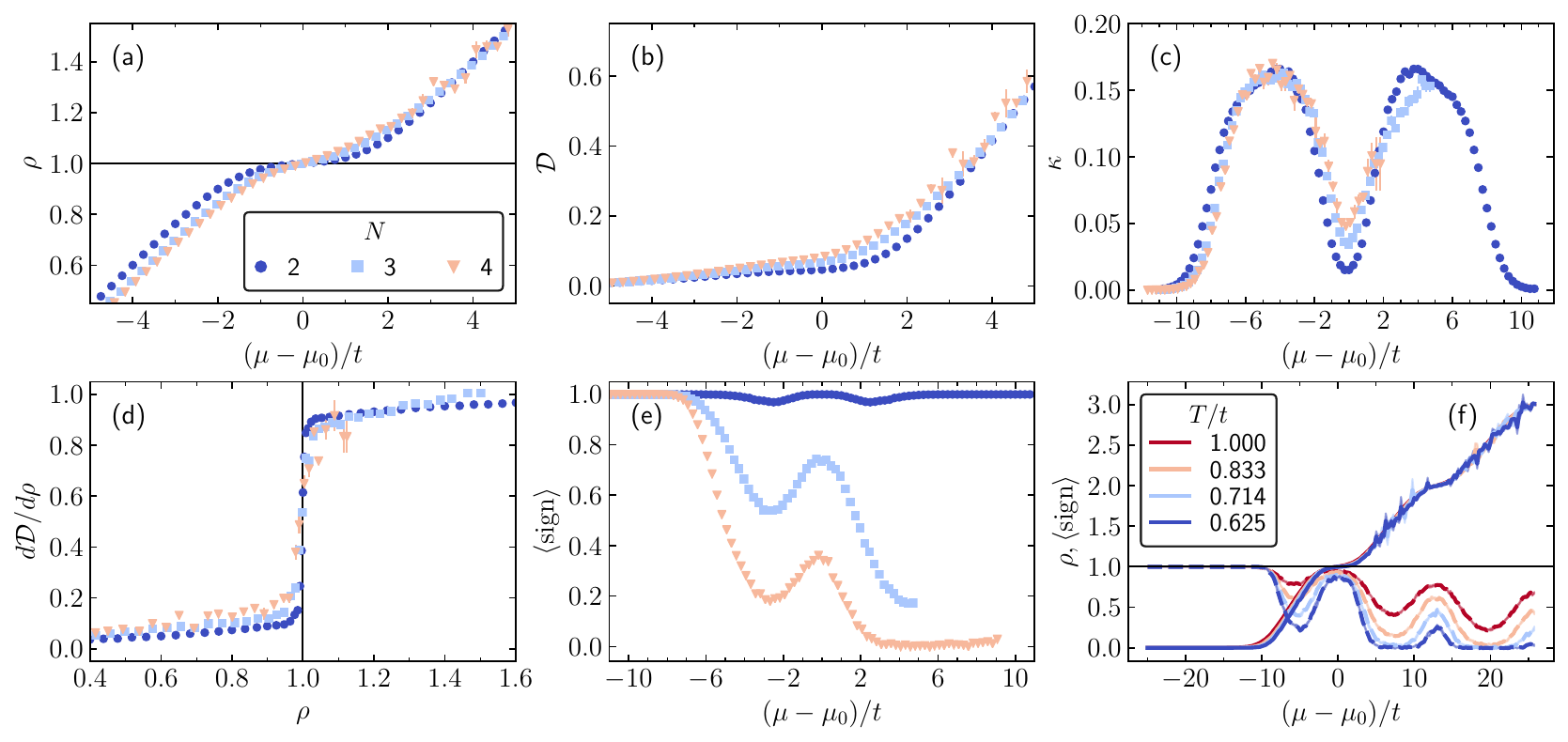}
\caption{\textbf{Density's, number of on-site pairs', compressibility's, and determinantal sign's dependence on chemical potential.} Panels (a-e) compare observables for $N=2,3,4$ for $U/t=8$ at $T/t=0.5$ as functions of the chemical potential $(\mu -\mu_0)/t$, where $\rho(\mu_0) = 1$. (a) Density. There is a clear softening of the Mott plateau as $N$ increases. (b) Number of on-site pairs. (c) Compressibility. (d) Derivative of the number of on-site pairs with respect to the density as a function of density. (e) Average sign. (f) Density (solid) and average sign (dashed) vs $(\mu -\mu_0)/t$ for different values of $T/t$ for $N=6$ at $U/t=12$. Shaded regions correspond to error bars.}\label{fig::rhosignvsmu} 
\end{figure*}

\begin{figure}[hbtp!]
\includegraphics[width=0.72\linewidth]{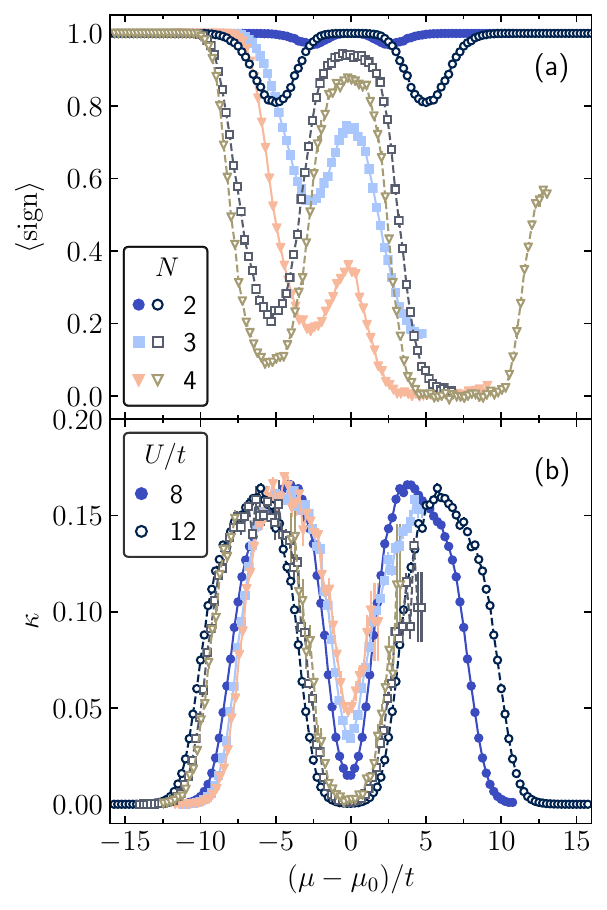}
\caption{\textbf{Determinantal signs' and compressibility's dependence on interaction strength.} (a) Average sign (b) Compressibility vs $(\mu -\mu_0)/t$, where $\rho(\mu_0) = 1$ for $U/t=8$ (full markers) and $U/t=12$ (open markers) for $N=2,3,4$ at $T/t=0.5$.}\label{fig::Udependence} 
\end{figure}

Figs.~\ref{fig::rhosignvsmu}(a-b) show the dependence of density $\rho = \langle n \rangle$ and number of on-site pairs $\mathcal D$ on the chemical potential. These are particularly important quantities because typical experiments on ultracold atoms use smooth traps, and the $\mu$-dependence of the observables is related to their spatial dependence by the local density approximation~\cite{PethickSmith}. These are also among the most straightforward observables to measure, and have been explored experimentally as a function of $U/t$, $N$, $\mu/t$, and $T/t$ in Refs.~\cite{Taie2012,Hofrichter2016}. 

The density as a function of chemical potential shows a Mott plateau -- a region of $\mu$ over which the density is nearly constant -- when the temperature is $T\lesssim U$, as shown in Fig.~\ref{fig::rhosignvsmu}(a), signaling the incompressible and insulating nature of the system. At fixed temperature, the Mott region becomes less sharply defined as $N$ increases. This is expected, as increasing $N$ allows for more density fluctuations at a given energy and thus a more compressible system at a fixed temperature [as corroborated by Fig.~\ref{fig::rhosignvsmu}(c)]. This behavior is also observed for $U/t=12$ at the same temperature (not shown here as to not overcrowd Fig. 1). The general trend is already seen in the second order HTSE~\cite{Hazzard2012} and was observed experimentally in Ref.~\cite{Hofrichter2016}.

Although the Mott plateau softens with increasing $N$, appearing only as a subtle shoulder for $N=4$ at $U/t=8$ and $T/t=0.5$, if one plots $d\mathcal{D}/d\rho$ as a function of $\rho$, there is a quite sharp and clear signature of the Mott plateau for all cases, as shown in Fig.~\ref{fig::rhosignvsmu}(d). 

We also show the average determinantal sign, which characterizes the sign problem, one of the fundamental limitations to quantum Monte Carlo calculations of interacting fermions~\cite{Iglovikov2015,Bautroni1990,Bautroni1993}. For the type of Hubbard-Stratonovich decomposition used in the current study for DQMC, we find the average sign decreases (i.e. the sign problem worsens) overall as $N$ increases and as the temperature is lowered [see Figs.~\ref{fig::rhosignvsmu}(e-f)]. On top of this, the sign problem is worse for the metallic phase than the Mott insulating phase at a fixed temperature. Figure~\ref{fig::Udependence}(a) shows that at fixed $T/t$, increasing $U/t$ worsens the sign problem in the metal, but improves it in the insulator in the currently studied temperature regime. The $N=2$ case is free of the sign problem at half-filling, and therefore $\langle \mathrm{sign} \rangle =1$ when $\langle n \rangle =1$ for all values of $U/t$.

\begin{figure*}[htbp!]
    \centering
    \includegraphics[width=\linewidth]{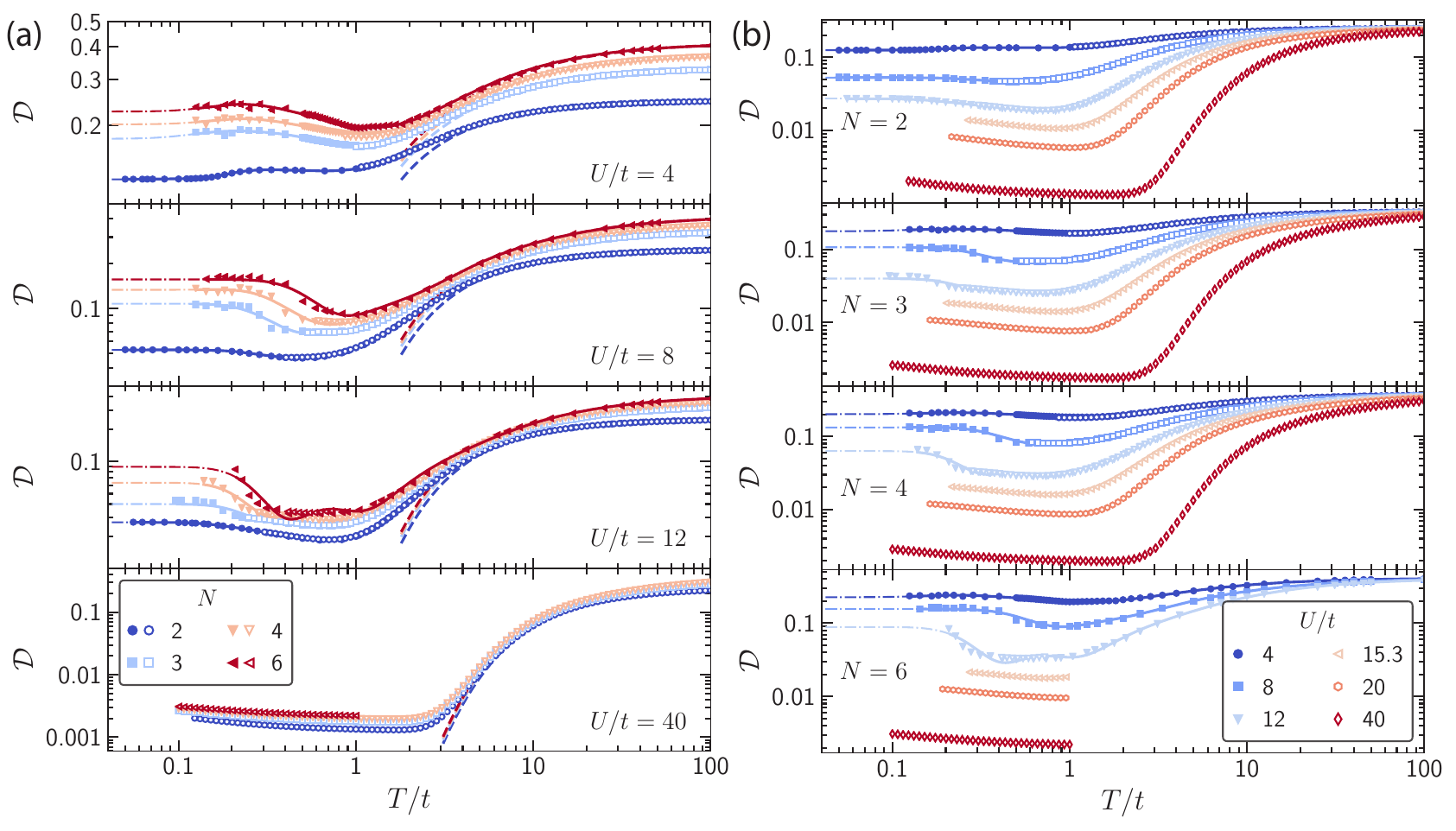}
    \caption{\textbf{Number of on-site pairs $\mathcal D$ versus temperature.} (a) Each panel compares $\mathcal D$ for $N=2,3,4,6$ for a fixed $U/t$ at $\langle n \rangle =1$. (b) Each panel compares $\mathcal D$ for $U/t=4,8,12,15.3,20,40$ for a fixed $N$. Solid markers are DQMC, open markers are NLCE, dashed lines are the zeroth order HTSE, and solid lines are the fits of  Eq.~\eqref{eq::E_fit} to the DQMC data down to the lowest $T_n$ point. Thinner dashdotted lines come from the fit in the extrapolated regime $T < \min(T_n)$, where $\{T_n\}$ is the dataset of temperatures where DQMC results are obtained.}\label{fig::Doublons}
\end{figure*}

Finally, the $U/t$ dependence of $\kappa$ for different $N$ is displayed in Fig.~\ref{fig::Udependence}(b). As the $U/t$ increases, the system becomes more incompressible where $\langle n \rangle =1$, highlighting the insulating nature of the system. Our results are in agreement with qualitative trends identified in previous dynamical mean-field theory (DMFT) results~\cite{Lee2018}.

\subsection{Number of on-site pairs at unit density: dependence on $U/t$, $T/t$, and $N$}\label{sub::DvsT}

The number of on-site pairs $\mathcal{D}$ decreases as temperature is lowered, almost always followed by an increase at the lowest temperatures. These features show clear trends with $U/t$ and $N$ as shown in~\cref{fig::Doublons}. The trends with $U/t$ are that, as the temperature is lowered, (1) $\mathcal{D}$ is suppressed from its high-temperature value at a temperature scale $T\sim U$, and (2) $\mathcal{D}$ increases at a much lower temperature that decreases with increasing $U/t$. Also, as expected, overall larger $U/t$ leads to smaller $\mathcal D$, most strongly in the temperature window between the two features discussed previously. The trends with $N$ are  also clear: (1) as $N$ increases, $\mathcal D$ increases, (2) the temperature at which the low-$T$ increase of $\mathcal D$ occurs is roughly independent from $N$ except for $U/t=8$, where is higher for larger $N$, and (3) the increase of $\mathcal D$ as the temperature is decreased through the lower temperature feature is larger for larger $N$. For sufficiently large $U/t$, the dependence on $N$ is weaker, as shown in Fig.~\ref{fig::Doublons}(a). These features will be explained below.

Although the temperatures are not extremely low, $T\gtrsim 0.1t$, the qualitative features are not captured with a low-order HTSE, as shown in Fig.~\ref{fig::Doublons}(a), which diverges from the true results at $T/t\sim 3$ or larger. Furthermore, for the temperature regions where NLCE and DQMC are well converged, both methods are in good agreement, supporting the validity and convergence of the different approaches.

The $T$-, $N$-, and $U$-dependence of $\mathcal{D}$ can be qualitatively understood by considering the two-site, two-particle (TSTP) system, which was employed to understand similar features in the $N=2$ anisotropic lattice calculations of Ref.~\cite{IbarraGarciaPadilla2020}. We begin by describing the $T$ dependence. For $T \gtrsim U$, eigenstates with energy $\sim U$ and a large fraction of double occupancies are occupied. As the temperature is lowered below $U$, the eigenstates dominated by one-particle-per-site configurations have the largest Boltzmann weight and have small admixture of doublons, thus explaining the high-temperature decrease of $\mathcal{D}$ upon cooling. The more interesting low-temperature increase of $\mathcal{D}$ is explained by considering the physics in this sector dominated by one-particle-per-site configurations. In this sector, these low-energy eigenstates are approximately ``SU(2) singlets"  on the two sites [$\propto(\vert \sigma,\tau \rangle - \vert \tau, \sigma \rangle)$ with $\sigma \ne \tau$] or ``SU(2) triplets" [$\propto(\vert \sigma,\tau \rangle +\vert \tau, \sigma \rangle)$ where $\tau$ and $\sigma$ may be equal]. The ``singlet" states include an admixture $\propto (t/U)^2$ of doublons, which allows for some delocalization, lowering the kinetic energy and therefore lowering the energy of singlet states relative to the triplet ones, which have no admixture of doublons. Therefore, as the temperature is lowered below the energy scale splitting the singlet and triplet configurations, the system populates the singlet states and the number of double occupancies increases until it saturates. This low-temperature population of SU(2) singlets also leads to the antiferromagnetic correlations observed in Ref.~\cite{Taie2020}.

%The dependence of $\mathcal{D}$ on $N$ can also be understood in this picture, by considering the number of available ways to form double occupancies. Since the number of possible configurations of $m$  particles on a single site is $\binom{N}{m}$, the number of double occupancies is enhanced for $N>2$ for all values of the interaction strength and temperature due to thermal fluctuations and quantum fluctuations (tunneling)~\bibnote[Dinf]{In the $T \to \infty$ limit, at $\langle n \rangle =1$, the number of on-site pairs is given by
The dependence of $\mathcal{D}$ on $N$ can also be understood in this picture, by considering the number of available ways to form an on-site pair. Since the number of possible configurations of $m$  particles on a single site is $\binom{N}{m}$, the number of on-site pairs is enhanced for $N>2$ for all values of the interaction strength and temperature due to thermal fluctuations and quantum fluctuations (tunneling)~\bibnote[Dinf]{In the $T \to \infty$ limit, at $\langle n \rangle =1$, the number of on-site pairs is given by
\begin{equation*}
\mathcal{D}_\infty = \frac{1}{2} \sum_{\sigma \neq \tau} \langle n_\sigma \rangle \langle n_\tau \rangle = \binom{N}{2} \frac{1}{N^2}= \frac{1}{2}\left(1-\frac{1}{N}\right),
\end{equation*}
where we used that $\langle n_\sigma \rangle = \langle n \rangle /N$ because of the SU($N$) symmetry.}.

\begin{figure}[tbp]
\includegraphics[width=0.8\linewidth]{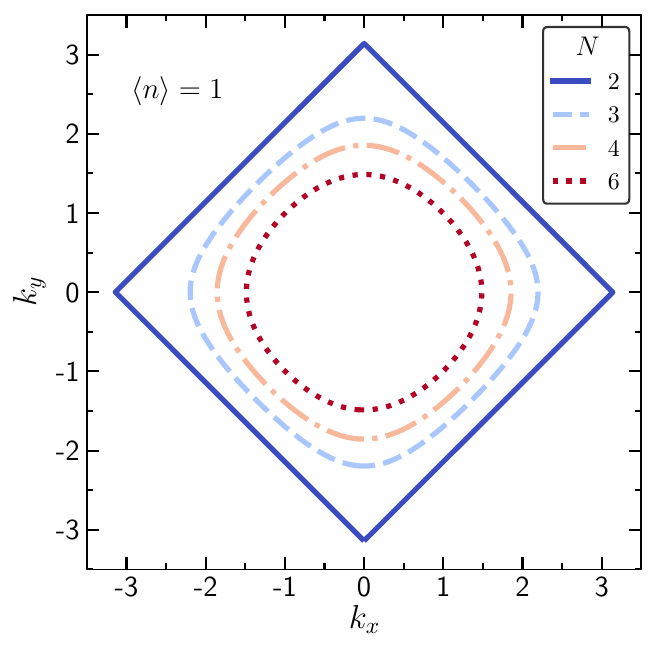}
\caption{\textbf{Fermi surface for $N=2,3,4,6$ in the  $U=0$ 2D square lattice at $\langle n \rangle =1$.}}\label{fig::FS} 
\end{figure}

This argument provides an understanding of the overall trends of $\mathcal{D}$ with $T$ and $N$, but the $U=4t$, $N=2$ curve is worth further consideration as the sole curve that does not show the low-temperature increase. The reason for this is not obvious: that a low-temperature rise would be smaller for small $N$ is explained above, but that it actually turns from a rise to a decrease is not. We note that this is likely a special feature of not only $N=2$ and small $U/t$, but also 2D systems, as when the system is perturbed away from 2D a low-temperature rise in $\mathcal D$ appears \cite{IbarraGarciaPadilla2020}. As such, it is natural to conjecture it is related to Fermi surface nesting (see Fig.~\ref{fig::FS}), which is most important at small $U/t$, and which is perfect only for $N=2$. 

\subsection{Kinetic energy at unit density: dependence on $U/t$, $T/t$, and $N$}\label{sub::KvsT}

\begin{figure}[tbp]
\includegraphics[width=\linewidth]{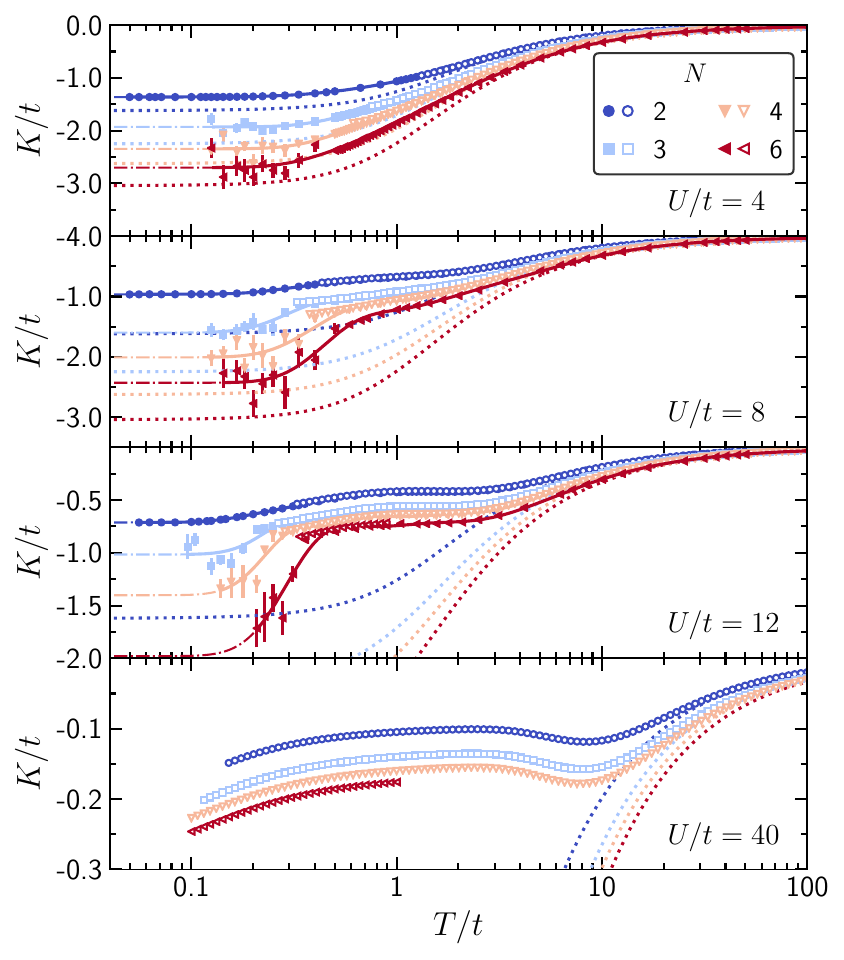}
\caption{\textbf{Kinetic energy vs temperature}. Each panel compares $K$ for $N=2,3,4,6$ for a fixed $U/t$ at $\langle n \rangle =1$. Solid markers are DQMC, open markers are NLCE, dotted lines correspond to the non-interacting limit, and solid lines are the fits of Eq.~\eqref{eq::E_fit} to the DQMC data down to the lowest $T_n$ point. Thinner dashdotted lines come from the fit in the extrapolated regime $T < \min(T_n)$, where $\{T_n\}$ is the dataset of temperatures where DQMC results are obtained.}\label{fig::KvsT_N} 
\end{figure}

The kinetic energy $K$ shows features at similar energy scales as $\mathcal D$, as shown in Fig.~\ref{fig::KvsT_N}. At high temperatures, the kinetic energy vanishes, and decreases as the temperature is lowered, in close agreement with the non-interacting calculations (described momentarily) until $T\sim U$. At $T\lesssim U$ the kinetic energy becomes smaller in magnitude than the non-interacting limit by an amount that increases with $U$. Finally, at the lower temperature scale on which $\mathcal{D}$ rises again, the kinetic energy drops significantly, signaling the same tunneling processes that create doublons, explained at the end of Sec.~\ref{sub::DvsT}.

The non-interacting limit's behavior is straightforward to understand: for $N=2$ and $\langle n \rangle=1$, the Fermi surface is a perfect square (Fig.~\ref{fig::FS}), and as $N$ is increased this shrinks and becomes circular. Thus the kinetic energy decreases as $N$ increases. Fig.~\ref{fig::KvsT_N} shows the non-interacting limit results (dotted lined)
\begin{equation}
    K = \frac{1}{(2\pi)^2}\int_{\text{BZ}} \!  \,  \frac{\epsilon_{\vec{k}} d^2k}{e^{\beta(\epsilon_{\vec{k}}-\mu)}+1},
\end{equation}
where the integral is over the Brillouin zone and ${\epsilon_{\vec{k}}=-2t(\cos k_x + \cos k_y)}$ is the non-interacting dispersion (setting the lattice constant to unity). The chemical potential $\mu$ is determined numerically to give $\langle n \rangle= 1/(2\pi)^2 \int_{\text{BZ}}\! d^2k/(e^{\beta (\epsilon_{\vec{k}}-\mu)}+1) =1$.

\subsection{Total energy at unit density: dependence on $U/t$, $T/t$, and $N$}\label{sub::EvsT}

\begin{figure}[tbp]
\includegraphics[width=\linewidth]{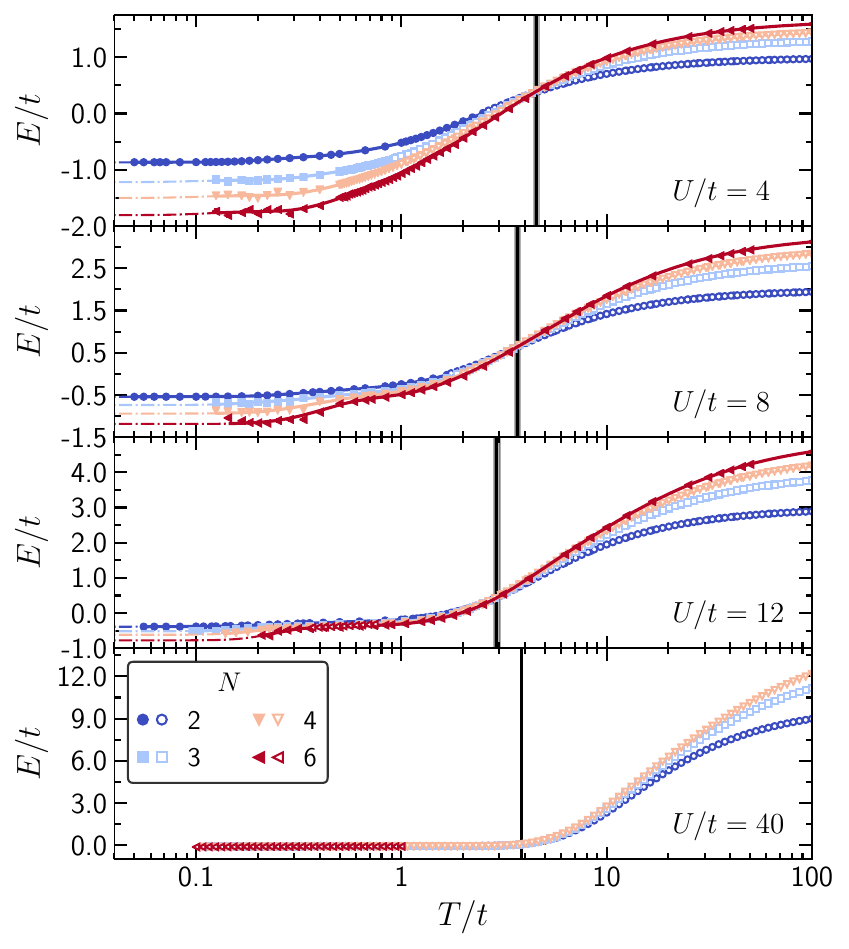}
\caption{\textbf{Energy vs temperature.} Each panel compares $E$ for $N=2,3,4,6$ for a fixed $U/t$ at $\langle n \rangle =1$. Solid markers are DQMC, open markers are NLCE, solid lines are the fits of Eq.~\eqref{eq::E_fit} to the DQMC data down to the lowest $T_n$ point. Thinner dashdotted lines come from the fit in the extrapolated regime $T < \min(T_n)$, where $\{T_n\}$ is the dataset of temperatures where DQMC results are obtained. Vertical regions in black indicate the temperature window where the different $N$ curves intersect.}\label{fig::EvsT_N} 
\end{figure}

The total energy $E= U\mathcal D + K$ (Fig.~\ref{fig::EvsT_N}) shows features simply related to $D$ and $K$.

\begin{figure}[tbp]
\includegraphics[width=\linewidth]{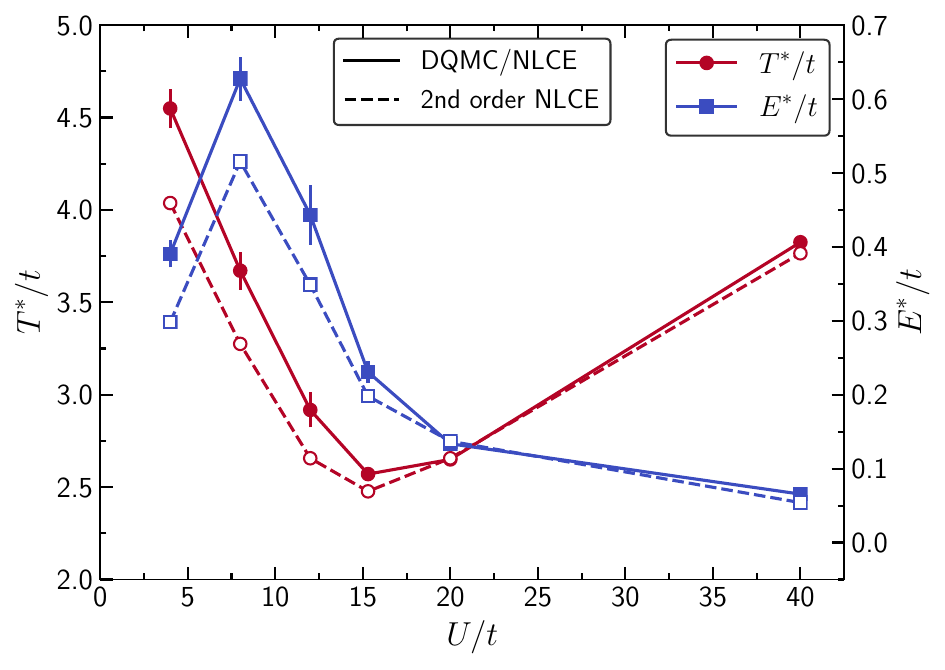}
\caption{\textbf{Interaction dependence of the energy crossing}. Temperature (red circles) and energy (blue squares) where the curves for different $N$ cross in Fig.~\ref{fig::EvsT_N}. Error bars correspond to the width of the crossings. Dashed lines correspond to the second order NLCE.}\label{fig::Crossing} 
\end{figure}

However, a new and surprising feature appears in $E$: the curves for different $N$ cross at a temperature and energy $(T^*,E^*)$ with $t < T^* < U$. Fig.~\ref{fig::Crossing} shows that $T^*$ first decreases then increases as a function of $U/t$, while $E^*$ first increases, then decreases. In Section~\ref{sub::universalN}, we will see that this crossing is a consequence of an even more dramatic phenomena --- a universal collapse upon rescaling over a broad temperature range.

The existence and qualitative trends of the crossing can be understood again by the system with two sites and two particles (TSTP), and can be quite accurately described by the second order NLCE, whose only inputs are the one- and two-site exact diagonalization calculations (a point we will revisit in Sec.~\ref{sub::universalN}).

Within the TSTP, the crossing occurs when $E^*=0$, indicating that for all $N$'s, their kinetic and interaction energies cancel each other at the same $T^*$ (see Appendix \ref{App::Crossing} for details). When we include higher particle numbers in the two-site problem, there is a small contribution to the energy from eigenstates that present multiple double occupancies and/or higher-than-double occupancies. Their contribution accounts for a constant positive shift in the energy for all $N$'s, implying that the crossing occurs at $E^*>0$. The second order NLCE is a linear combination of the one-site and two-site results. The one-site result contributes another constant positive shift for all $N$'s to $E$. Together, the second order NLCE clearly reproduces the trends displayed in Fig.~\ref{fig::Crossing}, where we present $E^*$ and $T^*$ as a function of the interaction strength.

\subsection{Universal $N$-dependence of energy, number of on-site pairs, and kinetic energy}\label{sub::universalN}

\begin{figure}[tbp]
\includegraphics[width=\linewidth]{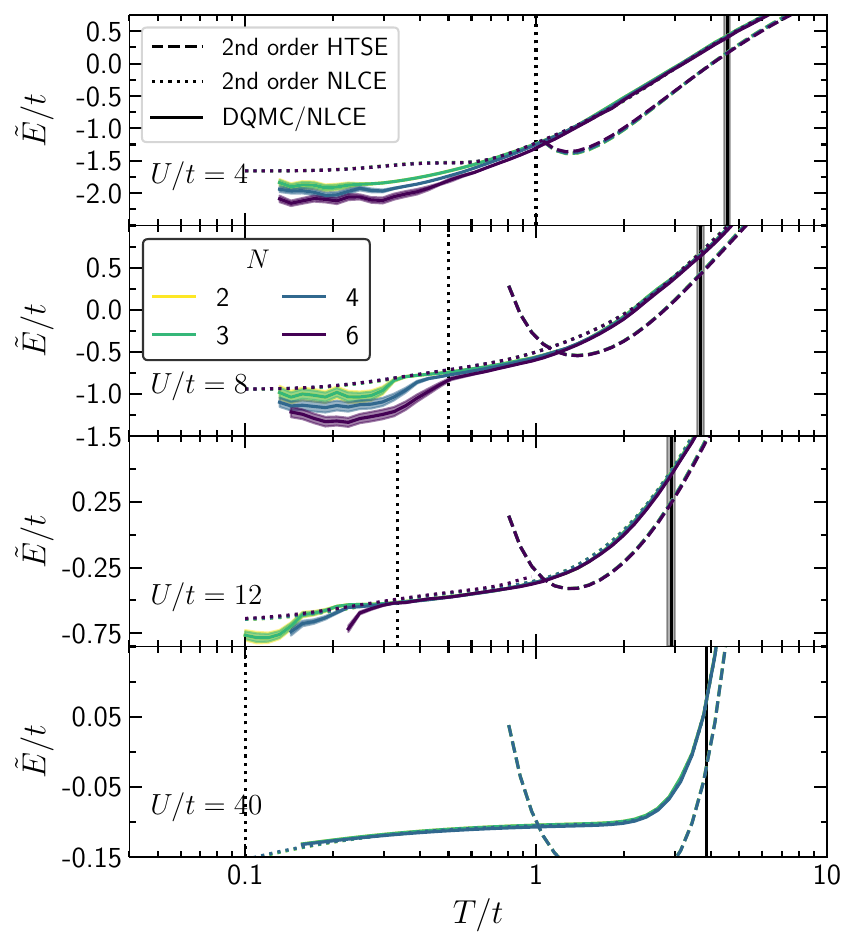}
\caption{\textbf{Universal dependence of energy on $N$.} $\tilde{E}$ vs temperature for several $N$ at fixed $U/t=4,8,12$ at $\langle n \rangle =1$. Solid lines correspond to numerical data: DQMC for $U/t=4,8,12$ and NLCE for $U/t=40$. Shaded regions correspond to error bars obtained by error propagation in Eq.~\eqref{eq:Etilde}. Dashed lines correspond to 2nd order HTSE calculations and dotted lines correspond to 2nd order NLCE. Solid vertical lines indicate the temperature where the different $N$ curves intersect and dotted vertical lines indicate the superexchange energy $J$.}\label{fig:EtildevsT}
\end{figure}

\begin{figure*}[tbp!]
    \centering
    \includegraphics[width=\linewidth]{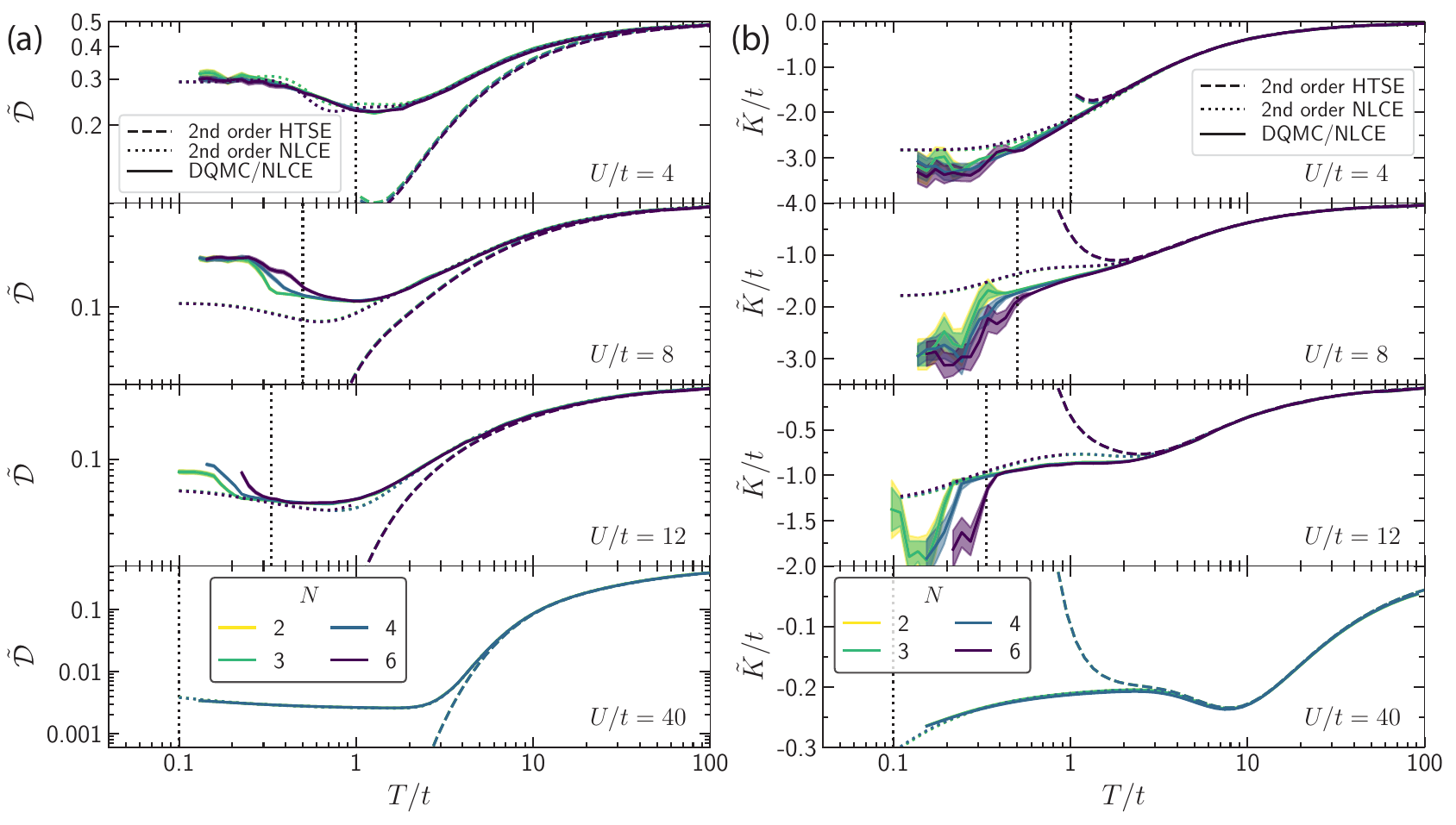}
    \caption{\textbf{Universal dependence of number of on-site pairs and kinetic energy on $N$.} (a) $\tilde{\mathcal{D}}$ (b) $\tilde{K}$ vs temperature for several $N$ at fixed $U/t=4,8,12$ at $\langle n \rangle =1$. Solid lines correspond to numerical data: DQMC for $U/t=4,8,12$ and NLCE for $U/t=40$. Shaded regions correspond to error bars obtained by error propagation in analogs of Eq.~\eqref{eq:Etilde}. Dashed lines correspond to 2nd order HTSE calculations and dotted lines correspond to 2nd order NLCE. Dotted vertical lines indicate the superexchange energy $J$.}
    \label{fig:collapse}
\end{figure*}

In this section, we show that the crossing point of $E$ vs $T$ for all $N$ in Fig.~\ref{fig::EvsT_N} is actually a consequence of a much stronger universal scaling relation that determines the $N$-dependence of all the observables studied here to temperatures well below the crossing temperature (though not arbitrarily low), down to a temperature comparable to the superexchange energy $4t^2/U$. We find that the energy satisfies
\begin{equation} \label{eq:E_TN}
E(T,N) = E(T,\infty) + (1/N)E_1(T)
\end{equation}
for some $E_1(T)$ independent of $N$ over a broad range of temperature. This is shown in Fig.~\ref{fig:EtildevsT} by a universal collapse of appropriately constructed quantity $\tilde{E}$, and we will discuss the features of this collapse more momentarily. First, to understand $\tilde{E}$'s construction, note that Eq.~\eqref{eq:E_TN} is equivalent to 
\begin{equation}\label{eq:Etilde}
\tilde{E}(T,N) \equiv E(T,N) - (1/N)E_1(T)
\end{equation}
being independent of $N$, since the right hand side is simply $E(T,\infty)$. Fig.~\ref{fig:EtildevsT} plots this $\tilde{E}$, taking 
\begin{equation}
E_1(T) = \frac{E(T,N_1)- E(T,N_2)}{(1/N_1)-(1/N_2)} \label{eq:E1_T} 
\end{equation}
for $N_1=2$ and $N_2=3$. When Eq.~\eqref{eq:E_TN} is satisfied, the $E_1(T)$ obtained would be the same for all choices of $N_1$ and $N_2$; we choose $N_1=2$ and $N_2=3$ as they are the least noisy datasets and span the largest range of temperatures, but the overall collapse is observed independent of this choice. The analysis of scaling is inspired by similar scalings discovered in the spectra of strongly correlated materials in Ref.~\cite{Greger2013}. We observe that Eq.~\eqref{eq:E_TN} has the form of a first order Taylor expansion of $E(T,N)$ in $1/N$; from this point of view, the remarkable aspect of the data collapse is that (in an appropriate temperature window) it accurately describes the physics even when $1/N$ is not small (e.g. for $N=2$). 

Figure~\ref{fig:EtildevsT} shows that $\tilde{E}$ is independent of $N$ at temperatures $T \gtrsim J\equiv 4t^2/U$ for all $U$ studied here, and therefore $E(T,N)$ has the simple $N$-dependence given by Eq.~\eqref{eq:E_TN}. Below $T \sim J$, $\tilde{E}$ no longer collapses, signaling a more complicated $N$-dependence. One consequence of the universal scaling is that the thermodynamics in this temperature regime can be obtained for any $N$ from the results for $N=2$ and 3 (or any two $N$). This is convenient for several reasons: The Hilbert space of SU(2) is more manageable for numerical calculations, and because numerical methods such as DQMC are free of the sign problem at $\langle n \rangle =1$ for SU(2).

\begin{figure*}[tbp]
\includegraphics[width=\linewidth]{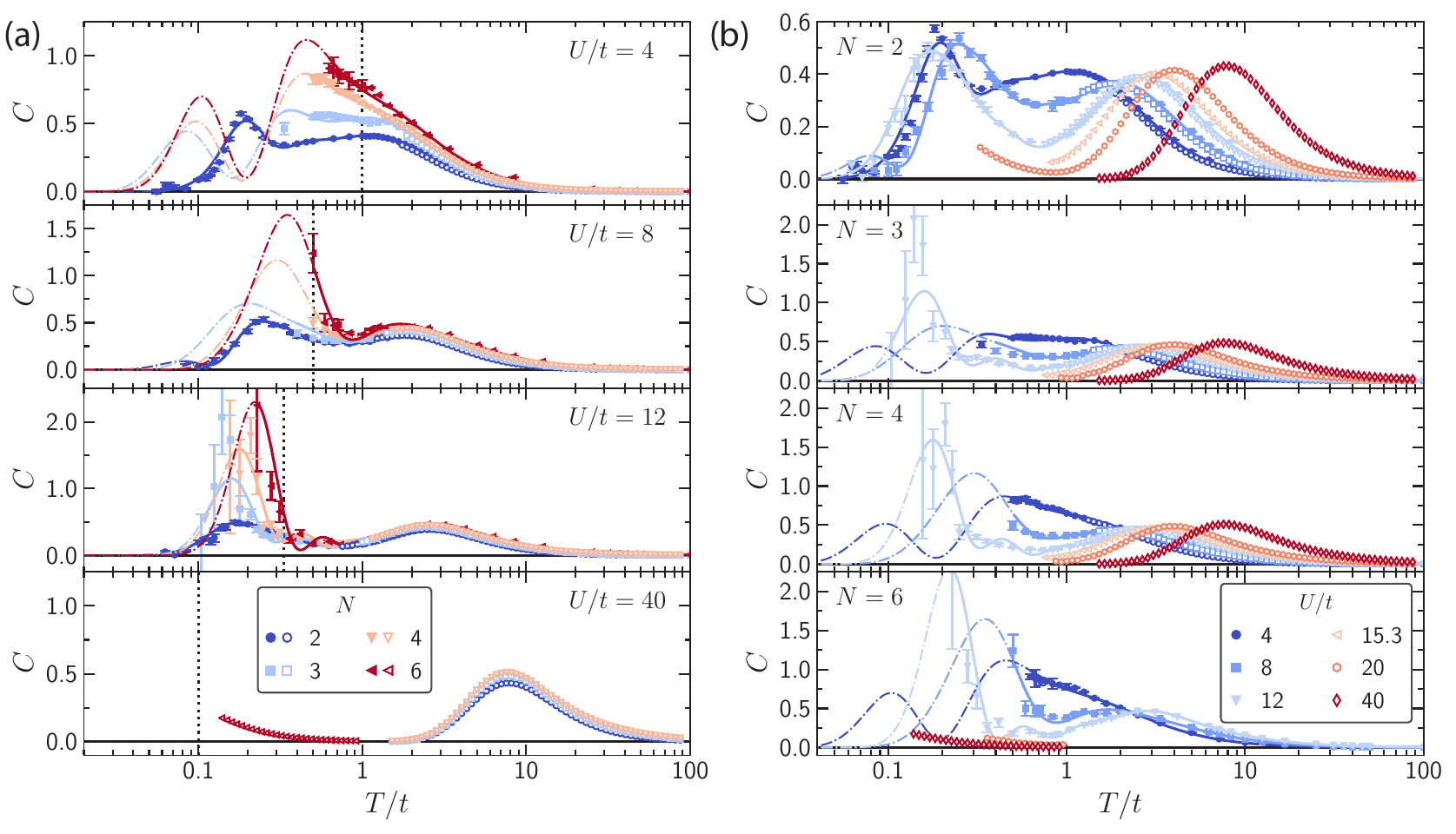}
\caption{\textbf{Specific heat versus temperature.} (a) Each panel compares $C$ for $N=2,3,4,6$ for a fixed $U/t$. (b) Each panel compares $C$ for $U/t=4,8,12,15.3,20,40$ for a fixed $N$. Solid markers are DQMC, open markers are NLCE, dotted lines correspond to the non-interacting limit, dashed lines are the zeroth order HTSE, and solid lines come from the fits of Eq.~\eqref{eq::E_fit} to the DQMC data in Fig.~\ref{fig::EvsT_N} down to the lowest $T_n$ point. Thinner dashdotted lines come from the fit in the extrapolated regime $T < \min(T_n)$, where $\{T_n\}$ is the dataset of temperatures where DQMC results are obtained. Dotted vertical lines indicate the superexchange energy $J$.}\label{fig::specific_heat} 
\end{figure*}

One natural attempt to explain the observed scaling would be the HTSE, since this is expected to be accurate at high temperatures; however, although $\tilde E$ calculated with the second order HTSE collapses, it deviates strongly from the data at $T\lesssim 5t$ [Fig.~\ref{fig:EtildevsT}], so it cannot explain the collapse to the lowest temperatures observed ($0.1t$ to $t$, depending on the value of $U$). In contrast, as Fig.~\ref{fig:EtildevsT} shows, the second order NLCE's $\tilde{E}$ not only collapses, but accurately reproduces the numerical results for all temperatures where the collapse occurs, thus providing a simple and effectively complete calculational tool to obtain the scaling, albeit not an analytic one. 
That the second order NLCE reproduces the data in the scaling regime allows us to infer characteristics of the physics. The first thing to notice is that the second order NLCE can capture one- and two-site nearest-neighbor correlations, but no longer-ranged correlations. Thus, one-site physics and nearest-neighbor correlations suffice to capture the physics in the regime where collapse occurs. This provides interesting insight into the physics, and explains why the collapse occurs at $T\gtrsim J$: this is the characteristic energy scale for correlations in the $\langle n \rangle=1$ system (at least when $U/t$ is large) and thus longer range correlations only develop at temperatures below $J$. Note that this also lets us understand why the collapse is not captured by the second order HTSE: this misses two-site correlations that are $\mathcal{O}(\beta t)^3$ or higher. Such non-perturbative effects are strong in the regime $4t^2/U \lesssim T \lesssim t$ and not easily captured at any order of the HTSE, which diverges for $T\lesssim t$. 

By examining the second order NLCE and simplifying it by taking advantage of the range of temperatures being considered, we can also arrive at an analytic explanation of the scaling phenomena. Although NLCE is typically used as a numerical method, at low-enough order and in simplified limits, it may provide simple analytic expressions. Indeed, in the present case, we show in Appendix~\ref{App::Crossing} that the energy in the second order NLCE in the temperature range $4t^2/U \ll T \ll U$ is given, to zeroth order in $\beta J$, by
\begin{equation}
    E(T,U,N) \approx -J + \frac{1}{N}J.
\end{equation}
We note the additional condition that $T \ll U$ not previously noted; indeed, there are small deviations of the data from collapsing in the $T \sim U$ regime. Finally, when $T \gg U$ the collapse is again recovered, since $K \to 0$ and $D \propto 1/N$ in that regime (see footnote~\bibnotemark[Dinf]). In summary, the parametrically accurate collapse for the two separate regimes $4t^2/U \ll T \ll U$ and $T\gg U$ is interpolated to a quite accurate collapse, though not parametrically so, for all $T \gg 4t^2/U$, as seen in the data.

Although we only analytically show the scaling of Eq.~\eqref{eq:E_TN} to leading order in $J/T$ and $T/U$ (i.e., deep in the $J\ll T \ll U$ regime), numerics seems to indicate the collapse holds beyond this. Explaining this is an open problem. Despite lacking a simple analytic formula, the second order NLCE reproduces all of the behavior, offering a simple predictive theory for the thermodynamics in the $T \gtrsim J$ regime.

The observables $\mathcal D$ and $K$ show a similar universal $N$-dependence, satisfying analogs of Eq.~\eqref{eq:E_TN}, as demonstrated in Fig.~\ref{fig:collapse}(a-b) by showing the collapse of $\tilde{\mathcal{D}}$ and $\tilde{K}$ defined analogously to $\tilde{E}$. These are also reproduced by the second order NLCE and its analytic simplifications in the temperature window of interest. The $U/t=4$ results for $\tilde{\mathcal{D}}$ exhibit a window around $T=t$ where the 2nd order NLCE weakly breaks the collapse ($< 4\%$), but is then recovered at lower temperatures around $T/t=0.2$, where the DQMC data collapses too. Why the $U/t=4$ results collapse even for $T\lesssim 4t^2/U$ remains an open question and merits further exploration.

\subsection{Temperature derivatives at unit density: $C = dE/dT$, $U d\mathcal{D}/dT$, and $dK/dT$}\label{sub::DerivativevsT}

\begin{figure*}[tbp]
\includegraphics[width=\linewidth]{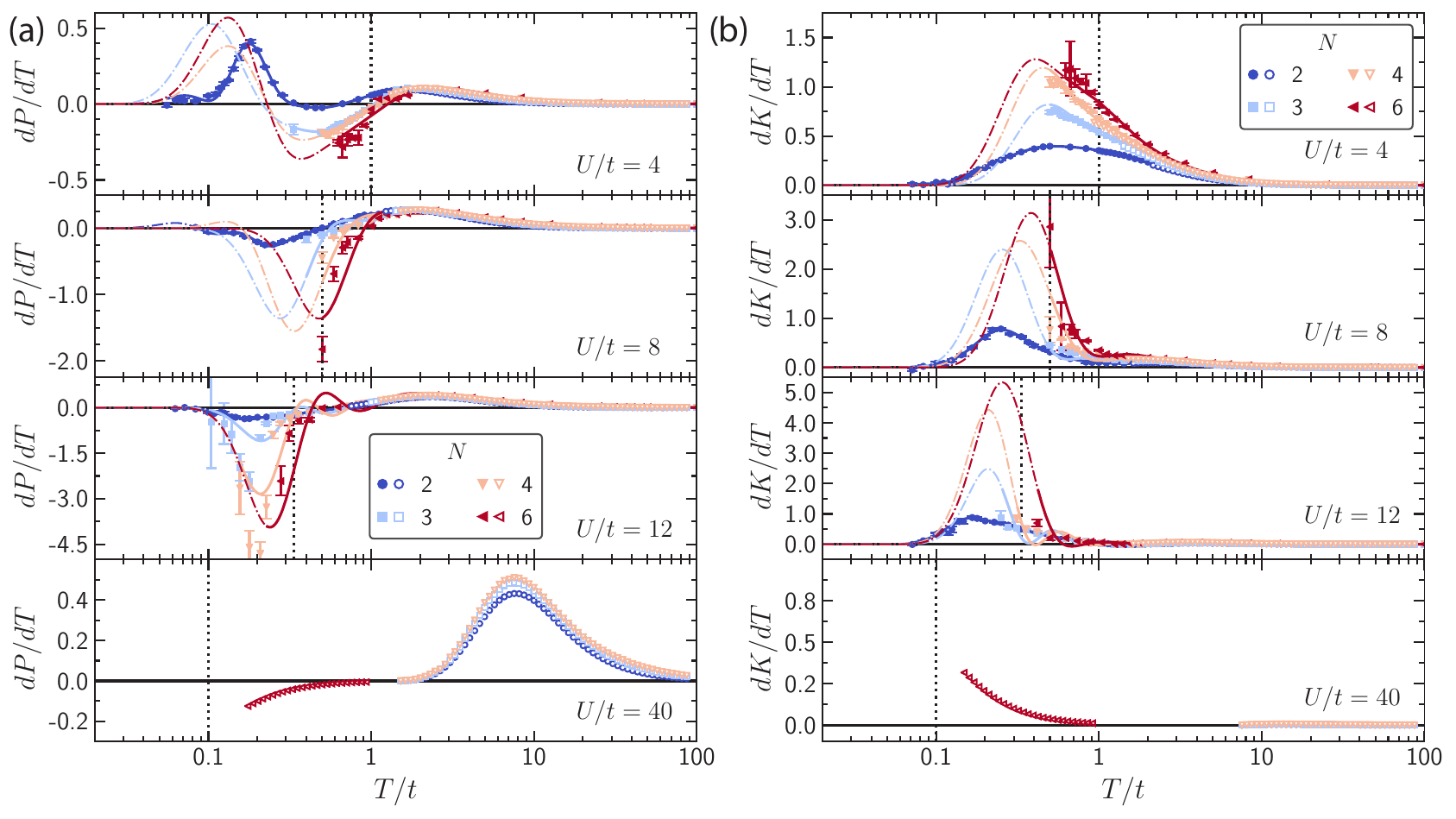}
\caption{\textbf{Contributions to the specific heat vs temperature.} Each panel compares (a) $dP/dT$ (b) $dK/dT$ for $N=2,3,4,6$ for a fixed $U/t$. Solid markers are DQMC, open markers are NLCE, and solid lines come from the fits of Eq.~\eqref{eq::E_fit} to the DQMC data in Figs.~\ref{fig::Doublons}(a) and \ref{fig::KvsT_N} down to the lowest $T_n$ point. Thinner dashdotted lines come from the fit in the extrapolated regime $T < \min(T_n)$, where $\{T_n\}$ is the dataset of temperatures where DQMC results are obtained. Dotted vertical lines indicate the superexchange energy $J$.}\label{fig::dPdKdT} 
\end{figure*}

We now present the derivatives of the energy $E$, interaction energy $P=U\mathcal{D}$, and the kinetic energy $K$. The specific heat ($dE/dT$) as a function of temperature is a valuable thermodynamic observable since its peaks indicate temperatures below which the entropy is significantly reduced as degrees of freedom reorganize and cease to fluctuate. 

The specific heat as a function of temperature [see Fig.~\ref{fig::specific_heat}(a)] presents a two-peak structure for $N=2$; for other $N$ a high-temperature peak is present in all cases, and in most an upturn occurs at lower temperatures, necessitating a second peak at lower temperatures beyond the range of our calculations since $C\to 0$ as $T\to 0$. At least at large $U/t$, the origin of the high-temperature peak is associated with freezing of the charge fluctuations as the temperature is lowered, while the low-temperature peak is associated with the onset of spin correlations, as has been shown for $N=2$ \cite{Paiva2001,IbarraGarciaPadilla2020} and will be evident from our results on $d\mathcal{P}/dT$ and $dK/dT$. For strong interactions, the high-temperature peak is closely in agreement with the results of the zeroth order HTSE and is roughly independent of $N$, with just small changes of amplitude at small $U/t$. The upturn of $C$ as $T/t$ is lowered towards a presumable low-temperature peak (though not directly accessible in the data for $N\geq 3$) depends on $N$ and $U/t$. The upturn seems to grow with $N$, and it generally decreases with $U/t$, although at the lowest temperatures, there may be a complicated non-monotonic dependence. The extent to which the trends of the upturn are either a reflection of the temperature at which the low-$T$ peak occurs or result from changes in the amplitude of the low-$T$ peak cannot be assessed with the current data, and is an interesting question for future theory and experiment. 

The final feature of the specific heat that we analyze is motivated by Ref.~\cite{Paiva2001}'s finding that the specific heat versus temperature curves cross around $T/t\approx 1.6$ for all $U/t\in [1,10]$ for $N=2$. Fig.~\ref{fig::specific_heat}(b) shows that this remains true for other values of $N$, with nearly the same value of the crossing temperature. However, we note that this crossing only occurs for $U/t \lesssim10$, and fails for $U=15.3t$ and larger. The physical significance of this crossing is unclear. Several Refs.~\cite{Vollhardt1997,Chandra1999,Duffy1997,Paiva2001,Macedo2002,Paiva2005} have seen this crossing in 2 dimensions (in square, honeycomb, and asymmetric [$t_\uparrow \neq t_\downarrow$] Hubbard models) at $(C^*,T^*)\approx (0.34,1.6t)$ but all are at relatively small $U/t$. For small $U/t$ Ref.~\cite{Chandra1999} shows that the presence of such high-$T$ crossing arises if one approximates two parameters as small: $1/d$ (where $d$ is the dimension) and the integral over the deviation of the density of states from a constant value~\cite{Greger2013}.

\begin{figure*}[tbp]
\includegraphics[width=\linewidth]{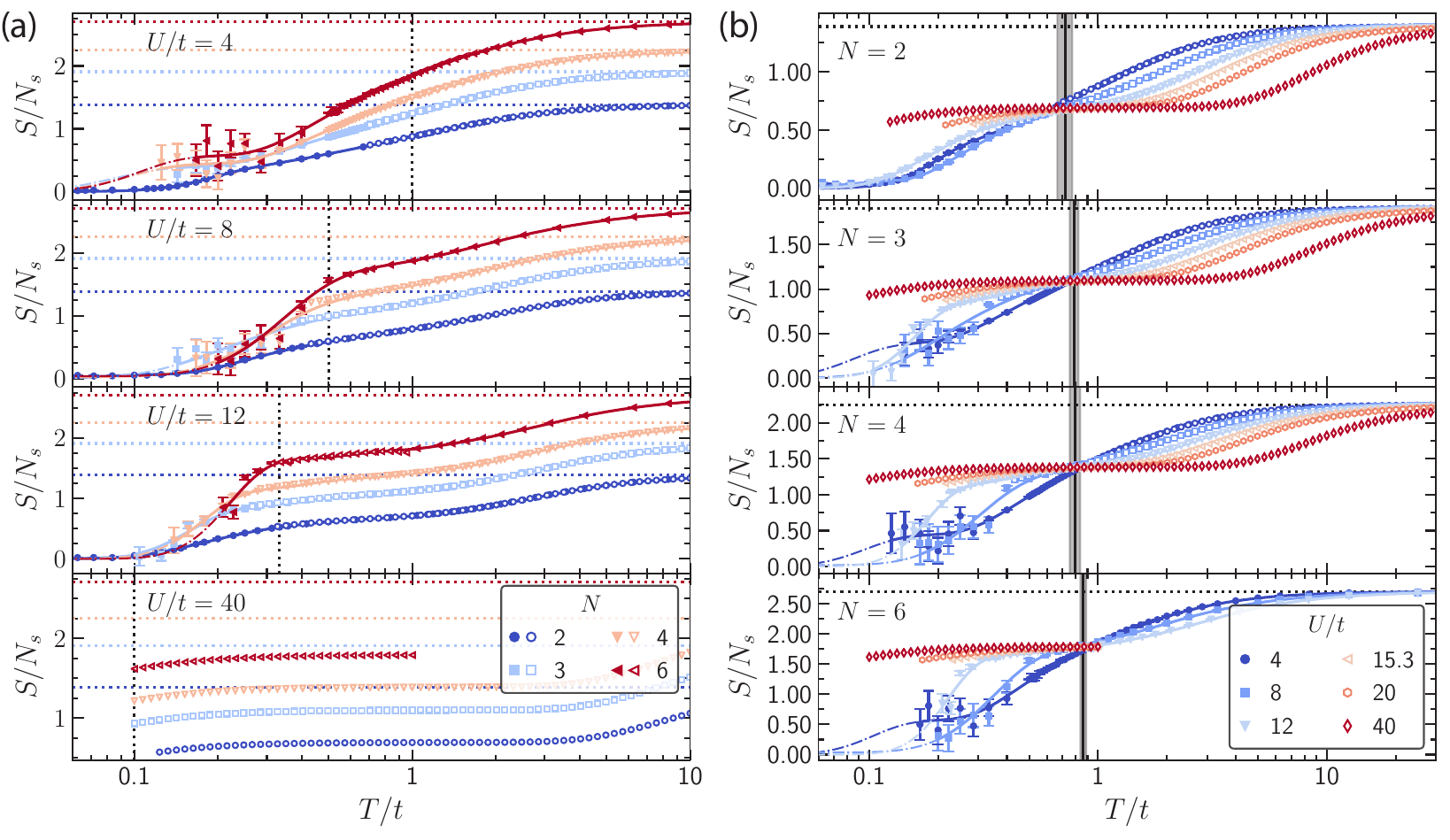}
\caption{\textbf{Entropy per site vs temperature.} (a) Each panel compares $S/N_s$ for $N=2,3,4,6$ for a fixed $U/t$. (b) Each panel compares $S/N_s$ for $U/t=4,8,12,15.3,20,40$ for a fixed $N$. Solid markers are DQMC, open markers are NLCE, and  solid lines come from the fits of Eq.~\eqref{eq::E_fit} to the DQMC data in~\cref{fig::EvsT_N} down to the lowest $T_n$ point. Thinner dashdotted lines come from the fit in the extrapolated regime $T < \min(T_n)$, where $\{T_n\}$ is the dataset of temperatures where DQMC results are obtained. Dotted vertical lines indicate the superexchange energy $J$ and solid vertical lines indicate the temperature where the different $U/t$ curves intersect.}\label{fig::Entropy} 
\end{figure*}

Examining the contributions $dP/dT$ and $dK/dT$ to the specific heat helps disentangle the contributions to the specific heat of the charge and spin degrees of freedom. In Fig.~\ref{fig::dPdKdT}(a) the $dP/dT$ data for $U/t \geq 8$ exhibit a high-$T$ charge peak and a negative dip at lower $T/t$ for all $N$. For such interactions, the high-$T$ peak in the specific heat comes from $dP/dT$. For $U/t=4$ and $N=2$ there is a low-$T$ peak in $dP/dT$, which gives rise to the low-$T$ peak in the specific heat. For $U/t=4$ and $N>2$, the fits suggest the existence of a dip and then a peak as temperature is lowered, however drawing firm conclusions here requires further studies.

In Fig.~\ref{fig::dPdKdT}(b), the $dK/dT$ data for all values of the ineraction strength are positive and exhibit a low-$T$ peak, or a low-$T$ upturn which implies the existence of a peak since $dK/dT \to 0$ as $T \to 0$. The magnitude of the upturn or peak increases with $N$. For $U/t\geq 8$ the low-$T$ peak (or upturn) in the specific heat arises from the spin degree of freedom, seen in $dK/dT$ [Fig.~\ref{fig::dPdKdT}(b)]. Together $dP/dT$ and $dK/dT$ give Fig.~\ref{fig::specific_heat}(a). These results complement the ones presented in Refs.~\cite{Paiva2001,IbarraGarciaPadilla2020}, which demonstrate that for $N=2$, at small $U/t$ the low-$T$ peak arises from $dP/dT$ as opposed to $dK/dT$ in the large $U/t$ limit. The results presented here imply the same conclusion for all $N$ studied in this work: at large $U/t$ the low-$T$ peak arises from $dK/dT$ and the high-$T$ peak from $dP/dT$, while at small $U/t$ the low-$T$ peak comes from $dP/dT$ and the high-$T$ peak from $dK/dT$.

\subsection{Entropy at unit density: dependence on $U/t$, $T/t$, and $N$}\label{sub::SvsT}

Figure~\ref{fig::Entropy}(a) shows the $N$-dependence of the  entropy per site as a function of $T$ for each $U/t$ studied. For all values of the interaction strength we observe that for temperatures above the superexchange energy, at fixed entropy, the system with larger $N$ is at a lower temperature. These results are in agreement with~\cite{Hazzard2012,Merlin2021}, highlighting that gases adiabatically loaded into an optical lattice in this regime will have a significantly lower temperature as $N$ is increased. For $U/t=4$ this cooling seems to occur for all values of $T/t$ and $N$. However, for $U/t \geq 8$, the curves roughly collapse below $T\lesssim 4t^2/U$, at least for $N>2$, suggesting that for 2D square lattices, the dramatic benefits in cooling to the superexchange energy scale will be less effective when cooling well below this scale. We note that this doesn't rule out the cooling with increasing $N$ persisting to arbitrarily low temperatures in other geometries, for example as been shown in 1D chains~\cite{Bonnes2012}. 

Figure~\ref{fig::Entropy}(b) shows the same entropy per site's $U$-dependence as a function of $T$ for each $N$ studied. For each $N$ there is a crossing at finite temperature for all $U/t$. The location of this crossing occurs at higher entropy and $T$ for larger $N$. The existence of a crossing in the entropy curves for different $U/t$ for $N=2$ follows from the presence of a crossing in the specific heat~\cite{Vollhardt1997,Chandra1999,Duffy1997,Paiva2001,Macedo2002,Paiva2005}, given that $C(T,U) = T [ \partial S(T,U)/\partial T ]$. Our results demonstrate that such behavior is still present for $N>2$.

%However, a similar analysis of the isosbestic point (or crossing) with $U/t$ can be done. This has been discussed for the specific heat crossing in SU(2) in references [94,106-110]. The results in Fig. 12 b) for SU(2) follow from the same discussion, given that $C(T,U) = T [ \partial S(T,U)/\partial T ]$. Our results demonstrate that such behavior is still present at $N>2$. Given that the main interest of our research is to understand the dependence on $N$, we refer to the reader to the appropriate literature for SU(2). 

\section{Conclusions} \label{sec::Conclusion}

We have explored the evolution of thermodynamic observables of the SU($N$) Fermi-Hubbard model as a function of temperature $T$, interaction strength $U/t$, and the number of flavors $N$ at $\langle n \rangle=1$. DQMC and NLCE provide accurate results over a wide range of temperatures, including temperatures roughly an order of magnitude below the tunneling $t$, with the exact value depending on $N$ and $U/t$. Neither method is able to access arbitrarily low-temperatures, but the obtained results are far beyond what is accessible to low-order HTSE methods or ED, which have serious inaccuracies even at $T\gtrsim 5t$. The DQMC and NLCE agree where their regimes of convergence overlap, further boosting confidence in the accuracy of the numerics. Some results were also presented in Fig.~\ref{fig::rhosignvsmu} for the dependence of $\langle n \rangle$, $\mathcal{D}$, and average determinantal sign as a function of $\mu/t$, as well as quantities derived from these.

A striking finding is the existence of a simple scaling law with $N$ for $T \gtrsim J$ for $E$, $\mathcal{D}$ and $K$. We show that this observed scaling can be reproduced by the second order NLCE, which takes as input only one- and two-site correlations and information about the lattice geometry, and in the appropriate regime this provides analytic expressions for the observed results. Furthermore, we show that this regime is well beyond the second order HTSE. Although the numerics cannot provide accurate results to arbitrarily low temperature, accurate results for $E$, $K$, and $\mathcal{D}$ are attained for all $N$ studied to temperatures where strong correlations are present. For example, the temperatures reached for all $N$ are slightly lower than recent experiments on the 2D SU(2) FHM~\cite{Mazurenko2017} that observed correlations that spanned the entire ($\sim 15$-site wide) system. Short-ranged correlations in the SU(6) FHM have been observed in Ref.~\cite{Taie2020}, and longer-ranged correlations will be an interesting subject for future work. For example, Ref.~\cite{Romen2020} found a unifying pattern for all $N$ in the Heisenberg limit at high temperatures: spin correlations are organized in shells of equal Manhattan distance and for $N=3$, they evolve from a two sublattice structure to a three sublattice structure as temperature is lowered. The thermodynamic results provided here provide a foundation for studying such phenomena.

Furthermore, the exploration of the specific heat and its contributions provided additional information about the $N$-dependence of the degrees of freedom that fluctuate in the temperature regime studied, and the specialness of the $N=2$ case, possibly due to the perfect nesting. Our results show that the behavior of $C$, $Ud\mathcal{D}/dT$, and $dK/dT$ are all qualitatively similar for all $N$, with only the location and height of peaks shifting. The high temperature peaks (at $T\propto U$) are roughly independent of $N$, while the low-temperature behavior shows a dependence on $N$. The details of the latter are difficult to resolve with current numerical capabilities, and point to interesting future numerical and experimental directions.

Finally, the results for the entropy have important implication for the observed dramatic cooling of SU($N$) FHM systems as $N$ is increased at fixed entropy \cite{Hazzard2012,Bonnes2012,Taie2012,Merlin2021}, which has been designated Pomeranchuk cooling. This has been important for achieving the lowest temperatures in Fermi-Hubbard models by using SU(6) gases~\cite{Taie2020}. Although this effect was shown theoretically at $T\gtrsim t$ using a HTSE~\cite{Hazzard2012} and experimentally~\cite{Taie2012,Taie2020} and theoretically in 1D down to much lower temperatures~\cite{Bonnes2012}, our results here indicate that as one reaches very low temperatures, the cooling as $N$ increases becomes less pronounced in 2D square lattices. In particular, Fig.~\ref{fig::Entropy}(a) suggests that when in the regime with $T$ well below the superexchange energy $4t^2/U$, the temperature may be nearly independent of $N$ at fixed entropy. However, this conclusion is reached in a regime where the noise in the numerical results is large and systematic effects may not be fully under control, so further work will be important to settle this question. Moreover, this is a lattice- and parameter-dependent phenomenon, as it is known in 1D chains that the cooling with increasing $N$ persists to arbitrarily low temperatures~\cite{Bonnes2012}.

\begin{acknowledgments}
The work of EIGP, SD, HW, and KRAH was supported by the NSF PHY-1848304 and the Robert A. Welch Foundation C-1872. This work was supported in part by the Big-Data Private-Cloud Research Cyberinfrastructure MRI-award funded by NSF under grant CNS-1338099 and by Rice University's Center for Research Computing (CRC). The work of RTS was supported by the grant DE-SC0014671 funded by the U.S. Department of Energy, Office of Science. The work of ST and YT was supported by the Grant-in-Aid for Scientific Research of JSPS (No. JP18H05228) and JST CREST (No. JP-MJCR1673).
\end{acknowledgments}

\appendix

\section{Chemical potential and entropy at fixed density when $T\to \infty$}\label{App::S_inf}

In order to compute the entropy using the results from DQMC at fixed density $\langle n \rangle$ using Eq.~\eqref{eq:S_DQMC}, we need to know a priori what is the entropy when $T \to \infty$. This depends on the chemical potential at $T\to \infty$, which we can analytically determine from the condition that $\langle n \rangle$ is fixed. As $T/t\to \infty$, the zeroth order HTSE captures the behavior of $\langle n \rangle$ and it can be used with the condition $\langle n \rangle =\rho$ to determine the chemical potential. When $T\gg U$, the density is 
$\rho = \frac{1}{Z}\sum_n n \binom{N}{n} e^{\beta \mu n}$, defining $Z= \sum_n \binom{N}{n} e^{\beta \mu n}$. Then
\begin{align}
    \rho   &= \frac{d\log Z}{d(\beta \mu)}  \\
    &=  \frac{d}{d(\beta \mu)}\left[ \log \left[ (1+e^{\beta \mu})^N \right]\right] \\
    &= N \frac{e^{\beta \mu}}{1+e^{\beta \mu}}.
\end{align}
Solving for $\beta \mu$, we obtain
\begin{equation}
    \beta \mu(N,\rho) = \ln\left(\frac{\rho}{N-\rho} \right).
\end{equation}
Using this result in the zeroth order HTSE for $S$ gives the $T\to \infty$ entropy per site $S_\infty(N,\rho)$:
\begin{align}
    S_\infty(N,\rho) &=  \ln \left[ \sum_{n=0}^N  \binom{N}{n} \left(\frac{\rho}{N-\rho} \right)^n \right] - \rho \ln\left(\frac{\rho}{N-\rho} \right), \\
     &= N  \ln\left(\frac{N}{N-\rho} \right) - \rho \ln\left(\frac{\rho}{N-\rho} \right).
\end{align}

\section{Convergence of NLCE as number of sites increases, and comparison with ED}\label{App::ed_compare}

\begin{figure}[tbp]
\includegraphics[width=0.9\linewidth]{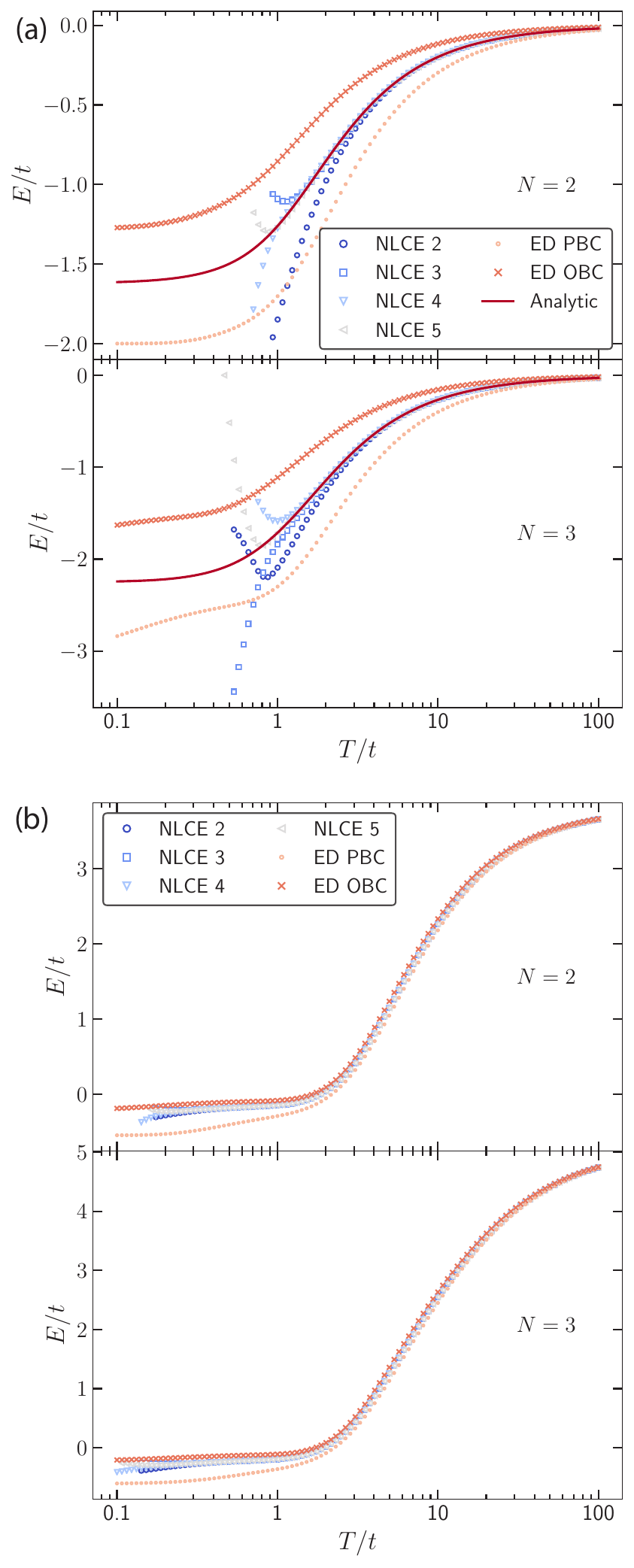}
\caption{\textbf{Convergence of NLCE with expansion order 5 and comparison with ED.} Energy vs $T/t$ at $\langle n \rangle = 1$ for (a) $U=0$ (b) $U=15.3t$ for SU(2) and SU(3). The ED is evaluated in a $3\times 2$ lattice for both open and periodic boundary conditions. (a) The NLCE converges to the analytic result (the solid line) to much lower $T/t$ than either of the ED results. (b) The NLCE curves converge to each other at much lower temperatures than the ED curves collapse on each other or on the NLCE results, signaling that the NLCE converges to significantly lower temperatures than the ED.}\label{fig::ed_compare} 
\end{figure}

We investigate the convergence of the NLCE with expansion order, and we demonstrate that it is significantly more accurate than ED, even when the ED is performed on larger clusters (and therefore requires more computational resources) than the NLCE. We focus on two cases: $U/t=0$ which offers an analytic solution for comparison [Fig.~\ref{fig::ed_compare}(a)] and $U/t=15.3$ [Fig.~\ref{fig::ed_compare}(b)], both for $\braket{n}=1$. Fig.~\ref{fig::ed_compare}(a) shows that the 6-site ($3\times 2$) ED calculations for $U=0$, whether with open-boundary or periodic boundary conditions, has noticeable deviations from the exact analytic result at temperatures $T/t \lesssim 20$. Even the very low-order 2-site NLCE converges accurately to much lower temperature, $T/t \lesssim 3$. Increasing the order of the NLCE calculation leads to results that converge down to still lower temperature. Note that the NLCE calculation is self-diagnosing: even without appealing to the analytic result, the NLCE demonstrates its accuracy when adjacent NLCE orders agree with each other. For example, when the order 4 and order 5 results closely agree with each other, then they also agree with the analytic result. This is consistent with earlier findings in other models~\cite{tang2013short,Khatami2011}. It is worth mentioning that ED results may still provide valuable information: at low-$T$ the NLCE fails dramatically, and while the ED may not be quantitatively accurate, it may still reproduce qualitative features. 

Now we show similar results for $U/t=15.3$ where no analytic result is available. The self-diagnosis of the NLCE demonstrates the convergence of 2-site NLCE to $T \sim 0.4t$, and lower temperatures upon increasing the order. Again, the NLCE converges down to a much lower temperature than the ED, which shows significant deviations due to finite-size effects already at $T/t\gtrsim 2$. These results show that even numerically inexpensive NLCE calculations (2- or 3-sites) accurately converge to much lower temperatures than the much more expensive 6-site ED.

\section{Basis truncation in the NLCE}\label{App::nlce_trunc}
The Hilbert space dimension for the SU($N$) system imposes a severe limit on ED and NLCE if implemented naively, and this difficulty increases dramatically with $N$: the Hilbert space dimension is $2^{N N_s}$, where $N_s$ is the number of lattice sites, reaching a nearly intractable dimension of $2^{24}$ already for SU(6) at 4 sites. Accounting for the SU($N$) symmetries ameliorates this considerably, but the basic difficulty remains.

To alleviate these problems for $N=6$ where the difficulties are worst, we employ a basis truncation scheme for the ED used in the NLCE; this truncation was first introduced for ED in Ref.~\cite{Taie2020}, and it can provide accurate results with negligible truncation error in the physical regime we consider, $\langle n \rangle \lesssim 1$, $U/t \gtrsim 1$, and $T/U$ not too large. To understand this scheme, note that eigenstates with significant weight on flavor-number basis states with large interaction energy will be highly suppressed in the thermal average by the Boltzmann factor for that eigenstate. Thus we restrict the basis states to those with interaction energy less than or equal to $pU$ for a constant $p$ that we choose to obtain sufficient accuracy while remaining computationally feasible. In addition, by a similar logic, we restrict the maximum number of particles in the cluster. In the main text, we choose $p=3$ and a maximum particle number of 6 (one more particle than the maximum number of sites used in the NLCE), and the truncation error is negligible at low-$T$ but increases as $T$ increases (details below).

\begin{figure}[tbp]
    \includegraphics[width=\linewidth]{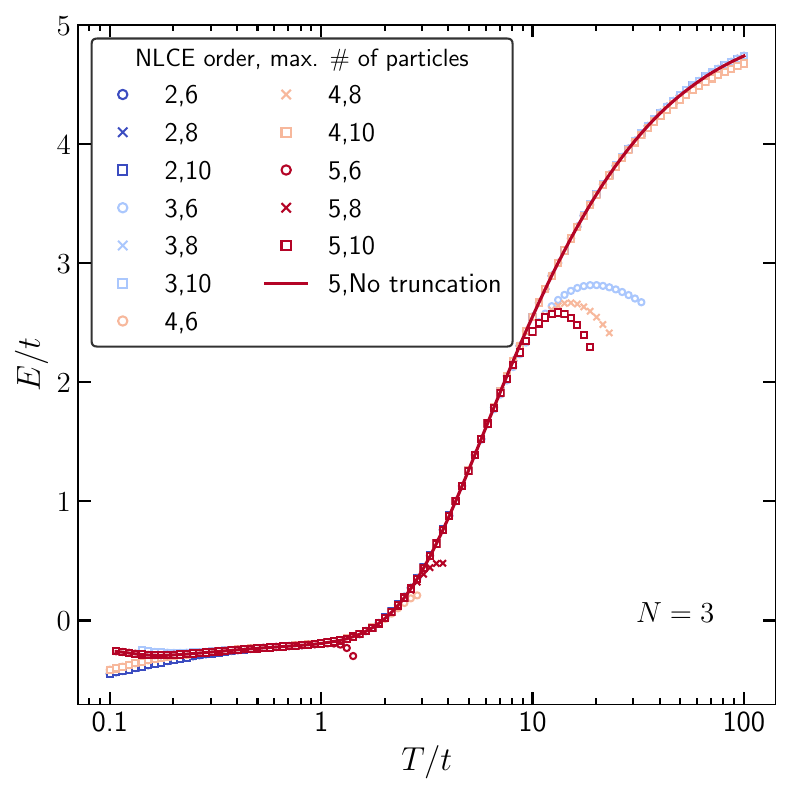}
    \caption{\textbf{Convergence of NLCE with restriction of maximum number of particles.} Energy vs $T/t$ plot at $U=15.3t$ for SU($3$). Different curves are different NLCE orders from $2$ to $5$, and restriction of maximum number of particles to $6$, $8$ and $10$, as indicated in the legend. The divergence of NLCE at low-temperature is due to the finite-order of the expansion, while the divergence at high-temperature is due to particle number truncation.}
    \label{fig::truncation_particles}
\end{figure}

Fig.~\ref{fig::truncation_particles} illustrates the accuracy of NLCE with maximum particle number restriction, and also the new numerical issues the truncation introduces, by comparing results for maximum number of particles $=\ 6,\ 8$ and $10$ and the unrestricted result for SU(3) at $U/t=15.3$ and $\braket{n}=1$. Results are plotted to temperatures a bit past where the truncations are accurate so that the effects of this restriction are visible. The feature apparent from the truncation is that as the temperature is increased, the results with particle number restriction deviate from the correct answer. This is expected: as temperature is increased, the Boltzmann weight on basis states with more particles increases. A less obvious feature is that the temperature above which the restriction fails to be accurate actually decreases as the NLCE order increases. This is because the NLCE relies on cancellation of finite-size errors when combining results from many clusters to obtain accurate results, and the number of clusters used increases with NLCE order, and thus so does the required level of cancellation. The truncation of maximum number of particles interferes with the exact cancellation, and is magnified by the NLCE procedure by an amount that grows with the number of contributing clusters. Thus, there is a finite window over which the NLCE results are highly converged: the particle number truncation constrains the results to being accurate below some temperature, while the finite-size clusters used in the  NLCE constraint the results to being accurate above some temperature. For the SU(3) results [Fig.~\ref{fig::truncation_particles}] this window is roughly from $T/t=0.2$ to 1 for maximum number of particles of 6 and 5-site NLCE, as seen through comparison to the results without the restriction. We also observe that the NLCE self-diagnoses its failure due to this restriction similar to how it diagnosed the failure due to the finite number of clusters used: when results with different particle number truncations agree, the calculation is accurately converged.

\begin{figure}[tbp]
    \includegraphics[width=\linewidth]{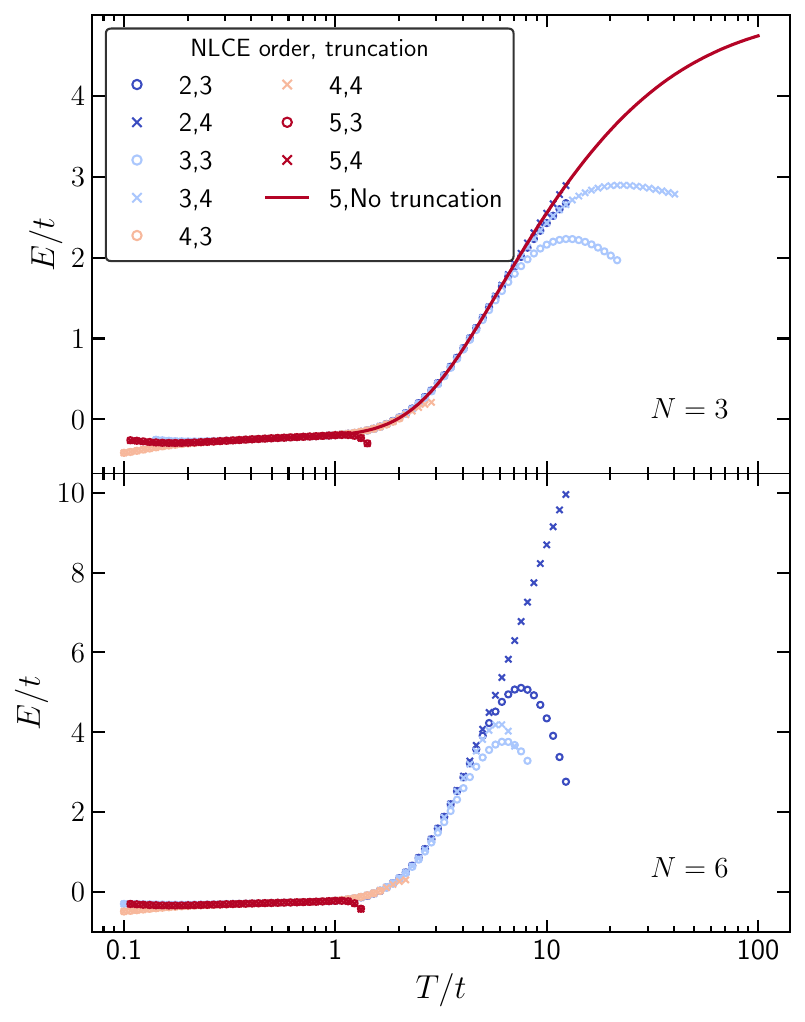}
    \caption{\textbf{Convergence of NLCE with Hilbert space truncation.} Energy vs $T/t$ plot at $U=15.3t$ for SU(3) [top] and SU(6) [bottom]. The total number of particles in the cluster was restricted to 6 or less and basis states with interaction energy
    $U \mathcal{D}$ greater than $pU$ were discarded. Different curves are different NLCE orders from $2$ to $5$, and truncations $p=3$ and $4$, as indicated in the legend. The divergence of NLCE at low-temperature is due to the finite-order of the expansion, while the divergence at high-temperature is due to the basis truncation. }
    \label{fig::truncation}
\end{figure}

Fig.~\ref{fig::truncation} illustrates the effects of the interaction-energy-based basis truncation on top of maximum particle number restriction to $6$ by comparing the results for $p=3$ and $p=4$ truncations to the non-truncated result for SU(3) at $U/t=15.3$. The additional effects are negligible for NLCE orders 4 and 5. This is not surprising since for a 5-site cluster a particle number restriction of 6 already discards most states with highly occupied sites (doublons and higher) and a further restriction of basis states with interaction energy $<3U$ serves mainly to discard triplon and and higher states which have very small Boltzmann weights in the region of interest. Thus, this additional truncation significantly reduces computational time, while introducing negligible additional numerical errors. The self-diagnosis of the NLCE is apparent here as well, which we use to analyze the $N=6$ results,  where results without truncation are unavailable [Fig.~\ref{fig::truncation} (bottom)]. When adjacent orders and different truncations agree, the NLCE is converged. We see the same trends for $N=6$ as for $N=3$, and a similar region of convergence for the 5-site NLCE. The results in the main text thus use $p=3$ and particle number restricted to 6 in the results for $N=6$. Naturally, the value of $U$ affects the region of convergence significantly. The size of the temperature region of convergence increases with $U$. For $U\leq 8t$, there is barely any region of convergence for these choices of the truncation parameters, and hence we cannot get converged results for the SU(6) system from the fifth-order NLCE with our truncation. 

\section{Second order NLCE calculation for $J \ll T \ll U $. Energy crossing and $1/N$ dependence}\label{App::Crossing}

In this section we focus on two things: explaining the existence of an energy crossing (as seen in Fig.~\ref{fig::EvsT_N}) and demonstrating the $1/N$ scaling observed in the limit $J \ll T \ll U$. As mentioned in the main text, the second order NLCE captures such behavior, but this does not admit a general analytic formula. However, analytic formulas can be obtained in the $J \ll T \ll U $ regime.

The second order NLCE in the square lattice is $E = 4 E^{(2)} - 3 E^{(1)}$, where $E^{(x)}$ is the energy per site in an $x$-site system. First we demonstrate that for $J \ll T \ll U$ the one-site problem does not contribute to the energy, then we calculate the energy in the relevant particle sectors in the two-site problem, and finally we present results for their linear combination, i.e. the second order NLCE.

\subsection{One-site problem}

In the one-site problem, the partition function is given by
\begin{equation}
Z^{(1)} = \sum_{n=0}^N \binom{N}{n} e^{-\beta \epsilon_0(n)},
\end{equation}
where $\epsilon_0(n) = U \binom{n}{2} - \mu n$. The density $\rho^{(1)}$ is
\begin{equation}
\rho^{(1)} = \langle n \rangle  = \frac{1}{Z^{(1)}} \sum_{n=0}^N n \binom{N}{n} e^{-\beta \epsilon_0(n)},
\end{equation} 
while the energy $E^{(1)}$ is
\begin{equation}
E^{(1)} = \langle H + \mu n \rangle = \frac{1}{Z^{(1)}} \sum_{n=0}^N U  \binom{n}{2} \binom{N}{n} e^{-\beta \epsilon_0(n)}.
\end{equation}

Because $T \ll U$, we can obtain an analytical approximate expression for the chemical potential $\mu_0(T,U,N)$ that fixes the density to $\rho=1$ by only considering the 0-, 1-, and 2-particle sectors. This expression is exact for $N=2$, but is only true to leading order in $T/U$ for $N>2$, since it truncates eigenstates with triplons and higher occupancies. The solution for $\mu_0$ is given by
\begin{equation}\label{eq:mu}
\mu_0(T,U,N)  = \frac{U}{2} + \frac{1}{2} T \ln\left[\frac{2}{N(N-1)}\right],
\end{equation}
and the energy in this limit is
\begin{equation}
E^{(1)} = \frac{U  e^{- \beta U/2}}{2 + \sqrt{\frac{2 N }{N-1}}} \approx 0.
\end{equation}
Therefore we have shown that in the $\beta U \gg 1$ limit the second order NLCE in the square lattice is determined by the two-site result, $E = 4 E^{(2)}$.

\subsection{Two-site problem}

The Hilbert space of the two-site problem is $4^{N}$, and analytically diagonalizing such matrix --- even if exploiting particle number conservation for each spin component and spin permutation symmetry --- is not possible. However, not all particle sectors need to be considered since $\beta U \gg 1$ and $\langle n \rangle =1$. Under these two conditions, and to leading order in $\beta J$, we can use the chemical potential from Eq.~\eqref{eq:mu} in the two-site calculation. This ensures that at $\langle n \rangle = 1$ only the 2-site 2-particle sector (TSTP) contributes, since all other sectors are $\propto e^{-\beta U}$, and therefore negligible.

In the TSTP, there are $N$ states where two particles of the same flavor sit on sites 1 and 2. Since these are Pauli blocked from hopping, and there is no $U$ contribution, they constitute $N$ independent one dimensional subspaces of energy $\epsilon=0$, giving rise to a contribution $N$ in the partition function. Furthermore, there are $\binom{N}{2}$ choices where the flavors of the two particles are different. Since the hopping conserves flavor, these form independent four dimensional subspaces with levels identical to the usual $N=2$ spectrum in the one spin up and one spin down sector. Therefore, in the TSTP the partition function $Z^{(2)}$ and energy per site $E^{(2)}$ are given by:
\begin{align}
    Z^{(2)} &= N + \binom{N}{2} Z_2, \label{eq:TSTP_Z} \\
    E^{(2)} &= \frac{1}{2}\binom{N}{2} E_2, \label{eq:TSTP_E} \\
    Z_2 &= \sum_{n=1}^4 \exp(-\beta \epsilon_n), \label{eq:TSTP_Z2} \\
    E_2 &= \frac{1}{Z}\sum_{n=1}^4 \epsilon_n \exp(-\beta \epsilon_n),\label{eq:TSTP_E2}
\end{align}
where $\epsilon_n$ are the eigenvalues of the 2-particle sector with different spin component, i.e. $\sigma \neq \tau$. These eigenvalues are $ \epsilon_n = \{0, U, U/2 \pm \sqrt{16t^2 + U^2}\}$.

First, an energy crossing for different values of $N$ as a function of $T/t$ at a fixed $U/t$ %[in the second order NLCE in the limit $\beta U \gg 1$] 
in the TSTP occurs when Eq.~\eqref{eq:TSTP_E} is equal to zero, i.e. $E_2=0$, to demand the $N$-independence of the energy. The temperature at which the crossing occurs is the solution to the following trascendental equation:
\begin{align}
    0 &= Ue^{-\beta U} \nonumber \\
    &+ \frac{U}{2} \left[ 1 + \sqrt{ 1 + \left(\frac{4 t}{U}\right)^2}\right] e^{-\beta\frac{U}{2} \left[ 1 + \sqrt{ 1 + \left(\frac{4 t}{U}\right)^2}\right]} \nonumber \\
    &+ \frac{U}{2} \left[ 1 - \sqrt{ 1 + \left(\frac{4 t}{U}\right)^2}\right] e^{-\beta\frac{U}{2} \left[ 1 - \sqrt{ 1 + \left(\frac{4 t}{U}\right)^2}\right]}.
\end{align}
That this equation has solutions demonstrates the existence of a crossing point, and it qualitatively explains the trends of $T^*/t$ with $U/t$, although it deviates quantitatively from the results in Fig.~\ref{fig::Crossing}.

Now we demonstrate the $1/N$ scaling for $J\ll T \ll U$, where the second order NLCE shows unconditionally that the collapse occurs in this regime. We present results for $E$, but analogous results can be obtained for $\mathcal{D}$ and $K$. The energy in the TSTP is given by Eq.~\eqref{eq:TSTP_E},
\begin{equation}
    E^{(2)}\left(T,U,N\right) = \frac{1}{2} \frac{\binom{N}{2} \sum_{n=1}^4 \epsilon_n \exp(-\beta \epsilon_n)}{ N + \binom{N}{2}\sum_{n=1}^4 \exp(-\beta \epsilon_n)}.
\end{equation}
Since $U \gg t $, the $\epsilon_n$ have simple expressions $\epsilon_n = \{0, U, U+ J, -J\}$. Because  $\beta U \gg 1$, $E^{(2)}$ is given to leading order by
\begin{align}
    E^{(2)}\left(T,U,N\right) &\approx \frac{1}{2} \frac{\binom{N}{2}\left(-J e^{\beta J} \right)}{N + \binom{N}{2} \left(1 + e^{\beta J} \right)}, \nonumber \\
    &= \frac{1}{2}\frac{-J e^{\beta J}}{\frac{N+1}{N-1} + e^{\beta J}}.
\end{align}

Finally, since in the $J \ll T \ll U$ limit $E = 4 E^{(2)}$, the second order NLCE to zeroth order in $\beta J \ll 1$ is
\begin{equation}
    E\left(T,U,N\right) \approx -J  + \frac{1}{N} J.
\end{equation}
This demonstrates that the scaling Eq.~\eqref{eq:E_TN} holds in the regime $4t^2/U \ll T \ll U$, when $t/U \ll 1$ to zeroth order in $\beta J$.

\bibliography{SUN_references.bib}

\begin{thebibliography}{123}%
\makeatletter
\providecommand \@ifxundefined [1]{%
 \@ifx{#1\undefined}
}%
\providecommand \@ifnum [1]{%
 \ifnum #1\expandafter \@firstoftwo
 \else \expandafter \@secondoftwo
 \fi
}%
\providecommand \@ifx [1]{%
 \ifx #1\expandafter \@firstoftwo
 \else \expandafter \@secondoftwo
 \fi
}%
\providecommand \natexlab [1]{#1}%
\providecommand \enquote  [1]{``#1''}%
\providecommand \bibnamefont  [1]{#1}%
\providecommand \bibfnamefont [1]{#1}%
\providecommand \citenamefont [1]{#1}%
\providecommand \href@noop [0]{\@secondoftwo}%
\providecommand \href [0]{\begingroup \@sanitize@url \@href}%
\providecommand \@href[1]{\@@startlink{#1}\@@href}%
\providecommand \@@href[1]{\endgroup#1\@@endlink}%
\providecommand \@sanitize@url [0]{\catcode `\\12\catcode `\$12\catcode
  `\&12\catcode `\#12\catcode `\^12\catcode `\_12\catcode `\%12\relax}%
\providecommand \@@startlink[1]{}%
\providecommand \@@endlink[0]{}%
\providecommand \url  [0]{\begingroup\@sanitize@url \@url }%
\providecommand \@url [1]{\endgroup\@href {#1}{\urlprefix }}%
\providecommand \urlprefix  [0]{URL }%
\providecommand \Eprint [0]{\href }%
\providecommand \doibase [0]{https://doi.org/}%
\providecommand \selectlanguage [0]{\@gobble}%
\providecommand \bibinfo  [0]{\@secondoftwo}%
\providecommand \bibfield  [0]{\@secondoftwo}%
\providecommand \translation [1]{[#1]}%
\providecommand \BibitemOpen [0]{}%
\providecommand \bibitemStop [0]{}%
\providecommand \bibitemNoStop [0]{.\EOS\space}%
\providecommand \EOS [0]{\spacefactor3000\relax}%
\providecommand \BibitemShut  [1]{\csname bibitem#1\endcsname}%
\let\auto@bib@innerbib\@empty
%</preamble>
\bibitem [{\citenamefont {Montorsi}(1992)}]{Montorsi1992}%
  \BibitemOpen
  \bibinfo {editor} {\bibfnamefont {A.}~\bibnamefont {Montorsi}},\ ed.,\
  \href@noop {} {\emph {\bibinfo {title} {The {H}ubbard {Model}: {A} reprint
  volume}}}\ (\bibinfo  {publisher} {World Scientific Publishing Co. Pte.
  Ltd.},\ \bibinfo {year} {1992})\BibitemShut {NoStop}%
\bibitem [{\citenamefont {Tasaki}(1998)}]{Tasaki1998}%
  \BibitemOpen
  \bibfield  {author} {\bibinfo {author} {\bibfnamefont {H.}~\bibnamefont
  {Tasaki}},\ }\bibfield  {title} {\bibinfo {title} {The {H}ubbard model - an
  introduction and selected rigorous results},\ }\href@noop {} {\bibfield
  {journal} {\bibinfo  {journal} {J. Phys.: Condens. Matter}\ }\textbf
  {\bibinfo {volume} {10}},\ \bibinfo {pages} {4353} (\bibinfo {year}
  {1998})}\BibitemShut {NoStop}%
\bibitem [{\citenamefont {Arovas}\ \emph {et~al.}(2021)\citenamefont {Arovas},
  \citenamefont {Berg}, \citenamefont {Kivelson},\ and\ \citenamefont
  {Raghu}}]{Arovas2021}%
  \BibitemOpen
  \bibfield  {author} {\bibinfo {author} {\bibfnamefont {D.~P.}\ \bibnamefont
  {Arovas}}, \bibinfo {author} {\bibfnamefont {E.}~\bibnamefont {Berg}},
  \bibinfo {author} {\bibfnamefont {S.}~\bibnamefont {Kivelson}},\ and\
  \bibinfo {author} {\bibfnamefont {S.}~\bibnamefont {Raghu}},\ }\bibfield
  {title} {\bibinfo {title} {The {H}ubbard {M}odel},\ }\href@noop {} {\bibfield
   {journal} {\bibinfo  {journal} {arXiv:2103.12097v1}\ } (\bibinfo {year}
  {2021})}\BibitemShut {NoStop}%
\bibitem [{\citenamefont {Bohrdt}\ \emph {et~al.}(2021)\citenamefont {Bohrdt},
  \citenamefont {Homeier}, \citenamefont {Reinmoser}, \citenamefont {Demler},\
  and\ \citenamefont {Grusdt}}]{Bohrdt2021}%
  \BibitemOpen
  \bibfield  {author} {\bibinfo {author} {\bibfnamefont {A.}~\bibnamefont
  {Bohrdt}}, \bibinfo {author} {\bibfnamefont {L.}~\bibnamefont {Homeier}},
  \bibinfo {author} {\bibfnamefont {C.}~\bibnamefont {Reinmoser}}, \bibinfo
  {author} {\bibfnamefont {E.}~\bibnamefont {Demler}},\ and\ \bibinfo {author}
  {\bibfnamefont {F.}~\bibnamefont {Grusdt}},\ }\bibfield  {title} {\bibinfo
  {title} {Exploration of doped quantum magnets with ultracold atoms},\
  }\href@noop {} {\bibfield  {journal} {\bibinfo  {journal}
  {arXiv:2107.08043v1}\ } (\bibinfo {year} {2021})}\BibitemShut {NoStop}%
\bibitem [{\citenamefont {Imada}\ \emph {et~al.}(1998)\citenamefont {Imada},
  \citenamefont {Fujimori},\ and\ \citenamefont {Tokura}}]{Imada1998}%
  \BibitemOpen
  \bibfield  {author} {\bibinfo {author} {\bibfnamefont {M.}~\bibnamefont
  {Imada}}, \bibinfo {author} {\bibfnamefont {A.}~\bibnamefont {Fujimori}},\
  and\ \bibinfo {author} {\bibfnamefont {Y.}~\bibnamefont {Tokura}},\
  }\bibfield  {title} {\bibinfo {title} {Metal-insulator transitions},\ }\href
  {https://doi.org/10.1103/RevModPhys.70.1039} {\bibfield  {journal} {\bibinfo
  {journal} {Rev. Mod. Phys.}\ }\textbf {\bibinfo {volume} {70}},\ \bibinfo
  {pages} {1039} (\bibinfo {year} {1998})}\BibitemShut {NoStop}%
\bibitem [{\citenamefont {White}\ \emph {et~al.}(1989)\citenamefont {White},
  \citenamefont {Scalapino}, \citenamefont {Sugar}, \citenamefont {Loh},
  \citenamefont {Gubernatis},\ and\ \citenamefont {Scalettar}}]{White1989}%
  \BibitemOpen
  \bibfield  {author} {\bibinfo {author} {\bibfnamefont {S.~R.}\ \bibnamefont
  {White}}, \bibinfo {author} {\bibfnamefont {D.~J.}\ \bibnamefont
  {Scalapino}}, \bibinfo {author} {\bibfnamefont {R.~L.}\ \bibnamefont
  {Sugar}}, \bibinfo {author} {\bibfnamefont {E.~Y.}\ \bibnamefont {Loh}},
  \bibinfo {author} {\bibfnamefont {J.~E.}\ \bibnamefont {Gubernatis}},\ and\
  \bibinfo {author} {\bibfnamefont {R.~T.}\ \bibnamefont {Scalettar}},\
  }\bibfield  {title} {\bibinfo {title} {Numerical study of the two-dimensional
  {H}ubbard model},\ }\href {https://doi.org/10.1103/PhysRevB.40.506}
  {\bibfield  {journal} {\bibinfo  {journal} {Phys. Rev. B}\ }\textbf {\bibinfo
  {volume} {40}},\ \bibinfo {pages} {506} (\bibinfo {year} {1989})}\BibitemShut
  {NoStop}%
\bibitem [{\citenamefont {Sch\"afer}\ \emph {et~al.}(2021)\citenamefont
  {Sch\"afer}, \citenamefont {Wentzell}, \citenamefont {\ifmmode~\check{S}\else
  \v{S}\fi{}imkovic}, \citenamefont {He}, \citenamefont {Hille}, \citenamefont
  {Klett}, \citenamefont {Eckhardt}, \citenamefont {Arzhang}, \citenamefont
  {Harkov}, \citenamefont {Le~R\'egent}, \citenamefont {Kirsch}, \citenamefont
  {Wang}, \citenamefont {Kim}, \citenamefont {Kozik}, \citenamefont {Stepanov},
  \citenamefont {Kauch}, \citenamefont {Andergassen}, \citenamefont {Hansmann},
  \citenamefont {Rohe}, \citenamefont {Vilk}, \citenamefont {LeBlanc},
  \citenamefont {Zhang}, \citenamefont {Tremblay}, \citenamefont {Ferrero},
  \citenamefont {Parcollet},\ and\ \citenamefont {Georges}}]{Schafer2021}%
  \BibitemOpen
  \bibfield  {author} {\bibinfo {author} {\bibfnamefont {T.}~\bibnamefont
  {Sch\"afer}}, \bibinfo {author} {\bibfnamefont {N.}~\bibnamefont {Wentzell}},
  \bibinfo {author} {\bibfnamefont {F.}~\bibnamefont {\ifmmode~\check{S}\else
  \v{S}\fi{}imkovic}}, \bibinfo {author} {\bibfnamefont {Y.-Y.}\ \bibnamefont
  {He}}, \bibinfo {author} {\bibfnamefont {C.}~\bibnamefont {Hille}}, \bibinfo
  {author} {\bibfnamefont {M.}~\bibnamefont {Klett}}, \bibinfo {author}
  {\bibfnamefont {C.~J.}\ \bibnamefont {Eckhardt}}, \bibinfo {author}
  {\bibfnamefont {B.}~\bibnamefont {Arzhang}}, \bibinfo {author} {\bibfnamefont
  {V.}~\bibnamefont {Harkov}}, \bibinfo {author} {\bibfnamefont {F.~m. c.-M.}\
  \bibnamefont {Le~R\'egent}}, \bibinfo {author} {\bibfnamefont
  {A.}~\bibnamefont {Kirsch}}, \bibinfo {author} {\bibfnamefont
  {Y.}~\bibnamefont {Wang}}, \bibinfo {author} {\bibfnamefont {A.~J.}\
  \bibnamefont {Kim}}, \bibinfo {author} {\bibfnamefont {E.}~\bibnamefont
  {Kozik}}, \bibinfo {author} {\bibfnamefont {E.~A.}\ \bibnamefont {Stepanov}},
  \bibinfo {author} {\bibfnamefont {A.}~\bibnamefont {Kauch}}, \bibinfo
  {author} {\bibfnamefont {S.}~\bibnamefont {Andergassen}}, \bibinfo {author}
  {\bibfnamefont {P.}~\bibnamefont {Hansmann}}, \bibinfo {author}
  {\bibfnamefont {D.}~\bibnamefont {Rohe}}, \bibinfo {author} {\bibfnamefont
  {Y.~M.}\ \bibnamefont {Vilk}}, \bibinfo {author} {\bibfnamefont {J.~P.~F.}\
  \bibnamefont {LeBlanc}}, \bibinfo {author} {\bibfnamefont {S.}~\bibnamefont
  {Zhang}}, \bibinfo {author} {\bibfnamefont {A.-M.~S.}\ \bibnamefont
  {Tremblay}}, \bibinfo {author} {\bibfnamefont {M.}~\bibnamefont {Ferrero}},
  \bibinfo {author} {\bibfnamefont {O.}~\bibnamefont {Parcollet}},\ and\
  \bibinfo {author} {\bibfnamefont {A.}~\bibnamefont {Georges}},\ }\bibfield
  {title} {\bibinfo {title} {Tracking the {F}ootprints of {S}pin
  {F}luctuations: A {M}ulti{M}ethod, {M}ulti{M}essenger {S}tudy of the
  {T}wo-{D}imensional {H}ubbard {M}odel},\ }\href
  {https://doi.org/10.1103/PhysRevX.11.011058} {\bibfield  {journal} {\bibinfo
  {journal} {Phys. Rev. X}\ }\textbf {\bibinfo {volume} {11}},\ \bibinfo
  {pages} {011058} (\bibinfo {year} {2021})}\BibitemShut {NoStop}%
\bibitem [{\citenamefont {Qin}\ \emph {et~al.}(2020)\citenamefont {Qin},
  \citenamefont {Chung}, \citenamefont {Shi}, \citenamefont {Vitali},
  \citenamefont {Hubig}, \citenamefont {Schollw\"ock}, \citenamefont {White},\
  and\ \citenamefont {Zhang}}]{Qin2020}%
  \BibitemOpen
  \bibfield  {author} {\bibinfo {author} {\bibfnamefont {M.}~\bibnamefont
  {Qin}}, \bibinfo {author} {\bibfnamefont {C.-M.}\ \bibnamefont {Chung}},
  \bibinfo {author} {\bibfnamefont {H.}~\bibnamefont {Shi}}, \bibinfo {author}
  {\bibfnamefont {E.}~\bibnamefont {Vitali}}, \bibinfo {author} {\bibfnamefont
  {C.}~\bibnamefont {Hubig}}, \bibinfo {author} {\bibfnamefont
  {U.}~\bibnamefont {Schollw\"ock}}, \bibinfo {author} {\bibfnamefont {S.~R.}\
  \bibnamefont {White}},\ and\ \bibinfo {author} {\bibfnamefont
  {S.}~\bibnamefont {Zhang}} (\bibinfo {collaboration} {Simons Collaboration on
  the Many-Electron Problem}),\ }\bibfield  {title} {\bibinfo {title} {Absence
  of {S}uperconductivity in the {P}ure {T}wo-{D}imensional {H}ubbard {M}odel},\
  }\href {https://doi.org/10.1103/PhysRevX.10.031016} {\bibfield  {journal}
  {\bibinfo  {journal} {Phys. Rev. X}\ }\textbf {\bibinfo {volume} {10}},\
  \bibinfo {pages} {031016} (\bibinfo {year} {2020})}\BibitemShut {NoStop}%
\bibitem [{\citenamefont {Qin}\ \emph {et~al.}(2021)\citenamefont {Qin},
  \citenamefont {Sch{\"a}fer}, \citenamefont {Andergassen}, \citenamefont
  {Corboz},\ and\ \citenamefont {Gull}}]{Qin2021}%
  \BibitemOpen
  \bibfield  {author} {\bibinfo {author} {\bibfnamefont {M.}~\bibnamefont
  {Qin}}, \bibinfo {author} {\bibfnamefont {T.}~\bibnamefont {Sch{\"a}fer}},
  \bibinfo {author} {\bibfnamefont {S.}~\bibnamefont {Andergassen}}, \bibinfo
  {author} {\bibfnamefont {P.}~\bibnamefont {Corboz}},\ and\ \bibinfo {author}
  {\bibfnamefont {E.}~\bibnamefont {Gull}},\ }\bibfield  {title} {\bibinfo
  {title} {The {H}ubbard model: {A} computational perspective},\ }\href@noop {}
  {\bibfield  {journal} {\bibinfo  {journal} {arXiv:2104.00064v1}\ } (\bibinfo
  {year} {2021})}\BibitemShut {NoStop}%
\bibitem [{\citenamefont {Li}\ \emph {et~al.}(1998)\citenamefont {Li},
  \citenamefont {Ma}, \citenamefont {Shi},\ and\ \citenamefont
  {Zhang}}]{Li1998}%
  \BibitemOpen
  \bibfield  {author} {\bibinfo {author} {\bibfnamefont {Y.~Q.}\ \bibnamefont
  {Li}}, \bibinfo {author} {\bibfnamefont {M.}~\bibnamefont {Ma}}, \bibinfo
  {author} {\bibfnamefont {D.~N.}\ \bibnamefont {Shi}},\ and\ \bibinfo {author}
  {\bibfnamefont {F.~C.}\ \bibnamefont {Zhang}},\ }\bibfield  {title} {\bibinfo
  {title} {{SU}(4) {T}heory for {S}pin {S}ystems with {O}rbital {D}egeneracy},\
  }\href {https://doi.org/10.1103/physrevlett.81.3527} {\bibfield  {journal}
  {\bibinfo  {journal} {Phys. Rev. Lett.}\ }\textbf {\bibinfo {volume} {81}},\
  \bibinfo {pages} {3527} (\bibinfo {year} {1998})}\BibitemShut {NoStop}%
\bibitem [{\citenamefont {Tokura}(2000)}]{Tokura2000}%
  \BibitemOpen
  \bibfield  {author} {\bibinfo {author} {\bibfnamefont {Y.}~\bibnamefont
  {Tokura}},\ }\bibfield  {title} {\bibinfo {title} {{O}rbital {P}hysics in
  {T}ransition-{M}etal {O}xides},\ }\href
  {https://doi.org/10.1126/science.288.5465.462} {\bibfield  {journal}
  {\bibinfo  {journal} {Science}\ }\textbf {\bibinfo {volume} {288}},\ \bibinfo
  {pages} {462} (\bibinfo {year} {2000})}\BibitemShut {NoStop}%
\bibitem [{\citenamefont {Dagotto}\ \emph {et~al.}(2001)\citenamefont
  {Dagotto}, \citenamefont {Hotta},\ and\ \citenamefont {Moreo}}]{Dagotto2001}%
  \BibitemOpen
  \bibfield  {author} {\bibinfo {author} {\bibfnamefont {E.}~\bibnamefont
  {Dagotto}}, \bibinfo {author} {\bibfnamefont {T.}~\bibnamefont {Hotta}},\
  and\ \bibinfo {author} {\bibfnamefont {A.}~\bibnamefont {Moreo}},\ }\bibfield
   {title} {\bibinfo {title} {Colossal magnetoresistant materials: the key role
  of phase separation},\ }\href@noop {} {\bibfield  {journal} {\bibinfo
  {journal} {Physics reports}\ }\textbf {\bibinfo {volume} {344}},\ \bibinfo
  {pages} {1} (\bibinfo {year} {2001})}\BibitemShut {NoStop}%
\bibitem [{\citenamefont {Goerbig}(2011)}]{Goerbig2011}%
  \BibitemOpen
  \bibfield  {author} {\bibinfo {author} {\bibfnamefont {M.~O.}\ \bibnamefont
  {Goerbig}},\ }\bibfield  {title} {\bibinfo {title} {Electronic properties of
  graphene in a strong magnetic field},\ }\href
  {https://doi.org/10.1103/RevModPhys.83.1193} {\bibfield  {journal} {\bibinfo
  {journal} {Rev. Mod. Phys.}\ }\textbf {\bibinfo {volume} {83}},\ \bibinfo
  {pages} {1193} (\bibinfo {year} {2011})}\BibitemShut {NoStop}%
\bibitem [{\citenamefont {Xu}\ \emph {et~al.}(2018{\natexlab{a}})\citenamefont
  {Xu}, \citenamefont {Law},\ and\ \citenamefont {Lee}}]{Xu2018_Kekule}%
  \BibitemOpen
  \bibfield  {author} {\bibinfo {author} {\bibfnamefont {X.~Y.}\ \bibnamefont
  {Xu}}, \bibinfo {author} {\bibfnamefont {K.~T.}\ \bibnamefont {Law}},\ and\
  \bibinfo {author} {\bibfnamefont {P.~A.}\ \bibnamefont {Lee}},\ }\bibfield
  {title} {\bibinfo {title} {Kekul\'e valence bond order in an extended hubbard
  model on the honeycomb lattice with possible applications to twisted bilayer
  graphene},\ }\href {https://doi.org/10.1103/PhysRevB.98.121406} {\bibfield
  {journal} {\bibinfo  {journal} {Phys. Rev. B}\ }\textbf {\bibinfo {volume}
  {98}},\ \bibinfo {pages} {121406(R)} (\bibinfo {year}
  {2018}{\natexlab{a}})}\BibitemShut {NoStop}%
\bibitem [{\citenamefont {You}\ and\ \citenamefont
  {Vishwanath}(2019)}]{You2019}%
  \BibitemOpen
  \bibfield  {author} {\bibinfo {author} {\bibfnamefont {Y.}~\bibnamefont
  {You}}\ and\ \bibinfo {author} {\bibfnamefont {A.}~\bibnamefont
  {Vishwanath}},\ }\bibfield  {title} {\bibinfo {title} {Superconductivity from
  valley fluctuations and approximate {SO}(4) symmetry in a weak coupling
  theory of twisted bilayer graphene},\ }\href
  {https://doi.org/10.1103/PhysRevB.100.205131} {\bibfield  {journal} {\bibinfo
   {journal} {npj Quantum Mater.}\ }\textbf {\bibinfo {volume} {4}},\ \bibinfo
  {pages} {16} (\bibinfo {year} {2019})}\BibitemShut {NoStop}%
\bibitem [{\citenamefont {Natori}\ \emph {et~al.}(2019)\citenamefont {Natori},
  \citenamefont {Nutakki}, \citenamefont {Pereira},\ and\ \citenamefont
  {Andrade}}]{Natori2019}%
  \BibitemOpen
  \bibfield  {author} {\bibinfo {author} {\bibfnamefont {W.~M.~H.}\
  \bibnamefont {Natori}}, \bibinfo {author} {\bibfnamefont {R.}~\bibnamefont
  {Nutakki}}, \bibinfo {author} {\bibfnamefont {R.~G.}\ \bibnamefont
  {Pereira}},\ and\ \bibinfo {author} {\bibfnamefont {E.~C.}\ \bibnamefont
  {Andrade}},\ }\bibfield  {title} {\bibinfo {title} {{SU}(4) {H}eisenberg
  model on the honeycomb lattice with exchange-frustrated perturbations:
  Implications for twistronics and {M}ott insulators},\ }\href
  {https://doi.org/10.1103/PhysRevB.100.205131} {\bibfield  {journal} {\bibinfo
   {journal} {Phys. Rev. B}\ }\textbf {\bibinfo {volume} {100}},\ \bibinfo
  {pages} {205131} (\bibinfo {year} {2019})}\BibitemShut {NoStop}%
\bibitem [{\citenamefont {Da~Liao}\ \emph {et~al.}(2021)\citenamefont
  {Da~Liao}, \citenamefont {Kang}, \citenamefont {Brei\o{}}, \citenamefont
  {Xu}, \citenamefont {Wu}, \citenamefont {Andersen}, \citenamefont
  {Fernandes},\ and\ \citenamefont {Meng}}]{DaLiao2021}%
  \BibitemOpen
  \bibfield  {author} {\bibinfo {author} {\bibfnamefont {Y.}~\bibnamefont
  {Da~Liao}}, \bibinfo {author} {\bibfnamefont {J.}~\bibnamefont {Kang}},
  \bibinfo {author} {\bibfnamefont {C.~N.}\ \bibnamefont {Brei\o{}}}, \bibinfo
  {author} {\bibfnamefont {X.~Y.}\ \bibnamefont {Xu}}, \bibinfo {author}
  {\bibfnamefont {H.-Q.}\ \bibnamefont {Wu}}, \bibinfo {author} {\bibfnamefont
  {B.~M.}\ \bibnamefont {Andersen}}, \bibinfo {author} {\bibfnamefont {R.~M.}\
  \bibnamefont {Fernandes}},\ and\ \bibinfo {author} {\bibfnamefont {Z.~Y.}\
  \bibnamefont {Meng}},\ }\bibfield  {title} {\bibinfo {title}
  {Correlation-induced insulating topological phases at charge neutrality in
  twisted bilayer graphene},\ }\href
  {https://doi.org/10.1103/PhysRevX.11.011014} {\bibfield  {journal} {\bibinfo
  {journal} {Phys. Rev. X}\ }\textbf {\bibinfo {volume} {11}},\ \bibinfo
  {pages} {011014} (\bibinfo {year} {2021})}\BibitemShut {NoStop}%
\bibitem [{\citenamefont {Liao}\ \emph {et~al.}(2021)\citenamefont {Liao},
  \citenamefont {Xu}, \citenamefont {Meng},\ and\ \citenamefont
  {Kang}}]{Liao2021}%
  \BibitemOpen
  \bibfield  {author} {\bibinfo {author} {\bibfnamefont {Y.-D.}\ \bibnamefont
  {Liao}}, \bibinfo {author} {\bibfnamefont {X.-Y.}\ \bibnamefont {Xu}},
  \bibinfo {author} {\bibfnamefont {Z.-Y.}\ \bibnamefont {Meng}},\ and\
  \bibinfo {author} {\bibfnamefont {J.}~\bibnamefont {Kang}},\ }\bibfield
  {title} {\bibinfo {title} {Correlated insulating phases in the twisted
  bilayer graphene},\ }\href {https://doi.org/10.1088/1674-1056/abcfa3}
  {\bibfield  {journal} {\bibinfo  {journal} {Chin. Phys. B}\ }\textbf
  {\bibinfo {volume} {30}},\ \bibinfo {pages} {017305} (\bibinfo {year}
  {2021})}\BibitemShut {NoStop}%
\bibitem [{\citenamefont {Chichinadze}\ \emph {et~al.}(2021)\citenamefont
  {Chichinadze}, \citenamefont {Classen}, \citenamefont {Wang},\ and\
  \citenamefont {Chubukov}}]{Chichinadze2021}%
  \BibitemOpen
  \bibfield  {author} {\bibinfo {author} {\bibfnamefont {D.~V.}\ \bibnamefont
  {Chichinadze}}, \bibinfo {author} {\bibfnamefont {L.}~\bibnamefont
  {Classen}}, \bibinfo {author} {\bibfnamefont {Y.}~\bibnamefont {Wang}},\ and\
  \bibinfo {author} {\bibfnamefont {A.~V.}\ \bibnamefont {Chubukov}},\
  }\bibfield  {title} {\bibinfo {title} {{SU}(4) symmetry in twisted bilayer
  graphene - an itinerant perspective},\ }\href@noop {} {\bibfield  {journal}
  {\bibinfo  {journal} {arXiv:2108.05334v1}\ } (\bibinfo {year}
  {2021})}\BibitemShut {NoStop}%
\bibitem [{\citenamefont {Yamashita}\ \emph {et~al.}(2013)\citenamefont
  {Yamashita}, \citenamefont {Tomura}, \citenamefont {Yanagi},\ and\
  \citenamefont {Ueda}}]{Yamashita2013}%
  \BibitemOpen
  \bibfield  {author} {\bibinfo {author} {\bibfnamefont {Y.}~\bibnamefont
  {Yamashita}}, \bibinfo {author} {\bibfnamefont {M.}~\bibnamefont {Tomura}},
  \bibinfo {author} {\bibfnamefont {Y.}~\bibnamefont {Yanagi}},\ and\ \bibinfo
  {author} {\bibfnamefont {K.}~\bibnamefont {Ueda}},\ }\bibfield  {title}
  {\bibinfo {title} {{SU} (3) {D}irac electrons in the 1/5-depleted
  square-lattice {H}ubbard model at 1/4 filling},\ }\href@noop {} {\bibfield
  {journal} {\bibinfo  {journal} {Phys. Rev. B}\ }\textbf {\bibinfo {volume}
  {88}},\ \bibinfo {pages} {195104} (\bibinfo {year} {2013})}\BibitemShut
  {NoStop}%
\bibitem [{\citenamefont {Titvinidze}\ \emph {et~al.}(2011)\citenamefont
  {Titvinidze}, \citenamefont {Privitera}, \citenamefont {Chang}, \citenamefont
  {Diehl}, \citenamefont {Baranov}, \citenamefont {Daley},\ and\ \citenamefont
  {Hofstetter}}]{Titvinidze2011}%
  \BibitemOpen
  \bibfield  {author} {\bibinfo {author} {\bibfnamefont {I.}~\bibnamefont
  {Titvinidze}}, \bibinfo {author} {\bibfnamefont {A.}~\bibnamefont
  {Privitera}}, \bibinfo {author} {\bibfnamefont {S.-Y.}\ \bibnamefont
  {Chang}}, \bibinfo {author} {\bibfnamefont {S.}~\bibnamefont {Diehl}},
  \bibinfo {author} {\bibfnamefont {M.~A.}\ \bibnamefont {Baranov}}, \bibinfo
  {author} {\bibfnamefont {A.}~\bibnamefont {Daley}},\ and\ \bibinfo {author}
  {\bibfnamefont {W.}~\bibnamefont {Hofstetter}},\ }\bibfield  {title}
  {\bibinfo {title} {Magnetism and domain formation in {SU}(3)-symmetric
  multi-species {F}ermi mixtures},\ }\href@noop {} {\bibfield  {journal}
  {\bibinfo  {journal} {New J. Phys.}\ }\textbf {\bibinfo {volume} {13}},\
  \bibinfo {pages} {035013} (\bibinfo {year} {2011})}\BibitemShut {NoStop}%
\bibitem [{\citenamefont {Sotnikov}\ and\ \citenamefont
  {Hofstetter}(2014)}]{Sotnikov2014}%
  \BibitemOpen
  \bibfield  {author} {\bibinfo {author} {\bibfnamefont {A.}~\bibnamefont
  {Sotnikov}}\ and\ \bibinfo {author} {\bibfnamefont {W.}~\bibnamefont
  {Hofstetter}},\ }\bibfield  {title} {\bibinfo {title} {Magnetic ordering of
  three-component ultracold fermionic mixtures in optical lattices},\ }\href
  {https://doi.org/10.1103/PhysRevA.89.063601} {\bibfield  {journal} {\bibinfo
  {journal} {Phys. Rev. A}\ }\textbf {\bibinfo {volume} {89}},\ \bibinfo
  {pages} {063601} (\bibinfo {year} {2014})}\BibitemShut {NoStop}%
\bibitem [{\citenamefont {Sotnikov}(2015)}]{Sotnikov2015}%
  \BibitemOpen
  \bibfield  {author} {\bibinfo {author} {\bibfnamefont {A.}~\bibnamefont
  {Sotnikov}},\ }\bibfield  {title} {\bibinfo {title} {Critical entropies and
  magnetic-phase-diagram analysis of ultracold three-component fermionic
  mixtures in optical lattices},\ }\href
  {https://doi.org/10.1103/PhysRevA.92.023633} {\bibfield  {journal} {\bibinfo
  {journal} {Phys. Rev. A}\ }\textbf {\bibinfo {volume} {92}},\ \bibinfo
  {pages} {023633} (\bibinfo {year} {2015})}\BibitemShut {NoStop}%
\bibitem [{\citenamefont {Hafez-Torbati}\ and\ \citenamefont
  {Hofstetter}(2018)}]{Hafez2018}%
  \BibitemOpen
  \bibfield  {author} {\bibinfo {author} {\bibfnamefont {M.}~\bibnamefont
  {Hafez-Torbati}}\ and\ \bibinfo {author} {\bibfnamefont {W.}~\bibnamefont
  {Hofstetter}},\ }\bibfield  {title} {\bibinfo {title} {Artificial {SU}(3)
  spin-orbit coupling and exotic {M}ott insulators},\ }\href
  {https://doi.org/10.1103/PhysRevB.98.245131} {\bibfield  {journal} {\bibinfo
  {journal} {Phys. Rev. B}\ }\textbf {\bibinfo {volume} {98}},\ \bibinfo
  {pages} {245131} (\bibinfo {year} {2018})}\BibitemShut {NoStop}%
\bibitem [{\citenamefont {Hafez-Torbati}\ and\ \citenamefont
  {Hofstetter}(2019)}]{Hafez2019}%
  \BibitemOpen
  \bibfield  {author} {\bibinfo {author} {\bibfnamefont {M.}~\bibnamefont
  {Hafez-Torbati}}\ and\ \bibinfo {author} {\bibfnamefont {W.}~\bibnamefont
  {Hofstetter}},\ }\bibfield  {title} {\bibinfo {title} {Competing charge and
  magnetic order in fermionic multicomponent systems},\ }\href
  {https://doi.org/10.1103/PhysRevB.100.035133} {\bibfield  {journal} {\bibinfo
   {journal} {Phys. Rev. B}\ }\textbf {\bibinfo {volume} {100}},\ \bibinfo
  {pages} {035133} (\bibinfo {year} {2019})}\BibitemShut {NoStop}%
\bibitem [{\citenamefont {Hafez-Torbati}\ \emph {et~al.}(2020)\citenamefont
  {Hafez-Torbati}, \citenamefont {Zheng}, \citenamefont {Irsigler},\ and\
  \citenamefont {Hofstetter}}]{Hafez2020}%
  \BibitemOpen
  \bibfield  {author} {\bibinfo {author} {\bibfnamefont {M.}~\bibnamefont
  {Hafez-Torbati}}, \bibinfo {author} {\bibfnamefont {J.-H.}\ \bibnamefont
  {Zheng}}, \bibinfo {author} {\bibfnamefont {B.}~\bibnamefont {Irsigler}},\
  and\ \bibinfo {author} {\bibfnamefont {W.}~\bibnamefont {Hofstetter}},\
  }\bibfield  {title} {\bibinfo {title} {Interaction-driven topological phase
  transitions in fermionic {SU}(3) systems},\ }\href
  {https://doi.org/10.1103/PhysRevB.101.245159} {\bibfield  {journal} {\bibinfo
   {journal} {Phys. Rev. B}\ }\textbf {\bibinfo {volume} {101}},\ \bibinfo
  {pages} {245159} (\bibinfo {year} {2020})}\BibitemShut {NoStop}%
\bibitem [{\citenamefont {Nie}\ \emph {et~al.}(2017)\citenamefont {Nie},
  \citenamefont {Zhang},\ and\ \citenamefont {Zhang}}]{Nie2017}%
  \BibitemOpen
  \bibfield  {author} {\bibinfo {author} {\bibfnamefont {W.}~\bibnamefont
  {Nie}}, \bibinfo {author} {\bibfnamefont {D.}~\bibnamefont {Zhang}},\ and\
  \bibinfo {author} {\bibfnamefont {W.}~\bibnamefont {Zhang}},\ }\bibfield
  {title} {\bibinfo {title} {Ferromagnetic ground state of the {SU}(3)
  {H}ubbard model on the {L}ieb lattice},\ }\href@noop {} {\bibfield  {journal}
  {\bibinfo  {journal} {Phys. Rev. A}\ }\textbf {\bibinfo {volume} {96}},\
  \bibinfo {pages} {053616} (\bibinfo {year} {2017})}\BibitemShut {NoStop}%
\bibitem [{\citenamefont {Honerkamp}\ and\ \citenamefont
  {Hofstetter}(2004)}]{Honerkamp2004}%
  \BibitemOpen
  \bibfield  {author} {\bibinfo {author} {\bibfnamefont {C.}~\bibnamefont
  {Honerkamp}}\ and\ \bibinfo {author} {\bibfnamefont {W.}~\bibnamefont
  {Hofstetter}},\ }\bibfield  {title} {\bibinfo {title} {{U}ltracold fermions
  and the {SU($N$)} {H}ubbard {M}odel},\ }\href
  {https://doi.org/10.1103/physrevlett.92.170403} {\bibfield  {journal}
  {\bibinfo  {journal} {Phys. Rev. Lett.}\ }\textbf {\bibinfo {volume} {92}},\
  \bibinfo {pages} {170403} (\bibinfo {year} {2004})}\BibitemShut {NoStop}%
\bibitem [{\citenamefont {Hofstetter}(2005)}]{Hofstetter_chapter}%
  \BibitemOpen
  \bibfield  {author} {\bibinfo {author} {\bibfnamefont {W.}~\bibnamefont
  {Hofstetter}},\ }\href {https://doi.org/10.1007/11423256_9} {\emph {\bibinfo
  {title} {Advances in Solid State Physics: Flavor Degeneracy and Effects of
  Disorder in Ultracold Atom Systems}}}\ (\bibinfo  {publisher} {Springer,
  Berlin, Heidelberg},\ \bibinfo {year} {2005})\BibitemShut {NoStop}%
\bibitem [{\citenamefont {Unukovych}\ and\ \citenamefont
  {Sotnikov}(2021)}]{Unukovych2021}%
  \BibitemOpen
  \bibfield  {author} {\bibinfo {author} {\bibfnamefont {V.}~\bibnamefont
  {Unukovych}}\ and\ \bibinfo {author} {\bibfnamefont {A.}~\bibnamefont
  {Sotnikov}},\ }\bibfield  {title} {\bibinfo {title} {{SU}(4)-symmetric
  {H}ubbard model at quarter filling: insights from the dynamical mean-field
  approach},\ }\href@noop {} {\bibfield  {journal} {\bibinfo  {journal}
  {arXiv:2107.11219v1}\ } (\bibinfo {year} {2021})}\BibitemShut {NoStop}%
\bibitem [{\citenamefont {Chen}\ \emph {et~al.}(2016)\citenamefont {Chen},
  \citenamefont {Hazzard}, \citenamefont {Rey},\ and\ \citenamefont
  {Hermele}}]{Chen2016}%
  \BibitemOpen
  \bibfield  {author} {\bibinfo {author} {\bibfnamefont {G.}~\bibnamefont
  {Chen}}, \bibinfo {author} {\bibfnamefont {K.~R.~A.}\ \bibnamefont
  {Hazzard}}, \bibinfo {author} {\bibfnamefont {A.~M.}\ \bibnamefont {Rey}},\
  and\ \bibinfo {author} {\bibfnamefont {M.}~\bibnamefont {Hermele}},\
  }\bibfield  {title} {\bibinfo {title} {Synthetic-gauge-field stabilization of
  the chiral-spin-liquid phase},\ }\href
  {https://doi.org/10.1103/PhysRevA.93.061601} {\bibfield  {journal} {\bibinfo
  {journal} {Phys. Rev. A}\ }\textbf {\bibinfo {volume} {93}},\ \bibinfo
  {pages} {061601(R)} (\bibinfo {year} {2016})}\BibitemShut {NoStop}%
\bibitem [{\citenamefont {Wang}\ \emph {et~al.}(2014)\citenamefont {Wang},
  \citenamefont {Li}, \citenamefont {Cai}, \citenamefont {Zhou}, \citenamefont
  {Wang},\ and\ \citenamefont {Wu}}]{Wang2014}%
  \BibitemOpen
  \bibfield  {author} {\bibinfo {author} {\bibfnamefont {D.}~\bibnamefont
  {Wang}}, \bibinfo {author} {\bibfnamefont {Y.}~\bibnamefont {Li}}, \bibinfo
  {author} {\bibfnamefont {Z.}~\bibnamefont {Cai}}, \bibinfo {author}
  {\bibfnamefont {Z.}~\bibnamefont {Zhou}}, \bibinfo {author} {\bibfnamefont
  {Y.}~\bibnamefont {Wang}},\ and\ \bibinfo {author} {\bibfnamefont
  {C.}~\bibnamefont {Wu}},\ }\bibfield  {title} {\bibinfo {title} {Competing
  orders in the 2{D} half-filled {SU}(2{$N$}) {H}ubbard model through the
  pinning-field quantum {M}onte {C}arlo simulations},\ }\href
  {https://doi.org/10.1103/PhysRevLett.112.156403} {\bibfield  {journal}
  {\bibinfo  {journal} {Phys. Rev. Lett.}\ }\textbf {\bibinfo {volume} {112}},\
  \bibinfo {pages} {156403} (\bibinfo {year} {2014})}\BibitemShut {NoStop}%
\bibitem [{\citenamefont {Zhou}\ \emph {et~al.}(2014)\citenamefont {Zhou},
  \citenamefont {Cai}, \citenamefont {Wu},\ and\ \citenamefont
  {Wang}}]{Zhou2014}%
  \BibitemOpen
  \bibfield  {author} {\bibinfo {author} {\bibfnamefont {Z.}~\bibnamefont
  {Zhou}}, \bibinfo {author} {\bibfnamefont {Z.}~\bibnamefont {Cai}}, \bibinfo
  {author} {\bibfnamefont {C.}~\bibnamefont {Wu}},\ and\ \bibinfo {author}
  {\bibfnamefont {Y.}~\bibnamefont {Wang}},\ }\bibfield  {title} {\bibinfo
  {title} {Quantum {M}onte {C}arlo simulations of thermodynamic properties of
  {SU(2$N$)} ultracold fermions in optical lattices},\ }\href
  {https://doi.org/10.1103/PhysRevB.90.235139} {\bibfield  {journal} {\bibinfo
  {journal} {Phys. Rev. B}\ }\textbf {\bibinfo {volume} {90}},\ \bibinfo
  {pages} {235139} (\bibinfo {year} {2014})}\BibitemShut {NoStop}%
\bibitem [{\citenamefont {Wang}\ \emph {et~al.}(2019)\citenamefont {Wang},
  \citenamefont {Wang},\ and\ \citenamefont {Wu}}]{Wang2019}%
  \BibitemOpen
  \bibfield  {author} {\bibinfo {author} {\bibfnamefont {D.}~\bibnamefont
  {Wang}}, \bibinfo {author} {\bibfnamefont {L.}~\bibnamefont {Wang}},\ and\
  \bibinfo {author} {\bibfnamefont {C.}~\bibnamefont {Wu}},\ }\bibfield
  {title} {\bibinfo {title} {Slater and {M}ott insulating states in the {SU}(6)
  {H}ubbard model},\ }\href {https://doi.org/10.1103/PhysRevB.100.115155}
  {\bibfield  {journal} {\bibinfo  {journal} {Phys. Rev. B}\ }\textbf {\bibinfo
  {volume} {100}},\ \bibinfo {pages} {115155} (\bibinfo {year}
  {2019})}\BibitemShut {NoStop}%
\bibitem [{\citenamefont {Golubeva}\ \emph {et~al.}(2017)\citenamefont
  {Golubeva}, \citenamefont {Sotnikov}, \citenamefont {Cichy}, \citenamefont
  {Kune\ifmmode~\check{s}\else \v{s}\fi{}},\ and\ \citenamefont
  {Hofstetter}}]{Goubeva2017}%
  \BibitemOpen
  \bibfield  {author} {\bibinfo {author} {\bibfnamefont {A.}~\bibnamefont
  {Golubeva}}, \bibinfo {author} {\bibfnamefont {A.}~\bibnamefont {Sotnikov}},
  \bibinfo {author} {\bibfnamefont {A.}~\bibnamefont {Cichy}}, \bibinfo
  {author} {\bibfnamefont {J.}~\bibnamefont {Kune\ifmmode~\check{s}\else
  \v{s}\fi{}}},\ and\ \bibinfo {author} {\bibfnamefont {W.}~\bibnamefont
  {Hofstetter}},\ }\bibfield  {title} {\bibinfo {title} {Breaking of {SU}(4)
  symmetry and interplay between strongly correlated phases in the {H}ubbard
  model},\ }\href {https://doi.org/10.1103/PhysRevB.95.125108} {\bibfield
  {journal} {\bibinfo  {journal} {Phys. Rev. B}\ }\textbf {\bibinfo {volume}
  {95}},\ \bibinfo {pages} {125108} (\bibinfo {year} {2017})}\BibitemShut
  {NoStop}%
\bibitem [{\citenamefont {Xu}\ \emph {et~al.}(2019)\citenamefont {Xu},
  \citenamefont {Wang}, \citenamefont {Zhou},\ and\ \citenamefont
  {Wu}}]{Xu2019}%
  \BibitemOpen
  \bibfield  {author} {\bibinfo {author} {\bibfnamefont {H.}~\bibnamefont
  {Xu}}, \bibinfo {author} {\bibfnamefont {Y.}~\bibnamefont {Wang}}, \bibinfo
  {author} {\bibfnamefont {Z.}~\bibnamefont {Zhou}},\ and\ \bibinfo {author}
  {\bibfnamefont {C.}~\bibnamefont {Wu}},\ }\bibfield  {title} {\bibinfo
  {title} {Mott insulating states of the anisotropic {SU}(4) {D}irac
  fermions},\ }\href@noop {} {\bibfield  {journal} {\bibinfo  {journal}
  {arXiv:1912.11791v1}\ } (\bibinfo {year} {2019})}\BibitemShut {NoStop}%
\bibitem [{\citenamefont {Zhou}\ \emph {et~al.}(2018)\citenamefont {Zhou},
  \citenamefont {Wu},\ and\ \citenamefont {Wang}}]{Zhou2018}%
  \BibitemOpen
  \bibfield  {author} {\bibinfo {author} {\bibfnamefont {Z.}~\bibnamefont
  {Zhou}}, \bibinfo {author} {\bibfnamefont {C.}~\bibnamefont {Wu}},\ and\
  \bibinfo {author} {\bibfnamefont {Y.}~\bibnamefont {Wang}},\ }\bibfield
  {title} {\bibinfo {title} {Mott transition in the $\ensuremath{\pi}$-flux
  {SU}(4) {H}ubbard model on a square lattice},\ }\href
  {https://doi.org/10.1103/PhysRevB.97.195122} {\bibfield  {journal} {\bibinfo
  {journal} {Phys. Rev. B}\ }\textbf {\bibinfo {volume} {97}},\ \bibinfo
  {pages} {195122} (\bibinfo {year} {2018})}\BibitemShut {NoStop}%
\bibitem [{\citenamefont {Zhou}\ \emph {et~al.}(2017)\citenamefont {Zhou},
  \citenamefont {Wang}, \citenamefont {Wu},\ and\ \citenamefont
  {Wang}}]{Zhou2017}%
  \BibitemOpen
  \bibfield  {author} {\bibinfo {author} {\bibfnamefont {Z.}~\bibnamefont
  {Zhou}}, \bibinfo {author} {\bibfnamefont {D.}~\bibnamefont {Wang}}, \bibinfo
  {author} {\bibfnamefont {C.}~\bibnamefont {Wu}},\ and\ \bibinfo {author}
  {\bibfnamefont {Y.}~\bibnamefont {Wang}},\ }\bibfield  {title} {\bibinfo
  {title} {Finite-temperature valence-bond-solid transitions and thermodynamic
  properties of interacting {SU$(2N)$} {D}irac fermions},\ }\href
  {https://doi.org/10.1103/PhysRevB.95.085128} {\bibfield  {journal} {\bibinfo
  {journal} {Phys. Rev. B}\ }\textbf {\bibinfo {volume} {95}},\ \bibinfo
  {pages} {085128} (\bibinfo {year} {2017})}\BibitemShut {NoStop}%
\bibitem [{\citenamefont {Zhou}\ \emph {et~al.}(2016)\citenamefont {Zhou},
  \citenamefont {Wang}, \citenamefont {Meng}, \citenamefont {Wang},\ and\
  \citenamefont {Wu}}]{Zhou2016}%
  \BibitemOpen
  \bibfield  {author} {\bibinfo {author} {\bibfnamefont {Z.}~\bibnamefont
  {Zhou}}, \bibinfo {author} {\bibfnamefont {D.}~\bibnamefont {Wang}}, \bibinfo
  {author} {\bibfnamefont {Z.~Y.}\ \bibnamefont {Meng}}, \bibinfo {author}
  {\bibfnamefont {Y.}~\bibnamefont {Wang}},\ and\ \bibinfo {author}
  {\bibfnamefont {C.}~\bibnamefont {Wu}},\ }\bibfield  {title} {\bibinfo
  {title} {Mott insulating states and quantum phase transitions of correlated
  {SU$(2N)$} {D}irac fermions},\ }\href
  {https://doi.org/10.1103/PhysRevB.93.245157} {\bibfield  {journal} {\bibinfo
  {journal} {Phys. Rev. B}\ }\textbf {\bibinfo {volume} {93}},\ \bibinfo
  {pages} {245157} (\bibinfo {year} {2016})}\BibitemShut {NoStop}%
\bibitem [{\citenamefont {Ouyang}\ and\ \citenamefont {Xu}(2021)}]{Ouyang2021}%
  \BibitemOpen
  \bibfield  {author} {\bibinfo {author} {\bibfnamefont {Y.}~\bibnamefont
  {Ouyang}}\ and\ \bibinfo {author} {\bibfnamefont {X.~Y.}\ \bibnamefont
  {Xu}},\ }\bibfield  {title} {\bibinfo {title} {Projection of infinite-$u$
  {H}ubbard model and algebraic sign structure},\ }\href@noop {} {\bibfield
  {journal} {\bibinfo  {journal} {arXiv:2108.04832v2}\ } (\bibinfo {year}
  {2021})}\BibitemShut {NoStop}%
\bibitem [{\citenamefont {Read}\ and\ \citenamefont {Newns}(1983)}]{Read1983}%
  \BibitemOpen
  \bibfield  {author} {\bibinfo {author} {\bibfnamefont {N.}~\bibnamefont
  {Read}}\ and\ \bibinfo {author} {\bibfnamefont {D.~M.}\ \bibnamefont
  {Newns}},\ }\bibfield  {title} {\bibinfo {title} {On the solution of the
  {C}oqblin-{S}chreiffer {H}amiltonian by the large-{N} expansion technique},\
  }\href {https://doi.org/10.1088/0022-3719/16/17/014} {\bibfield  {journal}
  {\bibinfo  {journal} {J. Phys. C: Solid State Phys.}\ }\textbf {\bibinfo
  {volume} {16}},\ \bibinfo {pages} {3273} (\bibinfo {year}
  {1983})}\BibitemShut {NoStop}%
\bibitem [{\citenamefont {Affleck}(1985)}]{Affleck1985}%
  \BibitemOpen
  \bibfield  {author} {\bibinfo {author} {\bibfnamefont {I.}~\bibnamefont
  {Affleck}},\ }\bibfield  {title} {\bibinfo {title} {Large-{N} {L}imit of
  {SU($N$)} {Q}uantum ``{S}pin" {C}hains},\ }\href
  {https://doi.org/10.1103/physrevlett.54.966} {\bibfield  {journal} {\bibinfo
  {journal} {Phys. Rev. Lett.}\ }\textbf {\bibinfo {volume} {54}},\ \bibinfo
  {pages} {966} (\bibinfo {year} {1985})}\BibitemShut {NoStop}%
\bibitem [{\citenamefont {Bickers}(1987)}]{Bickers1987}%
  \BibitemOpen
  \bibfield  {author} {\bibinfo {author} {\bibfnamefont {N.~E.}\ \bibnamefont
  {Bickers}},\ }\bibfield  {title} {\bibinfo {title} {Review of techniques in
  the large-{N} expansion for dilute magnetic alloys},\ }\href
  {https://doi.org/10.1103/revmodphys.59.845} {\bibfield  {journal} {\bibinfo
  {journal} {Rev. Mod. Phys.}\ }\textbf {\bibinfo {volume} {59}},\ \bibinfo
  {pages} {845} (\bibinfo {year} {1987})}\BibitemShut {NoStop}%
\bibitem [{\citenamefont {Affleck}\ and\ \citenamefont
  {Marston}(1988)}]{Affleck1988}%
  \BibitemOpen
  \bibfield  {author} {\bibinfo {author} {\bibfnamefont {I.}~\bibnamefont
  {Affleck}}\ and\ \bibinfo {author} {\bibfnamefont {J.~B.}\ \bibnamefont
  {Marston}},\ }\bibfield  {title} {\bibinfo {title} {Large-$n$ limit of the
  {H}eisenberg-{H}ubbard model: {I}mplications for high-${T}_c$
  superconductors},\ }\href {https://doi.org/10.1103/physrevb.37.3774}
  {\bibfield  {journal} {\bibinfo  {journal} {Phys. Rev. B}\ }\textbf {\bibinfo
  {volume} {37}},\ \bibinfo {pages} {3774} (\bibinfo {year}
  {1988})}\BibitemShut {NoStop}%
\bibitem [{\citenamefont {Auerbach}(1994)}]{Auerbach2012}%
  \BibitemOpen
  \bibfield  {author} {\bibinfo {author} {\bibfnamefont {A.}~\bibnamefont
  {Auerbach}},\ }\href@noop {} {\emph {\bibinfo {title} {{I}nteracting
  {E}lectrons and {Q}uantum {M}agnetism}}}\ (\bibinfo  {publisher}
  {Springer-Verlag New York},\ \bibinfo {year} {1994})\BibitemShut {NoStop}%
\bibitem [{\citenamefont {Yamashita}\ \emph {et~al.}(1998)\citenamefont
  {Yamashita}, \citenamefont {Shibata},\ and\ \citenamefont
  {Ueda}}]{Yamashita1998}%
  \BibitemOpen
  \bibfield  {author} {\bibinfo {author} {\bibfnamefont {Y.}~\bibnamefont
  {Yamashita}}, \bibinfo {author} {\bibfnamefont {N.}~\bibnamefont {Shibata}},\
  and\ \bibinfo {author} {\bibfnamefont {K.}~\bibnamefont {Ueda}},\ }\bibfield
  {title} {\bibinfo {title} {{SU}(4) spin-orbit critical state in one
  dimension},\ }\href@noop {} {\bibfield  {journal} {\bibinfo  {journal} {Phys.
  Rev. B}\ }\textbf {\bibinfo {volume} {58}},\ \bibinfo {pages} {9114}
  (\bibinfo {year} {1998})}\BibitemShut {NoStop}%
\bibitem [{\citenamefont {Assaraf}\ \emph {et~al.}(1999)\citenamefont
  {Assaraf}, \citenamefont {Azaria}, \citenamefont {Caffarel},\ and\
  \citenamefont {Lecheminant}}]{Assaraf1999}%
  \BibitemOpen
  \bibfield  {author} {\bibinfo {author} {\bibfnamefont {R.}~\bibnamefont
  {Assaraf}}, \bibinfo {author} {\bibfnamefont {P.}~\bibnamefont {Azaria}},
  \bibinfo {author} {\bibfnamefont {M.}~\bibnamefont {Caffarel}},\ and\
  \bibinfo {author} {\bibfnamefont {P.}~\bibnamefont {Lecheminant}},\
  }\bibfield  {title} {\bibinfo {title} {Metal-insulator transition in the
  one-dimensional {SU($N$)} {H}ubbard model},\ }\href@noop {} {\bibfield
  {journal} {\bibinfo  {journal} {Phys. Rev. B}\ }\textbf {\bibinfo {volume}
  {60}},\ \bibinfo {pages} {2299} (\bibinfo {year} {1999})}\BibitemShut
  {NoStop}%
\bibitem [{\citenamefont {Buchta}\ \emph {et~al.}(2007)\citenamefont {Buchta},
  \citenamefont {Legeza}, \citenamefont {Szirmai},\ and\ \citenamefont
  {S{\'o}lyom}}]{Buchta2007}%
  \BibitemOpen
  \bibfield  {author} {\bibinfo {author} {\bibfnamefont {K.}~\bibnamefont
  {Buchta}}, \bibinfo {author} {\bibfnamefont {{\"O}.}~\bibnamefont {Legeza}},
  \bibinfo {author} {\bibfnamefont {E.}~\bibnamefont {Szirmai}},\ and\ \bibinfo
  {author} {\bibfnamefont {J.}~\bibnamefont {S{\'o}lyom}},\ }\bibfield  {title}
  {\bibinfo {title} {Mott transition and dimerization in the one-dimensional
  {SU($N$)} {H}ubbard model},\ }\href@noop {} {\bibfield  {journal} {\bibinfo
  {journal} {Phys. Rev. B}\ }\textbf {\bibinfo {volume} {75}},\ \bibinfo
  {pages} {155108} (\bibinfo {year} {2007})}\BibitemShut {NoStop}%
\bibitem [{\citenamefont {Manmana}\ \emph {et~al.}(2011)\citenamefont
  {Manmana}, \citenamefont {Hazzard}, \citenamefont {Chen}, \citenamefont
  {Feiguin},\ and\ \citenamefont {Rey}}]{Manmana2011}%
  \BibitemOpen
  \bibfield  {author} {\bibinfo {author} {\bibfnamefont {S.~R.}\ \bibnamefont
  {Manmana}}, \bibinfo {author} {\bibfnamefont {K.~R.~A.}\ \bibnamefont
  {Hazzard}}, \bibinfo {author} {\bibfnamefont {G.}~\bibnamefont {Chen}},
  \bibinfo {author} {\bibfnamefont {A.~E.}\ \bibnamefont {Feiguin}},\ and\
  \bibinfo {author} {\bibfnamefont {A.~M.}\ \bibnamefont {Rey}},\ }\bibfield
  {title} {\bibinfo {title} {{SU}{($N$)} magnetism in chains of ultracold
  alkaline-earth-metal atoms: {M}ott transitions and quantum correlations},\
  }\href {https://doi.org/10.1103/PhysRevA.84.043601} {\bibfield  {journal}
  {\bibinfo  {journal} {Phys. Rev. A}\ }\textbf {\bibinfo {volume} {84}},\
  \bibinfo {pages} {043601} (\bibinfo {year} {2011})}\BibitemShut {NoStop}%
\bibitem [{\citenamefont {Bonnes}\ \emph {et~al.}(2012)\citenamefont {Bonnes},
  \citenamefont {Hazzard}, \citenamefont {Manmana}, \citenamefont {Rey},\ and\
  \citenamefont {Wessel}}]{Bonnes2012}%
  \BibitemOpen
  \bibfield  {author} {\bibinfo {author} {\bibfnamefont {L.}~\bibnamefont
  {Bonnes}}, \bibinfo {author} {\bibfnamefont {K.~R.~A.}\ \bibnamefont
  {Hazzard}}, \bibinfo {author} {\bibfnamefont {S.~R.}\ \bibnamefont
  {Manmana}}, \bibinfo {author} {\bibfnamefont {A.~M.}\ \bibnamefont {Rey}},\
  and\ \bibinfo {author} {\bibfnamefont {S.}~\bibnamefont {Wessel}},\
  }\bibfield  {title} {\bibinfo {title} {Adiabatic loading of one-dimensional
  {SU}({$N$}) alkaline-earth-atom fermions in optical lattices},\ }\href
  {https://doi.org/10.1103/PhysRevLett.109.205305} {\bibfield  {journal}
  {\bibinfo  {journal} {Phys. Rev. Lett.}\ }\textbf {\bibinfo {volume} {109}},\
  \bibinfo {pages} {205305} (\bibinfo {year} {2012})}\BibitemShut {NoStop}%
\bibitem [{\citenamefont {Messio}\ and\ \citenamefont
  {Mila}(2012)}]{Messio2012}%
  \BibitemOpen
  \bibfield  {author} {\bibinfo {author} {\bibfnamefont {L.}~\bibnamefont
  {Messio}}\ and\ \bibinfo {author} {\bibfnamefont {F.}~\bibnamefont {Mila}},\
  }\bibfield  {title} {\bibinfo {title} {Entropy dependence of correlations in
  one-dimensional {SU}({$N$}) antiferromagnets},\ }\href
  {https://doi.org/10.1103/PhysRevLett.109.205306} {\bibfield  {journal}
  {\bibinfo  {journal} {Phys. Rev. Lett.}\ }\textbf {\bibinfo {volume} {109}},\
  \bibinfo {pages} {205306} (\bibinfo {year} {2012})}\BibitemShut {NoStop}%
\bibitem [{\citenamefont {Xu}\ \emph {et~al.}(2018{\natexlab{b}})\citenamefont
  {Xu}, \citenamefont {Barreiro}, \citenamefont {Wang},\ and\ \citenamefont
  {Wu}}]{Xu2018}%
  \BibitemOpen
  \bibfield  {author} {\bibinfo {author} {\bibfnamefont {S.}~\bibnamefont
  {Xu}}, \bibinfo {author} {\bibfnamefont {J.~T.}\ \bibnamefont {Barreiro}},
  \bibinfo {author} {\bibfnamefont {Y.}~\bibnamefont {Wang}},\ and\ \bibinfo
  {author} {\bibfnamefont {C.}~\bibnamefont {Wu}},\ }\bibfield  {title}
  {\bibinfo {title} {Interaction effects with varying {$N$} in {SU$(N)$}
  symmetric fermion lattice systems},\ }\href
  {https://doi.org/10.1103/PhysRevLett.121.167205} {\bibfield  {journal}
  {\bibinfo  {journal} {Phys. Rev. Lett.}\ }\textbf {\bibinfo {volume} {121}},\
  \bibinfo {pages} {167205} (\bibinfo {year} {2018}{\natexlab{b}})}\BibitemShut
  {NoStop}%
\bibitem [{\citenamefont {Hermele}\ \emph {et~al.}(2009)\citenamefont
  {Hermele}, \citenamefont {Gurarie},\ and\ \citenamefont {Rey}}]{Hermele2009}%
  \BibitemOpen
  \bibfield  {author} {\bibinfo {author} {\bibfnamefont {M.}~\bibnamefont
  {Hermele}}, \bibinfo {author} {\bibfnamefont {V.}~\bibnamefont {Gurarie}},\
  and\ \bibinfo {author} {\bibfnamefont {A.~M.}\ \bibnamefont {Rey}},\
  }\bibfield  {title} {\bibinfo {title} {Mott {I}nsulators of ultracold
  fermionic alkaline earth atoms: Underconstrained magnetism and chiral spin
  liquid},\ }\href {https://doi.org/10.1103/PhysRevLett.103.135301} {\bibfield
  {journal} {\bibinfo  {journal} {Phys. Rev. Lett.}\ }\textbf {\bibinfo
  {volume} {103}},\ \bibinfo {pages} {135301} (\bibinfo {year}
  {2009})}\BibitemShut {NoStop}%
\bibitem [{\citenamefont {T{\'{o}}th}\ \emph {et~al.}(2010)\citenamefont
  {T{\'{o}}th}, \citenamefont {L{\"a}uchli}, \citenamefont {Mila},\ and\
  \citenamefont {Penc}}]{Toth2010}%
  \BibitemOpen
  \bibfield  {author} {\bibinfo {author} {\bibfnamefont {T.~A.}\ \bibnamefont
  {T{\'{o}}th}}, \bibinfo {author} {\bibfnamefont {A.~M.}\ \bibnamefont
  {L{\"a}uchli}}, \bibinfo {author} {\bibfnamefont {F.}~\bibnamefont {Mila}},\
  and\ \bibinfo {author} {\bibfnamefont {K.}~\bibnamefont {Penc}},\ }\bibfield
  {title} {\bibinfo {title} {Three-sublattice ordering of the {SU}(3)
  {H}eisenberg model of three-flavor fermions on the square and cubic
  lattices},\ }\href {https://doi.org/10.1103/physrevlett.105.265301}
  {\bibfield  {journal} {\bibinfo  {journal} {Phys. Rev. Lett.}\ }\textbf
  {\bibinfo {volume} {105}},\ \bibinfo {pages} {265301} (\bibinfo {year}
  {2010})}\BibitemShut {NoStop}%
\bibitem [{\citenamefont {Hermele}\ and\ \citenamefont
  {Gurarie}(2011)}]{Hermele2011}%
  \BibitemOpen
  \bibfield  {author} {\bibinfo {author} {\bibfnamefont {M.}~\bibnamefont
  {Hermele}}\ and\ \bibinfo {author} {\bibfnamefont {V.}~\bibnamefont
  {Gurarie}},\ }\bibfield  {title} {\bibinfo {title} {Topological liquids and
  valence cluster states in two-dimensional {SU($N$)} magnets},\ }\href
  {https://doi.org/10.1103/physrevb.84.174441} {\bibfield  {journal} {\bibinfo
  {journal} {Phys. Rev. B}\ }\textbf {\bibinfo {volume} {84}},\ \bibinfo
  {pages} {174441} (\bibinfo {year} {2011})}\BibitemShut {NoStop}%
\bibitem [{\citenamefont {Nataf}\ and\ \citenamefont {Mila}(2014)}]{Nataf2014}%
  \BibitemOpen
  \bibfield  {author} {\bibinfo {author} {\bibfnamefont {P.}~\bibnamefont
  {Nataf}}\ and\ \bibinfo {author} {\bibfnamefont {F.}~\bibnamefont {Mila}},\
  }\bibfield  {title} {\bibinfo {title} {Exact {D}iagonalization of
  {H}eisenberg {SU($N$)} {M}odels},\ }\href
  {https://doi.org/10.1103/physrevlett.113.127204} {\bibfield  {journal}
  {\bibinfo  {journal} {Phys. Rev. Lett.}\ }\textbf {\bibinfo {volume} {113}},\
  \bibinfo {pages} {127204} (\bibinfo {year} {2014})}\BibitemShut {NoStop}%
\bibitem [{\citenamefont {Corboz}\ \emph {et~al.}(2011)\citenamefont {Corboz},
  \citenamefont {L{\"a}uchli}, \citenamefont {Penc}, \citenamefont {Troyer},\
  and\ \citenamefont {Mila}}]{Corboz2011}%
  \BibitemOpen
  \bibfield  {author} {\bibinfo {author} {\bibfnamefont {P.}~\bibnamefont
  {Corboz}}, \bibinfo {author} {\bibfnamefont {A.~M.}\ \bibnamefont
  {L{\"a}uchli}}, \bibinfo {author} {\bibfnamefont {K.}~\bibnamefont {Penc}},
  \bibinfo {author} {\bibfnamefont {M.}~\bibnamefont {Troyer}},\ and\ \bibinfo
  {author} {\bibfnamefont {F.}~\bibnamefont {Mila}},\ }\bibfield  {title}
  {\bibinfo {title} {Simultaneous dimerization and {SU}(4) symmetry breaking of
  4-color fermions on the square lattice},\ }\href
  {https://doi.org/10.1103/physrevlett.107.215301} {\bibfield  {journal}
  {\bibinfo  {journal} {Phys. Rev. Lett.}\ }\textbf {\bibinfo {volume} {107}},\
  \bibinfo {pages} {215301} (\bibinfo {year} {2011})}\BibitemShut {NoStop}%
\bibitem [{\citenamefont {Bauer}\ \emph {et~al.}(2012)\citenamefont {Bauer},
  \citenamefont {Corboz}, \citenamefont {L{\"a}uchli}, \citenamefont {Messio},
  \citenamefont {Penc}, \citenamefont {Troyer},\ and\ \citenamefont
  {Mila}}]{Bauer2012}%
  \BibitemOpen
  \bibfield  {author} {\bibinfo {author} {\bibfnamefont {B.}~\bibnamefont
  {Bauer}}, \bibinfo {author} {\bibfnamefont {P.}~\bibnamefont {Corboz}},
  \bibinfo {author} {\bibfnamefont {A.~M.}\ \bibnamefont {L{\"a}uchli}},
  \bibinfo {author} {\bibfnamefont {L.}~\bibnamefont {Messio}}, \bibinfo
  {author} {\bibfnamefont {K.}~\bibnamefont {Penc}}, \bibinfo {author}
  {\bibfnamefont {M.}~\bibnamefont {Troyer}},\ and\ \bibinfo {author}
  {\bibfnamefont {F.}~\bibnamefont {Mila}},\ }\bibfield  {title} {\bibinfo
  {title} {Three-sublattice order in the {SU}(3) {H}eisenberg model on the
  square and triangular lattice},\ }\href
  {https://doi.org/10.1103/physrevb.85.125116} {\bibfield  {journal} {\bibinfo
  {journal} {Phys. Rev. B}\ }\textbf {\bibinfo {volume} {85}},\ \bibinfo
  {pages} {125116} (\bibinfo {year} {2012})}\BibitemShut {NoStop}%
\bibitem [{\citenamefont {Yamamoto}\ \emph {et~al.}(2020)\citenamefont
  {Yamamoto}, \citenamefont {Suzuki}, \citenamefont {Marmorini}, \citenamefont
  {Okazaki},\ and\ \citenamefont {Furukawa}}]{Yamamoto2020}%
  \BibitemOpen
  \bibfield  {author} {\bibinfo {author} {\bibfnamefont {D.}~\bibnamefont
  {Yamamoto}}, \bibinfo {author} {\bibfnamefont {C.}~\bibnamefont {Suzuki}},
  \bibinfo {author} {\bibfnamefont {G.}~\bibnamefont {Marmorini}}, \bibinfo
  {author} {\bibfnamefont {S.}~\bibnamefont {Okazaki}},\ and\ \bibinfo {author}
  {\bibfnamefont {N.}~\bibnamefont {Furukawa}},\ }\bibfield  {title} {\bibinfo
  {title} {Quantum and thermal phase transitions of the triangular {SU}(3)
  {H}eisenberg model under magnetic fields},\ }\href
  {https://doi.org/10.1103/PhysRevLett.125.057204} {\bibfield  {journal}
  {\bibinfo  {journal} {Phys. Rev. Lett.}\ }\textbf {\bibinfo {volume} {125}},\
  \bibinfo {pages} {057204} (\bibinfo {year} {2020})}\BibitemShut {NoStop}%
\bibitem [{\citenamefont {Keselman}\ \emph {et~al.}(2020)\citenamefont
  {Keselman}, \citenamefont {Bauer}, \citenamefont {Xu},\ and\ \citenamefont
  {Jian}}]{Keselman2020}%
  \BibitemOpen
  \bibfield  {author} {\bibinfo {author} {\bibfnamefont {A.}~\bibnamefont
  {Keselman}}, \bibinfo {author} {\bibfnamefont {B.}~\bibnamefont {Bauer}},
  \bibinfo {author} {\bibfnamefont {C.}~\bibnamefont {Xu}},\ and\ \bibinfo
  {author} {\bibfnamefont {C.-M.}\ \bibnamefont {Jian}},\ }\bibfield  {title}
  {\bibinfo {title} {Emergent {F}ermi surface in a triangular-lattice {SU}(4)
  quantum antiferromagnet},\ }\href
  {https://doi.org/10.1103/PhysRevLett.125.117202} {\bibfield  {journal}
  {\bibinfo  {journal} {Phys. Rev. Lett.}\ }\textbf {\bibinfo {volume} {125}},\
  \bibinfo {pages} {117202} (\bibinfo {year} {2020})}\BibitemShut {NoStop}%
\bibitem [{\citenamefont {Yao}\ \emph {et~al.}(2021)\citenamefont {Yao},
  \citenamefont {Gao},\ and\ \citenamefont {Chen}}]{Yao2021}%
  \BibitemOpen
  \bibfield  {author} {\bibinfo {author} {\bibfnamefont {X.-P.}\ \bibnamefont
  {Yao}}, \bibinfo {author} {\bibfnamefont {Y.}~\bibnamefont {Gao}},\ and\
  \bibinfo {author} {\bibfnamefont {G.}~\bibnamefont {Chen}},\ }\bibfield
  {title} {\bibinfo {title} {Topological chiral spin liquids and competing
  states in triangular lattice {SU$(N)$} {M}ott insulators},\ }\href
  {https://doi.org/10.1103/PhysRevResearch.3.023138} {\bibfield  {journal}
  {\bibinfo  {journal} {Phys. Rev. Research}\ }\textbf {\bibinfo {volume}
  {3}},\ \bibinfo {pages} {023138} (\bibinfo {year} {2021})}\BibitemShut
  {NoStop}%
\bibitem [{\citenamefont {Wu}(2006)}]{Wu2006}%
  \BibitemOpen
  \bibfield  {author} {\bibinfo {author} {\bibfnamefont {C.}~\bibnamefont
  {Wu}},\ }\bibfield  {title} {\bibinfo {title} {{H}idden {S}ymmetry and
  {Q}uantum {P}hases in {S}pin-3/2 {C}old {A}tomic {S}ystems},\ }\href@noop {}
  {\bibfield  {journal} {\bibinfo  {journal} {Mod. Phys. Lett. B}\ }\textbf
  {\bibinfo {volume} {20}},\ \bibinfo {pages} {1707} (\bibinfo {year}
  {2006})}\BibitemShut {NoStop}%
\bibitem [{\citenamefont {Cazalilla}\ \emph {et~al.}(2009)\citenamefont
  {Cazalilla}, \citenamefont {Ho},\ and\ \citenamefont {Ueda}}]{Cazalilla2009}%
  \BibitemOpen
  \bibfield  {author} {\bibinfo {author} {\bibfnamefont {M.~A.}\ \bibnamefont
  {Cazalilla}}, \bibinfo {author} {\bibfnamefont {A.~F.}\ \bibnamefont {Ho}},\
  and\ \bibinfo {author} {\bibfnamefont {M.}~\bibnamefont {Ueda}},\ }\bibfield
  {title} {\bibinfo {title} {{U}ltracold gases of ytterbium: {F}erromagnetism
  and {M}ott states in an {SU}(6) {F}ermi system},\ }\href
  {https://doi.org/10.1088/1367-2630/11/10/103033} {\bibfield  {journal}
  {\bibinfo  {journal} {New J. Phys.}\ }\textbf {\bibinfo {volume} {11}},\
  \bibinfo {pages} {103033} (\bibinfo {year} {2009})}\BibitemShut {NoStop}%
\bibitem [{\citenamefont {Gorshkov}\ \emph {et~al.}(2010)\citenamefont
  {Gorshkov}, \citenamefont {Hermele}, \citenamefont {Gurarie}, \citenamefont
  {Xu}, \citenamefont {Julienne}, \citenamefont {Ye}, \citenamefont {Zoller},
  \citenamefont {Demler}, \citenamefont {Lukin},\ and\ \citenamefont
  {Rey}}]{Gorshkov2010}%
  \BibitemOpen
  \bibfield  {author} {\bibinfo {author} {\bibfnamefont {A.~V.}\ \bibnamefont
  {Gorshkov}}, \bibinfo {author} {\bibfnamefont {M.}~\bibnamefont {Hermele}},
  \bibinfo {author} {\bibfnamefont {V.}~\bibnamefont {Gurarie}}, \bibinfo
  {author} {\bibfnamefont {C.}~\bibnamefont {Xu}}, \bibinfo {author}
  {\bibfnamefont {P.~S.}\ \bibnamefont {Julienne}}, \bibinfo {author}
  {\bibfnamefont {J.}~\bibnamefont {Ye}}, \bibinfo {author} {\bibfnamefont
  {P.}~\bibnamefont {Zoller}}, \bibinfo {author} {\bibfnamefont
  {E.}~\bibnamefont {Demler}}, \bibinfo {author} {\bibfnamefont {M.~D.}\
  \bibnamefont {Lukin}},\ and\ \bibinfo {author} {\bibfnamefont {A.~M.}\
  \bibnamefont {Rey}},\ }\bibfield  {title} {\bibinfo {title} {Two-orbital
  {SU($N$)} magnetism with ultracold alkaline-earth atoms},\ }\href
  {https://doi.org/10.1038/nphys1535} {\bibfield  {journal} {\bibinfo
  {journal} {Nat. Phys.}\ }\textbf {\bibinfo {volume} {6}},\ \bibinfo {pages}
  {289} (\bibinfo {year} {2010})}\BibitemShut {NoStop}%
\bibitem [{\citenamefont {Cazalilla}\ and\ \citenamefont
  {Rey}(2014)}]{Cazalilla2014}%
  \BibitemOpen
  \bibfield  {author} {\bibinfo {author} {\bibfnamefont {M.~A.}\ \bibnamefont
  {Cazalilla}}\ and\ \bibinfo {author} {\bibfnamefont {A.~M.}\ \bibnamefont
  {Rey}},\ }\bibfield  {title} {\bibinfo {title} {{U}ltracold {F}ermi gases
  with emergent {SU($N$)} symmetry},\ }\href
  {https://doi.org/10.1088/0034-4885/77/12/124401} {\bibfield  {journal}
  {\bibinfo  {journal} {Rep. Prog. Phys.}\ }\textbf {\bibinfo {volume} {77}},\
  \bibinfo {pages} {124401} (\bibinfo {year} {2014})}\BibitemShut {NoStop}%
\bibitem [{\citenamefont {Stellmer}\ \emph {et~al.}(2014)\citenamefont
  {Stellmer}, \citenamefont {Schreck},\ and\ \citenamefont
  {Killian}}]{Stellmer2014}%
  \BibitemOpen
  \bibfield  {author} {\bibinfo {author} {\bibfnamefont {S.}~\bibnamefont
  {Stellmer}}, \bibinfo {author} {\bibfnamefont {F.}~\bibnamefont {Schreck}},\
  and\ \bibinfo {author} {\bibfnamefont {T.~C.}\ \bibnamefont {Killian}},\
  }\bibinfo {title} {Degenerate quantum gases of strontium},\ in\ \href
  {https://doi.org/10.1142/9789814590174_0001} {\emph {\bibinfo {booktitle}
  {Annual Review of Cold Atoms and Molecules}}}\ (\bibinfo {year} {2014})\
  Chap.~\bibinfo {chapter} {1}, pp.\ \bibinfo {pages} {1--80}\BibitemShut
  {NoStop}%
\bibitem [{\citenamefont {Taie}\ \emph {et~al.}(2012)\citenamefont {Taie},
  \citenamefont {Yamazaki}, \citenamefont {Sugawa},\ and\ \citenamefont
  {Takahashi}}]{Taie2012}%
  \BibitemOpen
  \bibfield  {author} {\bibinfo {author} {\bibfnamefont {S.}~\bibnamefont
  {Taie}}, \bibinfo {author} {\bibfnamefont {R.}~\bibnamefont {Yamazaki}},
  \bibinfo {author} {\bibfnamefont {S.}~\bibnamefont {Sugawa}},\ and\ \bibinfo
  {author} {\bibfnamefont {Y.}~\bibnamefont {Takahashi}},\ }\bibfield  {title}
  {\bibinfo {title} {An {SU}(6) {M}ott insulator of an atomic {F}ermi gas
  realized by large-spin {P}omeranchuk cooling},\ }\href@noop {} {\bibfield
  {journal} {\bibinfo  {journal} {Nat. Phys.}\ }\textbf {\bibinfo {volume}
  {8}},\ \bibinfo {pages} {825} (\bibinfo {year} {2012})}\BibitemShut {NoStop}%
\bibitem [{\citenamefont {Hofrichter}\ \emph {et~al.}(2016)\citenamefont
  {Hofrichter}, \citenamefont {Riegger}, \citenamefont {Scazza}, \citenamefont
  {H\"ofer}, \citenamefont {Fernandes}, \citenamefont {Bloch},\ and\
  \citenamefont {F\"olling}}]{Hofrichter2016}%
  \BibitemOpen
  \bibfield  {author} {\bibinfo {author} {\bibfnamefont {C.}~\bibnamefont
  {Hofrichter}}, \bibinfo {author} {\bibfnamefont {L.}~\bibnamefont {Riegger}},
  \bibinfo {author} {\bibfnamefont {F.}~\bibnamefont {Scazza}}, \bibinfo
  {author} {\bibfnamefont {M.}~\bibnamefont {H\"ofer}}, \bibinfo {author}
  {\bibfnamefont {D.~R.}\ \bibnamefont {Fernandes}}, \bibinfo {author}
  {\bibfnamefont {I.}~\bibnamefont {Bloch}},\ and\ \bibinfo {author}
  {\bibfnamefont {S.}~\bibnamefont {F\"olling}},\ }\bibfield  {title} {\bibinfo
  {title} {Direct probing of the {M}ott crossover in the {SU($N$)}
  {F}ermi-{H}ubbard model},\ }\href {https://doi.org/10.1103/PhysRevX.6.021030}
  {\bibfield  {journal} {\bibinfo  {journal} {Phys. Rev. X}\ }\textbf {\bibinfo
  {volume} {6}},\ \bibinfo {pages} {021030} (\bibinfo {year}
  {2016})}\BibitemShut {NoStop}%
\bibitem [{\citenamefont {Ozawa}\ \emph {et~al.}(2018)\citenamefont {Ozawa},
  \citenamefont {Taie}, \citenamefont {Takasu},\ and\ \citenamefont
  {Takahashi}}]{Ozawa2018}%
  \BibitemOpen
  \bibfield  {author} {\bibinfo {author} {\bibfnamefont {H.}~\bibnamefont
  {Ozawa}}, \bibinfo {author} {\bibfnamefont {S.}~\bibnamefont {Taie}},
  \bibinfo {author} {\bibfnamefont {Y.}~\bibnamefont {Takasu}},\ and\ \bibinfo
  {author} {\bibfnamefont {Y.}~\bibnamefont {Takahashi}},\ }\bibfield  {title}
  {\bibinfo {title} {Antiferromagnetic spin correlation of {SU($N$)} {F}ermi
  gas in an optical superlattice},\ }\href
  {https://doi.org/10.1103/PhysRevLett.121.225303} {\bibfield  {journal}
  {\bibinfo  {journal} {Phys. Rev. Lett.}\ }\textbf {\bibinfo {volume} {121}},\
  \bibinfo {pages} {225303} (\bibinfo {year} {2018})}\BibitemShut {NoStop}%
\bibitem [{\citenamefont {Taie}\ \emph {et~al.}(2020)\citenamefont {Taie},
  \citenamefont {Ibarra-Garc\'{\i}a-Padilla}, \citenamefont {Nishizawa},
  \citenamefont {Takasu}, \citenamefont {Kuno}, \citenamefont {Wei},
  \citenamefont {Scalettar}, \citenamefont {Hazzard},\ and\ \citenamefont
  {Takahashi}}]{Taie2020}%
  \BibitemOpen
  \bibfield  {author} {\bibinfo {author} {\bibfnamefont {S.}~\bibnamefont
  {Taie}}, \bibinfo {author} {\bibfnamefont {E.}~\bibnamefont
  {Ibarra-Garc\'{\i}a-Padilla}}, \bibinfo {author} {\bibfnamefont
  {N.}~\bibnamefont {Nishizawa}}, \bibinfo {author} {\bibfnamefont
  {Y.}~\bibnamefont {Takasu}}, \bibinfo {author} {\bibfnamefont
  {Y.}~\bibnamefont {Kuno}}, \bibinfo {author} {\bibfnamefont {H.-T.}\
  \bibnamefont {Wei}}, \bibinfo {author} {\bibfnamefont {R.~T.}\ \bibnamefont
  {Scalettar}}, \bibinfo {author} {\bibfnamefont {K.~R.~A.}\ \bibnamefont
  {Hazzard}},\ and\ \bibinfo {author} {\bibfnamefont {Y.}~\bibnamefont
  {Takahashi}},\ }\bibfield  {title} {\bibinfo {title} {Observation of
  antiferromagnetic correlations in an ultracold {SU($N$)} {H}ubbard model},\
  }\href@noop {} {\bibfield  {journal} {\bibinfo  {journal}
  {arXiv:2010.07730v1}\ } (\bibinfo {year} {2020})}\BibitemShut {NoStop}%
\bibitem [{\citenamefont {Tusi}\ \emph {et~al.}(2021)\citenamefont {Tusi},
  \citenamefont {Franchi}, \citenamefont {Livi}, \citenamefont {Baumann},
  \citenamefont {Orenes}, \citenamefont {Re}, \citenamefont {Barfknecht},
  \citenamefont {Zhou}, \citenamefont {Inguscio}, \citenamefont {Cappellini},
  \citenamefont {Capone}, \citenamefont {Catani},\ and\ \citenamefont
  {Fallani}}]{Tusi2021}%
  \BibitemOpen
  \bibfield  {author} {\bibinfo {author} {\bibfnamefont {D.}~\bibnamefont
  {Tusi}}, \bibinfo {author} {\bibfnamefont {L.}~\bibnamefont {Franchi}},
  \bibinfo {author} {\bibfnamefont {L.~F.}\ \bibnamefont {Livi}}, \bibinfo
  {author} {\bibfnamefont {K.}~\bibnamefont {Baumann}}, \bibinfo {author}
  {\bibfnamefont {D.~B.}\ \bibnamefont {Orenes}}, \bibinfo {author}
  {\bibfnamefont {L.~D.}\ \bibnamefont {Re}}, \bibinfo {author} {\bibfnamefont
  {R.~E.}\ \bibnamefont {Barfknecht}}, \bibinfo {author} {\bibfnamefont
  {T.}~\bibnamefont {Zhou}}, \bibinfo {author} {\bibfnamefont {M.}~\bibnamefont
  {Inguscio}}, \bibinfo {author} {\bibfnamefont {G.}~\bibnamefont
  {Cappellini}}, \bibinfo {author} {\bibfnamefont {M.}~\bibnamefont {Capone}},
  \bibinfo {author} {\bibfnamefont {J.}~\bibnamefont {Catani}},\ and\ \bibinfo
  {author} {\bibfnamefont {L.}~\bibnamefont {Fallani}},\ }\bibfield  {title}
  {\bibinfo {title} {Flavour-selective localization in interacting lattice
  fermions via {SU($N$)} symmetry breaking},\ }\href@noop {} {\bibfield
  {journal} {\bibinfo  {journal} {arXiv:2104.13338v1}\ } (\bibinfo {year}
  {2021})}\BibitemShut {NoStop}%
\bibitem [{\citenamefont {Altman}\ \emph {et~al.}(2021)\citenamefont {Altman},
  \citenamefont {Brown}, \citenamefont {Carleo}, \citenamefont {Carr},
  \citenamefont {Demler}, \citenamefont {Chin}, \citenamefont {DeMarco},
  \citenamefont {Economou}, \citenamefont {Eriksson}, \citenamefont {Fu},
  \citenamefont {Greiner}, \citenamefont {Hazzard}, \citenamefont {Hulet},
  \citenamefont {Koll\'ar}, \citenamefont {Lev}, \citenamefont {Lukin},
  \citenamefont {Ma}, \citenamefont {Mi}, \citenamefont {Misra}, \citenamefont
  {Monroe}, \citenamefont {Murch}, \citenamefont {Nazario}, \citenamefont {Ni},
  \citenamefont {Potter}, \citenamefont {Roushan}, \citenamefont {Saffman},
  \citenamefont {Schleier-Smith}, \citenamefont {Siddiqi}, \citenamefont
  {Simmonds}, \citenamefont {Singh}, \citenamefont {Spielman}, \citenamefont
  {Temme}, \citenamefont {Weiss}, \citenamefont {Vu\ifmmode \check{c}\else
  \v{c}\fi{}kovi\ifmmode~\acute{c}\else \'{c}\fi{}}, \citenamefont
  {Vuleti\ifmmode~\acute{c}\else \'{c}\fi{}}, \citenamefont {Ye},\ and\
  \citenamefont {Zwierlein}}]{Altman2021}%
  \BibitemOpen
  \bibfield  {author} {\bibinfo {author} {\bibfnamefont {E.}~\bibnamefont
  {Altman}}, \bibinfo {author} {\bibfnamefont {K.~R.}\ \bibnamefont {Brown}},
  \bibinfo {author} {\bibfnamefont {G.}~\bibnamefont {Carleo}}, \bibinfo
  {author} {\bibfnamefont {L.~D.}\ \bibnamefont {Carr}}, \bibinfo {author}
  {\bibfnamefont {E.}~\bibnamefont {Demler}}, \bibinfo {author} {\bibfnamefont
  {C.}~\bibnamefont {Chin}}, \bibinfo {author} {\bibfnamefont {B.}~\bibnamefont
  {DeMarco}}, \bibinfo {author} {\bibfnamefont {S.~E.}\ \bibnamefont
  {Economou}}, \bibinfo {author} {\bibfnamefont {M.~A.}\ \bibnamefont
  {Eriksson}}, \bibinfo {author} {\bibfnamefont {K.-M.~C.}\ \bibnamefont {Fu}},
  \bibinfo {author} {\bibfnamefont {M.}~\bibnamefont {Greiner}}, \bibinfo
  {author} {\bibfnamefont {K.~R.}\ \bibnamefont {Hazzard}}, \bibinfo {author}
  {\bibfnamefont {R.~G.}\ \bibnamefont {Hulet}}, \bibinfo {author}
  {\bibfnamefont {A.~J.}\ \bibnamefont {Koll\'ar}}, \bibinfo {author}
  {\bibfnamefont {B.~L.}\ \bibnamefont {Lev}}, \bibinfo {author} {\bibfnamefont
  {M.~D.}\ \bibnamefont {Lukin}}, \bibinfo {author} {\bibfnamefont
  {R.}~\bibnamefont {Ma}}, \bibinfo {author} {\bibfnamefont {X.}~\bibnamefont
  {Mi}}, \bibinfo {author} {\bibfnamefont {S.}~\bibnamefont {Misra}}, \bibinfo
  {author} {\bibfnamefont {C.}~\bibnamefont {Monroe}}, \bibinfo {author}
  {\bibfnamefont {K.}~\bibnamefont {Murch}}, \bibinfo {author} {\bibfnamefont
  {Z.}~\bibnamefont {Nazario}}, \bibinfo {author} {\bibfnamefont {K.-K.}\
  \bibnamefont {Ni}}, \bibinfo {author} {\bibfnamefont {A.~C.}\ \bibnamefont
  {Potter}}, \bibinfo {author} {\bibfnamefont {P.}~\bibnamefont {Roushan}},
  \bibinfo {author} {\bibfnamefont {M.}~\bibnamefont {Saffman}}, \bibinfo
  {author} {\bibfnamefont {M.}~\bibnamefont {Schleier-Smith}}, \bibinfo
  {author} {\bibfnamefont {I.}~\bibnamefont {Siddiqi}}, \bibinfo {author}
  {\bibfnamefont {R.}~\bibnamefont {Simmonds}}, \bibinfo {author}
  {\bibfnamefont {M.}~\bibnamefont {Singh}}, \bibinfo {author} {\bibfnamefont
  {I.}~\bibnamefont {Spielman}}, \bibinfo {author} {\bibfnamefont
  {K.}~\bibnamefont {Temme}}, \bibinfo {author} {\bibfnamefont {D.~S.}\
  \bibnamefont {Weiss}}, \bibinfo {author} {\bibfnamefont {J.}~\bibnamefont
  {Vu\ifmmode \check{c}\else \v{c}\fi{}kovi\ifmmode~\acute{c}\else
  \'{c}\fi{}}}, \bibinfo {author} {\bibfnamefont {V.}~\bibnamefont
  {Vuleti\ifmmode~\acute{c}\else \'{c}\fi{}}}, \bibinfo {author} {\bibfnamefont
  {J.}~\bibnamefont {Ye}},\ and\ \bibinfo {author} {\bibfnamefont
  {M.}~\bibnamefont {Zwierlein}},\ }\bibfield  {title} {\bibinfo {title}
  {Quantum simulators: Architectures and opportunities},\ }\href
  {https://doi.org/10.1103/PRXQuantum.2.017003} {\bibfield  {journal} {\bibinfo
   {journal} {PRX Quantum}\ }\textbf {\bibinfo {volume} {2}},\ \bibinfo {pages}
  {017003} (\bibinfo {year} {2021})}\BibitemShut {NoStop}%
\bibitem [{\citenamefont {Gross}\ and\ \citenamefont
  {Bloch}(2017)}]{Gross2017}%
  \BibitemOpen
  \bibfield  {author} {\bibinfo {author} {\bibfnamefont {C.}~\bibnamefont
  {Gross}}\ and\ \bibinfo {author} {\bibfnamefont {I.}~\bibnamefont {Bloch}},\
  }\bibfield  {title} {\bibinfo {title} {Quantum simulations with ultracold
  atoms in optical lattices},\ }\href {https://doi.org/10.1126/science.aal3837}
  {\bibfield  {journal} {\bibinfo  {journal} {Science}\ }\textbf {\bibinfo
  {volume} {357}},\ \bibinfo {pages} {995} (\bibinfo {year}
  {2017})}\BibitemShut {NoStop}%
\bibitem [{\citenamefont {Bloch}\ \emph {et~al.}(2012)\citenamefont {Bloch},
  \citenamefont {Dalibard},\ and\ \citenamefont
  {Nascimb{\`{e}}ne}}]{Bloch2012}%
  \BibitemOpen
  \bibfield  {author} {\bibinfo {author} {\bibfnamefont {I.}~\bibnamefont
  {Bloch}}, \bibinfo {author} {\bibfnamefont {J.}~\bibnamefont {Dalibard}},\
  and\ \bibinfo {author} {\bibfnamefont {S.}~\bibnamefont {Nascimb{\`{e}}ne}},\
  }\bibfield  {title} {\bibinfo {title} {Quantum simulations with ultracold
  quantum gases},\ }\href {https://doi.org/10.1038/nphys2259} {\bibfield
  {journal} {\bibinfo  {journal} {Nat. Phys.}\ }\textbf {\bibinfo {volume}
  {8}},\ \bibinfo {pages} {267} (\bibinfo {year} {2012})}\BibitemShut {NoStop}%
\bibitem [{\citenamefont {Parsons}\ \emph {et~al.}(2016)\citenamefont
  {Parsons}, \citenamefont {Mazurenko}, \citenamefont {Chiu}, \citenamefont
  {Ji}, \citenamefont {Greif},\ and\ \citenamefont {Greiner}}]{Parsons2016}%
  \BibitemOpen
  \bibfield  {author} {\bibinfo {author} {\bibfnamefont {M.~F.}\ \bibnamefont
  {Parsons}}, \bibinfo {author} {\bibfnamefont {A.}~\bibnamefont {Mazurenko}},
  \bibinfo {author} {\bibfnamefont {C.~S.}\ \bibnamefont {Chiu}}, \bibinfo
  {author} {\bibfnamefont {G.}~\bibnamefont {Ji}}, \bibinfo {author}
  {\bibfnamefont {D.}~\bibnamefont {Greif}},\ and\ \bibinfo {author}
  {\bibfnamefont {M.}~\bibnamefont {Greiner}},\ }\bibfield  {title} {\bibinfo
  {title} {Site-resolved measurement of the spin-correlation function in the
  {F}ermi-{H}ubbard model},\ }\href {https://doi.org/10.1126/science.aag1430}
  {\bibfield  {journal} {\bibinfo  {journal} {Science}\ }\textbf {\bibinfo
  {volume} {353}},\ \bibinfo {pages} {1253} (\bibinfo {year}
  {2016})}\BibitemShut {NoStop}%
\bibitem [{\citenamefont {Mazurenko}\ \emph {et~al.}(2017)\citenamefont
  {Mazurenko}, \citenamefont {Chiu}, \citenamefont {Ji}, \citenamefont
  {Parsons}, \citenamefont {Kan{\'{a}}sz-Nagy}, \citenamefont {Schmidt},
  \citenamefont {Grusdt}, \citenamefont {Demler}, \citenamefont {Greif},\ and\
  \citenamefont {Greiner}}]{Mazurenko2017}%
  \BibitemOpen
  \bibfield  {author} {\bibinfo {author} {\bibfnamefont {A.}~\bibnamefont
  {Mazurenko}}, \bibinfo {author} {\bibfnamefont {C.~S.}\ \bibnamefont {Chiu}},
  \bibinfo {author} {\bibfnamefont {G.}~\bibnamefont {Ji}}, \bibinfo {author}
  {\bibfnamefont {M.~F.}\ \bibnamefont {Parsons}}, \bibinfo {author}
  {\bibfnamefont {M.}~\bibnamefont {Kan{\'{a}}sz-Nagy}}, \bibinfo {author}
  {\bibfnamefont {R.}~\bibnamefont {Schmidt}}, \bibinfo {author} {\bibfnamefont
  {F.}~\bibnamefont {Grusdt}}, \bibinfo {author} {\bibfnamefont
  {E.}~\bibnamefont {Demler}}, \bibinfo {author} {\bibfnamefont
  {D.}~\bibnamefont {Greif}},\ and\ \bibinfo {author} {\bibfnamefont
  {M.}~\bibnamefont {Greiner}},\ }\bibfield  {title} {\bibinfo {title} {A
  cold-atom {F}ermi-{H}ubbard antiferromagnet},\ }\href
  {https://doi.org/10.1038/nature22362} {\bibfield  {journal} {\bibinfo
  {journal} {Nature}\ }\textbf {\bibinfo {volume} {545}},\ \bibinfo {pages}
  {462} (\bibinfo {year} {2017})}\BibitemShut {NoStop}%
\bibitem [{\citenamefont {Koepsell}\ \emph
  {et~al.}(2020{\natexlab{a}})\citenamefont {Koepsell}, \citenamefont
  {Bourgund}, \citenamefont {Sompet}, \citenamefont {Hirthe}, \citenamefont
  {Bohrdt}, \citenamefont {Wang}, \citenamefont {Grusdt}, \citenamefont
  {Demler}, \citenamefont {Salomon}, \citenamefont {Gross},\ and\ \citenamefont
  {Bloch}}]{Koepsell2020}%
  \BibitemOpen
  \bibfield  {author} {\bibinfo {author} {\bibfnamefont {J.}~\bibnamefont
  {Koepsell}}, \bibinfo {author} {\bibfnamefont {D.}~\bibnamefont {Bourgund}},
  \bibinfo {author} {\bibfnamefont {P.}~\bibnamefont {Sompet}}, \bibinfo
  {author} {\bibfnamefont {S.}~\bibnamefont {Hirthe}}, \bibinfo {author}
  {\bibfnamefont {A.}~\bibnamefont {Bohrdt}}, \bibinfo {author} {\bibfnamefont
  {Y.}~\bibnamefont {Wang}}, \bibinfo {author} {\bibfnamefont {F.}~\bibnamefont
  {Grusdt}}, \bibinfo {author} {\bibfnamefont {E.}~\bibnamefont {Demler}},
  \bibinfo {author} {\bibfnamefont {G.}~\bibnamefont {Salomon}}, \bibinfo
  {author} {\bibfnamefont {C.}~\bibnamefont {Gross}},\ and\ \bibinfo {author}
  {\bibfnamefont {I.}~\bibnamefont {Bloch}},\ }\bibfield  {title} {\bibinfo
  {title} {Microscopic evolution of doped {M}ott insulators from polaronic
  metal to {F}ermi liquid},\ }\href@noop {} {\bibfield  {journal} {\bibinfo
  {journal} {arXiv:2009.04440v1}\ } (\bibinfo {year}
  {2020}{\natexlab{a}})}\BibitemShut {NoStop}%
\bibitem [{\citenamefont {Ji}\ \emph {et~al.}(2021)\citenamefont {Ji},
  \citenamefont {Xu}, \citenamefont {Kendrick}, \citenamefont {Chiu},
  \citenamefont {Br\"uggenj\"urgen}, \citenamefont {Greif}, \citenamefont
  {Bohrdt}, \citenamefont {Grusdt}, \citenamefont {Demler}, \citenamefont
  {Lebrat},\ and\ \citenamefont {Greiner}}]{Ji2021}%
  \BibitemOpen
  \bibfield  {author} {\bibinfo {author} {\bibfnamefont {G.}~\bibnamefont
  {Ji}}, \bibinfo {author} {\bibfnamefont {M.}~\bibnamefont {Xu}}, \bibinfo
  {author} {\bibfnamefont {L.~H.}\ \bibnamefont {Kendrick}}, \bibinfo {author}
  {\bibfnamefont {C.~S.}\ \bibnamefont {Chiu}}, \bibinfo {author}
  {\bibfnamefont {J.~C.}\ \bibnamefont {Br\"uggenj\"urgen}}, \bibinfo {author}
  {\bibfnamefont {D.}~\bibnamefont {Greif}}, \bibinfo {author} {\bibfnamefont
  {A.}~\bibnamefont {Bohrdt}}, \bibinfo {author} {\bibfnamefont
  {F.}~\bibnamefont {Grusdt}}, \bibinfo {author} {\bibfnamefont
  {E.}~\bibnamefont {Demler}}, \bibinfo {author} {\bibfnamefont
  {M.}~\bibnamefont {Lebrat}},\ and\ \bibinfo {author} {\bibfnamefont
  {M.}~\bibnamefont {Greiner}},\ }\bibfield  {title} {\bibinfo {title}
  {Coupling a mobile hole to an antiferromagnetic spin background: Transient
  dynamics of a magnetic polaron},\ }\href
  {https://doi.org/10.1103/PhysRevX.11.021022} {\bibfield  {journal} {\bibinfo
  {journal} {Phys. Rev. X}\ }\textbf {\bibinfo {volume} {11}},\ \bibinfo
  {pages} {021022} (\bibinfo {year} {2021})}\BibitemShut {NoStop}%
\bibitem [{\citenamefont {Yamamoto}\ \emph {et~al.}(2016)\citenamefont
  {Yamamoto}, \citenamefont {Kobayashi}, \citenamefont {Kuno}, \citenamefont
  {Kato},\ and\ \citenamefont {Takahashi}}]{Yamamoto2016}%
  \BibitemOpen
  \bibfield  {author} {\bibinfo {author} {\bibfnamefont {R.}~\bibnamefont
  {Yamamoto}}, \bibinfo {author} {\bibfnamefont {J.}~\bibnamefont {Kobayashi}},
  \bibinfo {author} {\bibfnamefont {T.}~\bibnamefont {Kuno}}, \bibinfo {author}
  {\bibfnamefont {K.}~\bibnamefont {Kato}},\ and\ \bibinfo {author}
  {\bibfnamefont {Y.}~\bibnamefont {Takahashi}},\ }\bibfield  {title} {\bibinfo
  {title} {An ytterbium quantum gas microscope with narrow-line laser
  cooling},\ }\href {https://doi.org/10.1088/1367-2630/18/2/023016} {\bibfield
  {journal} {\bibinfo  {journal} {New J. Phys.}\ }\textbf {\bibinfo {volume}
  {18}},\ \bibinfo {pages} {023016} (\bibinfo {year} {2016})}\BibitemShut
  {NoStop}%
\bibitem [{\citenamefont {Sch{\"a}fer}\ \emph {et~al.}(2020)\citenamefont
  {Sch{\"a}fer}, \citenamefont {Fukuhara}, \citenamefont {Sugawa},
  \citenamefont {Takasu},\ and\ \citenamefont {Takahashi}}]{Schafer2020}%
  \BibitemOpen
  \bibfield  {author} {\bibinfo {author} {\bibfnamefont {F.}~\bibnamefont
  {Sch{\"a}fer}}, \bibinfo {author} {\bibfnamefont {T.}~\bibnamefont
  {Fukuhara}}, \bibinfo {author} {\bibfnamefont {S.}~\bibnamefont {Sugawa}},
  \bibinfo {author} {\bibfnamefont {Y.}~\bibnamefont {Takasu}},\ and\ \bibinfo
  {author} {\bibfnamefont {Y.}~\bibnamefont {Takahashi}},\ }\bibfield  {title}
  {\bibinfo {title} {Tools for quantum simulation with ultracold atoms in
  optical lattices},\ }\href
  {https://doi.org/https://doi.org/10.1038/s42254-020-0195-3} {\bibfield
  {journal} {\bibinfo  {journal} {Nat. Rev. Phys.}\ }\textbf {\bibinfo {volume}
  {2}},\ \bibinfo {pages} {411} (\bibinfo {year} {2020})}\BibitemShut {NoStop}%
\bibitem [{\citenamefont {Okuno}\ \emph {et~al.}(2020)\citenamefont {Okuno},
  \citenamefont {Amano}, \citenamefont {Enomoto}, \citenamefont {Takei},\ and\
  \citenamefont {Takahashi}}]{Okuno2020}%
  \BibitemOpen
  \bibfield  {author} {\bibinfo {author} {\bibfnamefont {D.}~\bibnamefont
  {Okuno}}, \bibinfo {author} {\bibfnamefont {Y.}~\bibnamefont {Amano}},
  \bibinfo {author} {\bibfnamefont {K.}~\bibnamefont {Enomoto}}, \bibinfo
  {author} {\bibfnamefont {N.}~\bibnamefont {Takei}},\ and\ \bibinfo {author}
  {\bibfnamefont {Y.}~\bibnamefont {Takahashi}},\ }\bibfield  {title} {\bibinfo
  {title} {Schemes for nondestructive quantum gas microscopy of single atoms in
  an optical lattice},\ }\href {https://doi.org/10.1088/1367-2630/ab6af9}
  {\bibfield  {journal} {\bibinfo  {journal} {New J. Phys.}\ }\textbf {\bibinfo
  {volume} {22}},\ \bibinfo {pages} {013041} (\bibinfo {year}
  {2020})}\BibitemShut {NoStop}%
\bibitem [{1()}]{1}%
  \BibitemOpen
  \href@noop {} {}\bibinfo {note} {The number of on-site pairs and the double
  occupancy are equivalent in the SU(2) Fermi-Hubbard model due to Pauli
  exclusion principle. However, for $N>2$ this is not the case. The number of
  on-site pairs $\mathcal {D}$ counts the pairs of particles per site, and is
  what controls the interaction energy $U \mathcal {D}$, while the double
  occupancy most naturally refers to the probability of configurations with
  exactly two particles per site, or to summing probabilities of all
  configurations with two or more particles per site. Either one of these is
  distinct from the number of on-site pairs $\mathcal {D}$. Computing the
  number of double occupancies thus requires the calculation of density
  fluctuations, i.e. terms of the order $\langle n^x\rangle $ with $x \in
  [2,N]$, which are computationally more expensive and experimentally harder to
  access.}\BibitemShut {Stop}%
\bibitem [{\citenamefont {Nakamura}\ \emph {et~al.}(2019)\citenamefont
  {Nakamura}, \citenamefont {Takasu}, \citenamefont {Kobayashi}, \citenamefont
  {Asaka}, \citenamefont {Fukushima}, \citenamefont {Inaba}, \citenamefont
  {Yamashita},\ and\ \citenamefont {Takahashi}}]{Nakamura2019}%
  \BibitemOpen
  \bibfield  {author} {\bibinfo {author} {\bibfnamefont {Y.}~\bibnamefont
  {Nakamura}}, \bibinfo {author} {\bibfnamefont {Y.}~\bibnamefont {Takasu}},
  \bibinfo {author} {\bibfnamefont {J.}~\bibnamefont {Kobayashi}}, \bibinfo
  {author} {\bibfnamefont {H.}~\bibnamefont {Asaka}}, \bibinfo {author}
  {\bibfnamefont {Y.}~\bibnamefont {Fukushima}}, \bibinfo {author}
  {\bibfnamefont {K.}~\bibnamefont {Inaba}}, \bibinfo {author} {\bibfnamefont
  {M.}~\bibnamefont {Yamashita}},\ and\ \bibinfo {author} {\bibfnamefont
  {Y.}~\bibnamefont {Takahashi}},\ }\bibfield  {title} {\bibinfo {title}
  {Experimental determination of {B}ose-{H}ubbard energies},\ }\href
  {https://doi.org/10.1103/PhysRevA.99.033609} {\bibfield  {journal} {\bibinfo
  {journal} {Phys. Rev. A}\ }\textbf {\bibinfo {volume} {99}},\ \bibinfo
  {pages} {033609} (\bibinfo {year} {2019})}\BibitemShut {NoStop}%
\bibitem [{\citenamefont {Koepsell}\ \emph
  {et~al.}(2020{\natexlab{b}})\citenamefont {Koepsell}, \citenamefont {Hirthe},
  \citenamefont {Bourgund}, \citenamefont {Sompet}, \citenamefont {Vijayan},
  \citenamefont {Salomon}, \citenamefont {Gross},\ and\ \citenamefont
  {Bloch}}]{Koepsell2020PRL}%
  \BibitemOpen
  \bibfield  {author} {\bibinfo {author} {\bibfnamefont {J.}~\bibnamefont
  {Koepsell}}, \bibinfo {author} {\bibfnamefont {S.}~\bibnamefont {Hirthe}},
  \bibinfo {author} {\bibfnamefont {D.}~\bibnamefont {Bourgund}}, \bibinfo
  {author} {\bibfnamefont {P.}~\bibnamefont {Sompet}}, \bibinfo {author}
  {\bibfnamefont {J.}~\bibnamefont {Vijayan}}, \bibinfo {author} {\bibfnamefont
  {G.}~\bibnamefont {Salomon}}, \bibinfo {author} {\bibfnamefont
  {C.}~\bibnamefont {Gross}},\ and\ \bibinfo {author} {\bibfnamefont
  {I.}~\bibnamefont {Bloch}},\ }\bibfield  {title} {\bibinfo {title} {Robust
  bilayer charge pumping for spin- and density-resolved quantum gas
  microscopy},\ }\href {https://doi.org/10.1103/PhysRevLett.125.010403}
  {\bibfield  {journal} {\bibinfo  {journal} {Phys. Rev. Lett.}\ }\textbf
  {\bibinfo {volume} {125}},\ \bibinfo {pages} {010403} (\bibinfo {year}
  {2020}{\natexlab{b}})}\BibitemShut {NoStop}%
\bibitem [{\citenamefont {Zhou}\ and\ \citenamefont {Ho}(2011)}]{Zhou2011}%
  \BibitemOpen
  \bibfield  {author} {\bibinfo {author} {\bibfnamefont {Q.}~\bibnamefont
  {Zhou}}\ and\ \bibinfo {author} {\bibfnamefont {T.-L.}\ \bibnamefont {Ho}},\
  }\bibfield  {title} {\bibinfo {title} {Universal thermometry for quantum
  simulation},\ }\href {https://doi.org/10.1103/PhysRevLett.106.225301}
  {\bibfield  {journal} {\bibinfo  {journal} {Phys. Rev. Lett.}\ }\textbf
  {\bibinfo {volume} {106}},\ \bibinfo {pages} {225301} (\bibinfo {year}
  {2011})}\BibitemShut {NoStop}%
\bibitem [{\citenamefont {Hartke}\ \emph {et~al.}(2020)\citenamefont {Hartke},
  \citenamefont {Oreg}, \citenamefont {Jia},\ and\ \citenamefont
  {Zwierlein}}]{Hartke2020}%
  \BibitemOpen
  \bibfield  {author} {\bibinfo {author} {\bibfnamefont {T.}~\bibnamefont
  {Hartke}}, \bibinfo {author} {\bibfnamefont {B.}~\bibnamefont {Oreg}},
  \bibinfo {author} {\bibfnamefont {N.}~\bibnamefont {Jia}},\ and\ \bibinfo
  {author} {\bibfnamefont {M.}~\bibnamefont {Zwierlein}},\ }\bibfield  {title}
  {\bibinfo {title} {Doublon-hole correlations and fluctuation thermometry in a
  fermi-hubbard gas},\ }\href {https://doi.org/10.1103/PhysRevLett.125.113601}
  {\bibfield  {journal} {\bibinfo  {journal} {Phys. Rev. Lett.}\ }\textbf
  {\bibinfo {volume} {125}},\ \bibinfo {pages} {113601} (\bibinfo {year}
  {2020})}\BibitemShut {NoStop}%
\bibitem [{\citenamefont {Blankenbecler}\ \emph {et~al.}(1981)\citenamefont
  {Blankenbecler}, \citenamefont {Scalapino},\ and\ \citenamefont
  {Sugar}}]{Blankenbecler1981}%
  \BibitemOpen
  \bibfield  {author} {\bibinfo {author} {\bibfnamefont {R.}~\bibnamefont
  {Blankenbecler}}, \bibinfo {author} {\bibfnamefont {D.~J.}\ \bibnamefont
  {Scalapino}},\ and\ \bibinfo {author} {\bibfnamefont {R.~L.}\ \bibnamefont
  {Sugar}},\ }\bibfield  {title} {\bibinfo {title} {Monte {C}arlo calculations
  of coupled boson-fermion systems. {I}},\ }\href
  {https://doi.org/10.1103/PhysRevD.24.2278} {\bibfield  {journal} {\bibinfo
  {journal} {Phys. Rev. D}\ }\textbf {\bibinfo {volume} {24}},\ \bibinfo
  {pages} {2278} (\bibinfo {year} {1981})}\BibitemShut {NoStop}%
\bibitem [{\citenamefont {Sorella}\ \emph {et~al.}(1989)\citenamefont
  {Sorella}, \citenamefont {Baroni}, \citenamefont {Car},\ and\ \citenamefont
  {Parrinello}}]{Sorella1989}%
  \BibitemOpen
  \bibfield  {author} {\bibinfo {author} {\bibfnamefont {S.}~\bibnamefont
  {Sorella}}, \bibinfo {author} {\bibfnamefont {S.}~\bibnamefont {Baroni}},
  \bibinfo {author} {\bibfnamefont {R.}~\bibnamefont {Car}},\ and\ \bibinfo
  {author} {\bibfnamefont {M.}~\bibnamefont {Parrinello}},\ }\bibfield  {title}
  {\bibinfo {title} {A novel technique for the simulation of interacting
  fermion systems},\ }\href {https://doi.org/10.1209/0295-5075/8/7/014}
  {\bibfield  {journal} {\bibinfo  {journal} {Europhys. Lett.}\ }\textbf
  {\bibinfo {volume} {8}},\ \bibinfo {pages} {663} (\bibinfo {year}
  {1989})}\BibitemShut {NoStop}%
\bibitem [{\citenamefont {Rigol}\ \emph {et~al.}(2006)\citenamefont {Rigol},
  \citenamefont {Bryant},\ and\ \citenamefont {Singh}}]{rigol2006numerical}%
  \BibitemOpen
  \bibfield  {author} {\bibinfo {author} {\bibfnamefont {M.}~\bibnamefont
  {Rigol}}, \bibinfo {author} {\bibfnamefont {T.}~\bibnamefont {Bryant}},\ and\
  \bibinfo {author} {\bibfnamefont {R.~R.~P.}\ \bibnamefont {Singh}},\
  }\bibfield  {title} {\bibinfo {title} {Numerical linked-cluster approach to
  quantum lattice models},\ }\href@noop {} {\bibfield  {journal} {\bibinfo
  {journal} {Phys. Rev. Lett.}\ }\textbf {\bibinfo {volume} {97}},\ \bibinfo
  {pages} {187202} (\bibinfo {year} {2006})}\BibitemShut {NoStop}%
\bibitem [{\citenamefont {Tang}\ \emph {et~al.}(2013)\citenamefont {Tang},
  \citenamefont {Khatami},\ and\ \citenamefont {Rigol}}]{tang2013short}%
  \BibitemOpen
  \bibfield  {author} {\bibinfo {author} {\bibfnamefont {B.}~\bibnamefont
  {Tang}}, \bibinfo {author} {\bibfnamefont {E.}~\bibnamefont {Khatami}},\ and\
  \bibinfo {author} {\bibfnamefont {M.}~\bibnamefont {Rigol}},\ }\bibfield
  {title} {\bibinfo {title} {A short introduction to numerical linked-cluster
  expansions},\ }\href
  {https://doi.org/https://doi.org/10.1016/j.cpc.2012.10.008} {\bibfield
  {journal} {\bibinfo  {journal} {Computer Physics Communications}\ }\textbf
  {\bibinfo {volume} {184}},\ \bibinfo {pages} {557} (\bibinfo {year}
  {2013})}\BibitemShut {NoStop}%
\bibitem [{\citenamefont {Hart}\ \emph {et~al.}(2015)\citenamefont {Hart},
  \citenamefont {Duarte}, \citenamefont {Yang}, \citenamefont {Liu},
  \citenamefont {Paiva}, \citenamefont {Khatami}, \citenamefont {Scalettar},
  \citenamefont {Trivedi}, \citenamefont {Huse},\ and\ \citenamefont
  {Hulet}}]{Hart2015}%
  \BibitemOpen
  \bibfield  {author} {\bibinfo {author} {\bibfnamefont {R.~A.}\ \bibnamefont
  {Hart}}, \bibinfo {author} {\bibfnamefont {P.~M.}\ \bibnamefont {Duarte}},
  \bibinfo {author} {\bibfnamefont {T.-L.}\ \bibnamefont {Yang}}, \bibinfo
  {author} {\bibfnamefont {X.}~\bibnamefont {Liu}}, \bibinfo {author}
  {\bibfnamefont {T.}~\bibnamefont {Paiva}}, \bibinfo {author} {\bibfnamefont
  {E.}~\bibnamefont {Khatami}}, \bibinfo {author} {\bibfnamefont {R.~T.}\
  \bibnamefont {Scalettar}}, \bibinfo {author} {\bibfnamefont {N.}~\bibnamefont
  {Trivedi}}, \bibinfo {author} {\bibfnamefont {D.~A.}\ \bibnamefont {Huse}},\
  and\ \bibinfo {author} {\bibfnamefont {R.~G.}\ \bibnamefont {Hulet}},\
  }\bibfield  {title} {\bibinfo {title} {Observation of antiferromagnetic
  correlations in the {H}ubbard model with ultracold atoms},\ }\href@noop {}
  {\bibfield  {journal} {\bibinfo  {journal} {Nature}\ }\textbf {\bibinfo
  {volume} {519}},\ \bibinfo {pages} {211} (\bibinfo {year}
  {2015})}\BibitemShut {NoStop}%
\bibitem [{\citenamefont {Greif}\ \emph {et~al.}(2013)\citenamefont {Greif},
  \citenamefont {Uehlinger}, \citenamefont {Jotzu}, \citenamefont {Tarruell},\
  and\ \citenamefont {Esslinger}}]{Greif2013}%
  \BibitemOpen
  \bibfield  {author} {\bibinfo {author} {\bibfnamefont {D.}~\bibnamefont
  {Greif}}, \bibinfo {author} {\bibfnamefont {T.}~\bibnamefont {Uehlinger}},
  \bibinfo {author} {\bibfnamefont {G.}~\bibnamefont {Jotzu}}, \bibinfo
  {author} {\bibfnamefont {L.}~\bibnamefont {Tarruell}},\ and\ \bibinfo
  {author} {\bibfnamefont {T.}~\bibnamefont {Esslinger}},\ }\bibfield  {title}
  {\bibinfo {title} {Short-range quantum magnetism of ultracold fermions in an
  optical lattice},\ }\href@noop {} {\bibfield  {journal} {\bibinfo  {journal}
  {Science}\ }\textbf {\bibinfo {volume} {340}},\ \bibinfo {pages} {1307}
  (\bibinfo {year} {2013})}\BibitemShut {NoStop}%
\bibitem [{\citenamefont {Cheuk}\ \emph {et~al.}(2016)\citenamefont {Cheuk},
  \citenamefont {Nichols}, \citenamefont {Lawrence}, \citenamefont {Okan},
  \citenamefont {Zhang}, \citenamefont {Khatami}, \citenamefont {Trivedi},
  \citenamefont {Paiva}, \citenamefont {Rigol},\ and\ \citenamefont
  {Zwierlein}}]{Cheuk2016}%
  \BibitemOpen
  \bibfield  {author} {\bibinfo {author} {\bibfnamefont {L.~W.}\ \bibnamefont
  {Cheuk}}, \bibinfo {author} {\bibfnamefont {M.~A.}\ \bibnamefont {Nichols}},
  \bibinfo {author} {\bibfnamefont {K.~R.}\ \bibnamefont {Lawrence}}, \bibinfo
  {author} {\bibfnamefont {M.}~\bibnamefont {Okan}}, \bibinfo {author}
  {\bibfnamefont {H.}~\bibnamefont {Zhang}}, \bibinfo {author} {\bibfnamefont
  {E.}~\bibnamefont {Khatami}}, \bibinfo {author} {\bibfnamefont
  {N.}~\bibnamefont {Trivedi}}, \bibinfo {author} {\bibfnamefont
  {T.}~\bibnamefont {Paiva}}, \bibinfo {author} {\bibfnamefont
  {M.}~\bibnamefont {Rigol}},\ and\ \bibinfo {author} {\bibfnamefont {M.~W.}\
  \bibnamefont {Zwierlein}},\ }\bibfield  {title} {\bibinfo {title}
  {Observation of spatial charge and spin correlations in the 2{D}
  {F}ermi-{H}ubbard model},\ }\href {https://doi.org/10.1126/science.aag3349}
  {\bibfield  {journal} {\bibinfo  {journal} {Science}\ }\textbf {\bibinfo
  {volume} {353}},\ \bibinfo {pages} {1260} (\bibinfo {year}
  {2016})}\BibitemShut {NoStop}%
\bibitem [{\citenamefont {Brown}\ \emph {et~al.}(2017)\citenamefont {Brown},
  \citenamefont {Mitra}, \citenamefont {Guardado-Sanchez}, \citenamefont
  {Schau{\ss}}, \citenamefont {Kondov}, \citenamefont {Khatami}, \citenamefont
  {Paiva}, \citenamefont {Trivedi}, \citenamefont {Huse},\ and\ \citenamefont
  {Bakr}}]{Brown2017}%
  \BibitemOpen
  \bibfield  {author} {\bibinfo {author} {\bibfnamefont {P.~T.}\ \bibnamefont
  {Brown}}, \bibinfo {author} {\bibfnamefont {D.}~\bibnamefont {Mitra}},
  \bibinfo {author} {\bibfnamefont {E.}~\bibnamefont {Guardado-Sanchez}},
  \bibinfo {author} {\bibfnamefont {P.}~\bibnamefont {Schau{\ss}}}, \bibinfo
  {author} {\bibfnamefont {S.~S.}\ \bibnamefont {Kondov}}, \bibinfo {author}
  {\bibfnamefont {E.}~\bibnamefont {Khatami}}, \bibinfo {author} {\bibfnamefont
  {T.}~\bibnamefont {Paiva}}, \bibinfo {author} {\bibfnamefont
  {N.}~\bibnamefont {Trivedi}}, \bibinfo {author} {\bibfnamefont {D.~A.}\
  \bibnamefont {Huse}},\ and\ \bibinfo {author} {\bibfnamefont {W.~S.}\
  \bibnamefont {Bakr}},\ }\bibfield  {title} {\bibinfo {title} {Spin-imbalance
  in a 2{D} {F}ermi-{H}ubbard system},\ }\href@noop {} {\bibfield  {journal}
  {\bibinfo  {journal} {Science}\ }\textbf {\bibinfo {volume} {357}},\ \bibinfo
  {pages} {1385} (\bibinfo {year} {2017})}\BibitemShut {NoStop}%
\bibitem [{\citenamefont {Brown}\ \emph {et~al.}(2018)\citenamefont {Brown},
  \citenamefont {Mitra}, \citenamefont {Guardado-Sanchez}, \citenamefont
  {Nourafkan}, \citenamefont {Reymbaut}, \citenamefont {H{\'{e}}bert},
  \citenamefont {Bergeron}, \citenamefont {Tremblay}, \citenamefont {Kokalj},
  \citenamefont {Huse}, \citenamefont {Schau{\ss}},\ and\ \citenamefont
  {Bakr}}]{Brown2018}%
  \BibitemOpen
  \bibfield  {author} {\bibinfo {author} {\bibfnamefont {P.~T.}\ \bibnamefont
  {Brown}}, \bibinfo {author} {\bibfnamefont {D.}~\bibnamefont {Mitra}},
  \bibinfo {author} {\bibfnamefont {E.}~\bibnamefont {Guardado-Sanchez}},
  \bibinfo {author} {\bibfnamefont {R.}~\bibnamefont {Nourafkan}}, \bibinfo
  {author} {\bibfnamefont {A.}~\bibnamefont {Reymbaut}}, \bibinfo {author}
  {\bibfnamefont {C.-D.}\ \bibnamefont {H{\'{e}}bert}}, \bibinfo {author}
  {\bibfnamefont {S.}~\bibnamefont {Bergeron}}, \bibinfo {author}
  {\bibfnamefont {A.-M.~S.}\ \bibnamefont {Tremblay}}, \bibinfo {author}
  {\bibfnamefont {J.}~\bibnamefont {Kokalj}}, \bibinfo {author} {\bibfnamefont
  {D.~A.}\ \bibnamefont {Huse}}, \bibinfo {author} {\bibfnamefont
  {P.}~\bibnamefont {Schau{\ss}}},\ and\ \bibinfo {author} {\bibfnamefont
  {W.~S.}\ \bibnamefont {Bakr}},\ }\bibfield  {title} {\bibinfo {title} {Bad
  metallic transport in a cold atom {F}ermi-{H}ubbard system},\ }\href@noop {}
  {\bibfield  {journal} {\bibinfo  {journal} {Science}\ }\textbf {\bibinfo
  {volume} {363}},\ \bibinfo {pages} {379} (\bibinfo {year}
  {2018})}\BibitemShut {NoStop}%
\bibitem [{2()}]{2}%
  \BibitemOpen
  \href@noop {} {}\bibinfo {note} {Previous work applied Determinant Quantum
  Monte Carlo for the SU($2N$) Fermi-Hubbard model at half-filling, i.e.
  $\langle n \rangle =N/2$, using an alternative Hubbard-Stratonovitch
  decomposition. This alternative decomposition is free of the sign problem at
  half-fillling for SU($2N$)~\cite
  {Assaad1998,Wang2014,Zhou2014,Assaad2005}}\BibitemShut {NoStop}%
\bibitem [{3()}]{3}%
  \BibitemOpen
  \href@noop {} {}\bibinfo {note} {A sweep updates all the auxiliary fields at
  every lattice site and imaginary time slice.}\BibitemShut {Stop}%
\bibitem [{\citenamefont {Scalettar}\ \emph {et~al.}(1991)\citenamefont
  {Scalettar}, \citenamefont {Noack},\ and\ \citenamefont
  {Singh}}]{Scalettar1991}%
  \BibitemOpen
  \bibfield  {author} {\bibinfo {author} {\bibfnamefont {R.~T.}\ \bibnamefont
  {Scalettar}}, \bibinfo {author} {\bibfnamefont {R.~M.}\ \bibnamefont
  {Noack}},\ and\ \bibinfo {author} {\bibfnamefont {R.~R.~P.}\ \bibnamefont
  {Singh}},\ }\bibfield  {title} {\bibinfo {title} {Ergodicity at large
  couplings with the determinant {M}onte {C}arlo algorithm},\ }\href
  {https://doi.org/10.1103/PhysRevB.44.10502} {\bibfield  {journal} {\bibinfo
  {journal} {Phys. Rev. B}\ }\textbf {\bibinfo {volume} {44}},\ \bibinfo
  {pages} {10502} (\bibinfo {year} {1991})}\BibitemShut {NoStop}%
\bibitem [{4()}]{4}%
  \BibitemOpen
  \href@noop {} {}\bibinfo {note} {We used the three-point differentiation rule
  \begin {align*}\label {eq:Deriv} f'(x) =& \left [\frac {x_i -
  x_{i+1}}{(x_{i-1} - x_i)(x_{i-1} - x_{i+1})} \right ] f(x_{i-1}) \\ + &\left
  [\frac {2x_i - x_{i-1} - x_{i+1}}{(x_i - x_{i-1})(x_i - x_{i+1})} \right ]
  f(x_{i}) \\ + &\left [\frac { x_i - x_{i-1} }{(x_{i+1} - x_{i-1})(x_{i+1} -
  x_i)} \right ] f(x_{i+1}), \end {align*} with error $\mathcal {O}(h^2)$ where
  $h$ is the maximum spacing of adjacent $x_i$. Statistical error bars are
  obtained by error propagation.}\BibitemShut {Stop}%
\bibitem [{\citenamefont {Paiva}\ \emph {et~al.}(2001)\citenamefont {Paiva},
  \citenamefont {Scalettar}, \citenamefont {Huscroft},\ and\ \citenamefont
  {McMahan}}]{Paiva2001}%
  \BibitemOpen
  \bibfield  {author} {\bibinfo {author} {\bibfnamefont {T.}~\bibnamefont
  {Paiva}}, \bibinfo {author} {\bibfnamefont {R.~T.}\ \bibnamefont
  {Scalettar}}, \bibinfo {author} {\bibfnamefont {C.}~\bibnamefont
  {Huscroft}},\ and\ \bibinfo {author} {\bibfnamefont {A.~K.}\ \bibnamefont
  {McMahan}},\ }\bibfield  {title} {\bibinfo {title} {Signatures of spin and
  charge energy scales in the local moment and specific heat of the half-filled
  two-dimensional {H}ubbard model},\ }\href
  {https://doi.org/10.1103/PhysRevB.63.125116} {\bibfield  {journal} {\bibinfo
  {journal} {Phys. Rev. B}\ }\textbf {\bibinfo {volume} {63}},\ \bibinfo
  {pages} {125116} (\bibinfo {year} {2001})}\BibitemShut {NoStop}%
\bibitem [{\citenamefont {McMahan}\ \emph {et~al.}(1998)\citenamefont
  {McMahan}, \citenamefont {Huscroft}, \citenamefont {Scalettar},\ and\
  \citenamefont {Pollock}}]{McMahan1998}%
  \BibitemOpen
  \bibfield  {author} {\bibinfo {author} {\bibfnamefont {A.}~\bibnamefont
  {McMahan}}, \bibinfo {author} {\bibfnamefont {C.}~\bibnamefont {Huscroft}},
  \bibinfo {author} {\bibfnamefont {R.}~\bibnamefont {Scalettar}},\ and\
  \bibinfo {author} {\bibfnamefont {E.}~\bibnamefont {Pollock}},\ }\bibfield
  {title} {\bibinfo {title} {Volume-collapse transitions in the rare earth
  metals},\ }\href {https://doi.org/https://doi.org/10.1023/A:1008698422183}
  {\bibfield  {journal} {\bibinfo  {journal} {J. Comput.-Aided Mater. Des.}\
  }\textbf {\bibinfo {volume} {5}},\ \bibinfo {pages} {131} (\bibinfo {year}
  {1998})}\BibitemShut {NoStop}%
\bibitem [{5()}]{5}%
  \BibitemOpen
  \href@noop {} {}\bibinfo {note} {An equal weight on regularization and
  fitting terms is enough to ensure that $S\to 0$ as $T\to 0$ with an error
  $\lesssim 10^{-2}$ for all $N$ and $U/t$.}\BibitemShut {Stop}%
\bibitem [{6()}]{6}%
  \BibitemOpen
  \href@noop {} {}\bibinfo {note} {The idea of truncating the Hilbert space is
  not new, see for example, Ref.~\cite {Bhattaram2019}, where authors limit the
  thermal averages to a fraction of low-lying states and use Lanczos algorithm
  to reduce the computational cost of full diagonalization.}\BibitemShut
  {Stop}%
\bibitem [{\citenamefont {Hazzard}\ \emph {et~al.}(2012)\citenamefont
  {Hazzard}, \citenamefont {Gurarie}, \citenamefont {Hermele},\ and\
  \citenamefont {Rey}}]{Hazzard2012}%
  \BibitemOpen
  \bibfield  {author} {\bibinfo {author} {\bibfnamefont {K.~R.~A.}\
  \bibnamefont {Hazzard}}, \bibinfo {author} {\bibfnamefont {V.}~\bibnamefont
  {Gurarie}}, \bibinfo {author} {\bibfnamefont {M.}~\bibnamefont {Hermele}},\
  and\ \bibinfo {author} {\bibfnamefont {A.~M.}\ \bibnamefont {Rey}},\
  }\bibfield  {title} {\bibinfo {title} {High-temperature properties of
  fermionic alkaline-earth-metal atoms in optical lattices},\ }\href
  {https://doi.org/10.1103/PhysRevA.85.041604} {\bibfield  {journal} {\bibinfo
  {journal} {Phys. Rev. A}\ }\textbf {\bibinfo {volume} {85}},\ \bibinfo
  {pages} {041604(R)} (\bibinfo {year} {2012})}\BibitemShut {NoStop}%
\bibitem [{\citenamefont {Pethick}\ and\ \citenamefont
  {Smith}(2008)}]{PethickSmith}%
  \BibitemOpen
  \bibfield  {author} {\bibinfo {author} {\bibfnamefont {C.~J.}\ \bibnamefont
  {Pethick}}\ and\ \bibinfo {author} {\bibfnamefont {H.}~\bibnamefont
  {Smith}},\ }\href@noop {} {\emph {\bibinfo {title} {Bose-{E}instein
  {C}ondensation in {D}ilute {G}ases}}}\ (\bibinfo  {publisher} {Cambridge
  University},\ \bibinfo {year} {2008})\BibitemShut {NoStop}%
\bibitem [{\citenamefont {Iglovikov}\ \emph {et~al.}(2015)\citenamefont
  {Iglovikov}, \citenamefont {Khatami},\ and\ \citenamefont
  {Scalettar}}]{Iglovikov2015}%
  \BibitemOpen
  \bibfield  {author} {\bibinfo {author} {\bibfnamefont {V.~I.}\ \bibnamefont
  {Iglovikov}}, \bibinfo {author} {\bibfnamefont {E.}~\bibnamefont {Khatami}},\
  and\ \bibinfo {author} {\bibfnamefont {R.~T.}\ \bibnamefont {Scalettar}},\
  }\bibfield  {title} {\bibinfo {title} {Geometry dependence of the sign
  problem in quantum {M}onte {C}arlo simulations},\ }\href
  {https://doi.org/10.1103/PhysRevB.92.045110} {\bibfield  {journal} {\bibinfo
  {journal} {Phys. Rev. B}\ }\textbf {\bibinfo {volume} {92}},\ \bibinfo
  {pages} {045110} (\bibinfo {year} {2015})}\BibitemShut {NoStop}%
\bibitem [{\citenamefont {Batrouni}\ and\ \citenamefont
  {Scalettar}(1990)}]{Bautroni1990}%
  \BibitemOpen
  \bibfield  {author} {\bibinfo {author} {\bibfnamefont {G.~G.}\ \bibnamefont
  {Batrouni}}\ and\ \bibinfo {author} {\bibfnamefont {R.~T.}\ \bibnamefont
  {Scalettar}},\ }\bibfield  {title} {\bibinfo {title} {Anomalous decouplings
  and the fermion sign problem},\ }\href
  {https://doi.org/10.1103/PhysRevB.42.2282} {\bibfield  {journal} {\bibinfo
  {journal} {Phys. Rev. B}\ }\textbf {\bibinfo {volume} {42}},\ \bibinfo
  {pages} {2282} (\bibinfo {year} {1990})}\BibitemShut {NoStop}%
\bibitem [{\citenamefont {Batrouni}\ and\ \citenamefont
  {de~Forcrand}(1993)}]{Bautroni1993}%
  \BibitemOpen
  \bibfield  {author} {\bibinfo {author} {\bibfnamefont {G.~G.}\ \bibnamefont
  {Batrouni}}\ and\ \bibinfo {author} {\bibfnamefont {P.}~\bibnamefont
  {de~Forcrand}},\ }\bibfield  {title} {\bibinfo {title} {Fermion sign problem:
  Decoupling transformation and simulation algorithm},\ }\href
  {https://doi.org/10.1103/PhysRevB.48.589} {\bibfield  {journal} {\bibinfo
  {journal} {Phys. Rev. B}\ }\textbf {\bibinfo {volume} {48}},\ \bibinfo
  {pages} {589} (\bibinfo {year} {1993})}\BibitemShut {NoStop}%
\bibitem [{\citenamefont {Lee}\ \emph {et~al.}(2018)\citenamefont {Lee},
  \citenamefont {von Delft},\ and\ \citenamefont {Weichselbaum}}]{Lee2018}%
  \BibitemOpen
  \bibfield  {author} {\bibinfo {author} {\bibfnamefont {S.-S.~B.}\
  \bibnamefont {Lee}}, \bibinfo {author} {\bibfnamefont {J.}~\bibnamefont {von
  Delft}},\ and\ \bibinfo {author} {\bibfnamefont {A.}~\bibnamefont
  {Weichselbaum}},\ }\bibfield  {title} {\bibinfo {title} {Filling-driven
  {M}ott transition in {SU}$({N})$ {H}ubbard models},\ }\href
  {https://doi.org/10.1103/PhysRevB.97.165143} {\bibfield  {journal} {\bibinfo
  {journal} {Phys. Rev. B}\ }\textbf {\bibinfo {volume} {97}},\ \bibinfo
  {pages} {165143} (\bibinfo {year} {2018})}\BibitemShut {NoStop}%
\bibitem [{\citenamefont {Ibarra-Garc\'{\i}a-Padilla}\ \emph
  {et~al.}(2020)\citenamefont {Ibarra-Garc\'{\i}a-Padilla}, \citenamefont
  {Mukherjee}, \citenamefont {Hulet}, \citenamefont {Hazzard}, \citenamefont
  {Paiva},\ and\ \citenamefont {Scalettar}}]{IbarraGarciaPadilla2020}%
  \BibitemOpen
  \bibfield  {author} {\bibinfo {author} {\bibfnamefont {E.}~\bibnamefont
  {Ibarra-Garc\'{\i}a-Padilla}}, \bibinfo {author} {\bibfnamefont
  {R.}~\bibnamefont {Mukherjee}}, \bibinfo {author} {\bibfnamefont {R.~G.}\
  \bibnamefont {Hulet}}, \bibinfo {author} {\bibfnamefont {K.~R.~A.}\
  \bibnamefont {Hazzard}}, \bibinfo {author} {\bibfnamefont {T.}~\bibnamefont
  {Paiva}},\ and\ \bibinfo {author} {\bibfnamefont {R.~T.}\ \bibnamefont
  {Scalettar}},\ }\bibfield  {title} {\bibinfo {title} {Thermodynamics and
  magnetism in the two-dimensional to three-dimensional crossover of the
  {H}ubbard model},\ }\href {https://doi.org/10.1103/PhysRevA.102.033340}
  {\bibfield  {journal} {\bibinfo  {journal} {Phys. Rev. A}\ }\textbf {\bibinfo
  {volume} {102}},\ \bibinfo {pages} {033340} (\bibinfo {year}
  {2020})}\BibitemShut {NoStop}%
\bibitem [{Din()}]{Dinf}%
  \BibitemOpen
  \href@noop {} {}\bibinfo {note} {In the $T \to \infty $ limit, at $\langle n
  \rangle =1$, the number of on-site pairs is given by \begin {equation*}
  \mathcal {D}_\infty = \frac {1}{2} \sum _{\sigma \neq \tau } \langle n_\sigma
  \rangle \langle n_\tau \rangle = \binom {N}{2} \frac {1}{N^2}= \frac
  {1}{2}\left (1-\frac {1}{N}\right ), \end {equation*} where we used that
  $\langle n_\sigma \rangle = \langle n \rangle /N$ because of the SU($N$)
  symmetry.}\BibitemShut {Stop}%
\bibitem [{\citenamefont {Greger}\ \emph {et~al.}(2013)\citenamefont {Greger},
  \citenamefont {Kollar},\ and\ \citenamefont {Vollhardt}}]{Greger2013}%
  \BibitemOpen
  \bibfield  {author} {\bibinfo {author} {\bibfnamefont {M.}~\bibnamefont
  {Greger}}, \bibinfo {author} {\bibfnamefont {M.}~\bibnamefont {Kollar}},\
  and\ \bibinfo {author} {\bibfnamefont {D.}~\bibnamefont {Vollhardt}},\
  }\bibfield  {title} {\bibinfo {title} {Isosbestic points: How a narrow
  crossing region of curves determines their leading parameter dependence},\
  }\href {https://doi.org/10.1103/PhysRevB.87.195140} {\bibfield  {journal}
  {\bibinfo  {journal} {Phys. Rev. B}\ }\textbf {\bibinfo {volume} {87}},\
  \bibinfo {pages} {195140} (\bibinfo {year} {2013})}\BibitemShut {NoStop}%
\bibitem [{\citenamefont {Vollhardt}(1997)}]{Vollhardt1997}%
  \BibitemOpen
  \bibfield  {author} {\bibinfo {author} {\bibfnamefont {D.}~\bibnamefont
  {Vollhardt}},\ }\bibfield  {title} {\bibinfo {title} {Characteristic crossing
  points in specific heat curves of correlated systems},\ }\href
  {https://doi.org/10.1103/PhysRevLett.78.1307} {\bibfield  {journal} {\bibinfo
   {journal} {Phys. Rev. Lett.}\ }\textbf {\bibinfo {volume} {78}},\ \bibinfo
  {pages} {1307} (\bibinfo {year} {1997})}\BibitemShut {NoStop}%
\bibitem [{\citenamefont {Chandra}\ \emph {et~al.}(1999)\citenamefont
  {Chandra}, \citenamefont {Kollar},\ and\ \citenamefont
  {Vollhardt}}]{Chandra1999}%
  \BibitemOpen
  \bibfield  {author} {\bibinfo {author} {\bibfnamefont {N.}~\bibnamefont
  {Chandra}}, \bibinfo {author} {\bibfnamefont {M.}~\bibnamefont {Kollar}},\
  and\ \bibinfo {author} {\bibfnamefont {D.}~\bibnamefont {Vollhardt}},\
  }\bibfield  {title} {\bibinfo {title} {Nearly universal crossing point of the
  specific heat curves of {H}ubbard models},\ }\href
  {https://doi.org/10.1103/PhysRevB.59.10541} {\bibfield  {journal} {\bibinfo
  {journal} {Phys. Rev. B}\ }\textbf {\bibinfo {volume} {59}},\ \bibinfo
  {pages} {10541} (\bibinfo {year} {1999})}\BibitemShut {NoStop}%
\bibitem [{\citenamefont {Duffy}\ and\ \citenamefont
  {Moreo}(1997)}]{Duffy1997}%
  \BibitemOpen
  \bibfield  {author} {\bibinfo {author} {\bibfnamefont {D.}~\bibnamefont
  {Duffy}}\ and\ \bibinfo {author} {\bibfnamefont {A.}~\bibnamefont {Moreo}},\
  }\bibfield  {title} {\bibinfo {title} {Specific heat of the two-dimensional
  {H}ubbard model},\ }\href {https://doi.org/10.1103/PhysRevB.55.12918}
  {\bibfield  {journal} {\bibinfo  {journal} {Phys. Rev. B}\ }\textbf {\bibinfo
  {volume} {55}},\ \bibinfo {pages} {12918} (\bibinfo {year}
  {1997})}\BibitemShut {NoStop}%
\bibitem [{\citenamefont {Macedo}\ and\ \citenamefont
  {de~Souza}(2002)}]{Macedo2002}%
  \BibitemOpen
  \bibfield  {author} {\bibinfo {author} {\bibfnamefont {C.~A.}\ \bibnamefont
  {Macedo}}\ and\ \bibinfo {author} {\bibfnamefont {A.~M.~C.}\ \bibnamefont
  {de~Souza}},\ }\bibfield  {title} {\bibinfo {title} {Crossing points in
  specific-heat curves of the asymmetric {H}ubbard model},\ }\href
  {https://doi.org/10.1103/PhysRevB.65.153109} {\bibfield  {journal} {\bibinfo
  {journal} {Phys. Rev. B}\ }\textbf {\bibinfo {volume} {65}},\ \bibinfo
  {pages} {153109} (\bibinfo {year} {2002})}\BibitemShut {NoStop}%
\bibitem [{\citenamefont {Paiva}\ \emph {et~al.}(2005)\citenamefont {Paiva},
  \citenamefont {Scalettar}, \citenamefont {Zheng}, \citenamefont {Singh},\
  and\ \citenamefont {Oitmaa}}]{Paiva2005}%
  \BibitemOpen
  \bibfield  {author} {\bibinfo {author} {\bibfnamefont {T.}~\bibnamefont
  {Paiva}}, \bibinfo {author} {\bibfnamefont {R.~T.}\ \bibnamefont
  {Scalettar}}, \bibinfo {author} {\bibfnamefont {W.}~\bibnamefont {Zheng}},
  \bibinfo {author} {\bibfnamefont {R.~R.~P.}\ \bibnamefont {Singh}},\ and\
  \bibinfo {author} {\bibfnamefont {J.}~\bibnamefont {Oitmaa}},\ }\bibfield
  {title} {\bibinfo {title} {Ground-state and finite-temperature signatures of
  quantum phase transitions in the half-filled {H}ubbard model on a honeycomb
  lattice},\ }\href {https://doi.org/10.1103/PhysRevB.72.085123} {\bibfield
  {journal} {\bibinfo  {journal} {Phys. Rev. B}\ }\textbf {\bibinfo {volume}
  {72}},\ \bibinfo {pages} {085123} (\bibinfo {year} {2005})}\BibitemShut
  {NoStop}%
\bibitem [{\citenamefont {M\"uller}\ \emph {et~al.}(2021)\citenamefont
  {M\"uller}, \citenamefont {Lajk\'o}, \citenamefont {Schreck}, \citenamefont
  {Mila},\ and\ \citenamefont {Min\'a\ifmmode~\check{r}\else
  \v{r}\fi{}}}]{Merlin2021}%
  \BibitemOpen
  \bibfield  {author} {\bibinfo {author} {\bibfnamefont {A.~M.}\ \bibnamefont
  {M\"uller}}, \bibinfo {author} {\bibfnamefont {M.}~\bibnamefont {Lajk\'o}},
  \bibinfo {author} {\bibfnamefont {F.}~\bibnamefont {Schreck}}, \bibinfo
  {author} {\bibfnamefont {F.}~\bibnamefont {Mila}},\ and\ \bibinfo {author}
  {\bibfnamefont {J.~c.~v.}\ \bibnamefont {Min\'a\ifmmode~\check{r}\else
  \v{r}\fi{}}},\ }\bibfield  {title} {\bibinfo {title} {State selective cooling
  of $\mathrm{SU}({N})$ {F}ermi gases},\ }\href
  {https://doi.org/10.1103/PhysRevA.104.013304} {\bibfield  {journal} {\bibinfo
   {journal} {Phys. Rev. A}\ }\textbf {\bibinfo {volume} {104}},\ \bibinfo
  {pages} {013304} (\bibinfo {year} {2021})}\BibitemShut {NoStop}%
\bibitem [{\citenamefont {Romen}\ and\ \citenamefont
  {L\"auchli}(2020)}]{Romen2020}%
  \BibitemOpen
  \bibfield  {author} {\bibinfo {author} {\bibfnamefont {C.}~\bibnamefont
  {Romen}}\ and\ \bibinfo {author} {\bibfnamefont {A.~M.}\ \bibnamefont
  {L\"auchli}},\ }\bibfield  {title} {\bibinfo {title} {Structure of spin
  correlations in high-temperature {SU}({$N$}) quantum magnets},\ }\href
  {https://doi.org/10.1103/PhysRevResearch.2.043009} {\bibfield  {journal}
  {\bibinfo  {journal} {Phys. Rev. Research}\ }\textbf {\bibinfo {volume}
  {2}},\ \bibinfo {pages} {043009} (\bibinfo {year} {2020})}\BibitemShut
  {NoStop}%
\bibitem [{\citenamefont {Khatami}\ and\ \citenamefont
  {Rigol}(2011)}]{Khatami2011}%
  \BibitemOpen
  \bibfield  {author} {\bibinfo {author} {\bibfnamefont {E.}~\bibnamefont
  {Khatami}}\ and\ \bibinfo {author} {\bibfnamefont {M.}~\bibnamefont
  {Rigol}},\ }\bibfield  {title} {\bibinfo {title} {Thermodynamics of strongly
  interacting fermions in two-dimensional optical lattices},\ }\href@noop {}
  {\bibfield  {journal} {\bibinfo  {journal} {Phys. Rev. A}\ }\textbf {\bibinfo
  {volume} {84}},\ \bibinfo {pages} {053611} (\bibinfo {year}
  {2011})}\BibitemShut {NoStop}%
\bibitem [{\citenamefont {Assaad}(1998)}]{Assaad1998}%
  \BibitemOpen
  \bibfield  {author} {\bibinfo {author} {\bibfnamefont {F.~F.}\ \bibnamefont
  {Assaad}},\ }\bibfield  {title} {\bibinfo {title} {{SU}(2)-spin invariant
  auxiliary field quantum {M}onte {C}arlo algorithm for {H}ubbard models},\
  }\href@noop {} {\bibfield  {journal} {\bibinfo  {journal}
  {arXiv:cond-mat/9806307v1}\ } (\bibinfo {year} {1998})}\BibitemShut {NoStop}%
\bibitem [{\citenamefont {Assaad}(2005)}]{Assaad2005}%
  \BibitemOpen
  \bibfield  {author} {\bibinfo {author} {\bibfnamefont {F.~F.}\ \bibnamefont
  {Assaad}},\ }\bibfield  {title} {\bibinfo {title} {Phase diagram of the
  half-filled two-dimensional {SU($N$)} {H}ubbard-{H}eisenberg model: {A}
  quantum {M}onte {C}arlo study},\ }\href
  {https://doi.org/10.1103/physrevb.71.075103} {\bibfield  {journal} {\bibinfo
  {journal} {Phys. Rev. B}\ }\textbf {\bibinfo {volume} {71}},\ \bibinfo
  {pages} {075103} (\bibinfo {year} {2005})}\BibitemShut {NoStop}%
\bibitem [{\citenamefont {Bhattaram}\ and\ \citenamefont
  {Khatami}(2019)}]{Bhattaram2019}%
  \BibitemOpen
  \bibfield  {author} {\bibinfo {author} {\bibfnamefont {K.}~\bibnamefont
  {Bhattaram}}\ and\ \bibinfo {author} {\bibfnamefont {E.}~\bibnamefont
  {Khatami}},\ }\bibfield  {title} {\bibinfo {title} {Lanczos-boosted numerical
  linked-cluster expansion for quantum lattice models},\ }\href
  {https://doi.org/10.1103/PhysRevE.100.013305} {\bibfield  {journal} {\bibinfo
   {journal} {Phys. Rev. E}\ }\textbf {\bibinfo {volume} {100}},\ \bibinfo
  {pages} {013305} (\bibinfo {year} {2019})}\BibitemShut {NoStop}%
\end{thebibliography}%

\end{document}